\documentclass[final,1p,times]{elsarticle}

\usepackage{comment}        
\usepackage{amsmath,amssymb,amsfonts}
\usepackage{mathabx}
\usepackage{algorithm}
\usepackage[noend]{algpseudocode}
\usepackage{graphicx}
\usepackage{textcomp}
\usepackage{xcolor}
\usepackage{booktabs}
\usepackage{hyperref}
\usepackage{xspace}
\usepackage{balance}
\usepackage{array}
\usepackage{caption}
\usepackage{subcaption}
\usepackage{amsthm}
\usepackage{enumitem} 
\newtheorem{theorem}{Theorem}[section]
\newtheorem{definition}{Definition}[section]
\newtheorem{lemma}[theorem]{Lemma}

\newcommand{\GithubCCBS}{\url{https://github.com/PathPlanning/Continuous-CBS}\xspace}
\newcommand{\GithubOurs}{\url{https://github.com/Adcombrink/Optimal-Continuous-CBS}\xspace}
\newcommand{\GithubOursOriginal}{\url{https://github.com/Adcombrink/Optimal-Continuous-CBS/tree/originalCCBS}\xspace}
\newcommand{\GithubSMTimplementation}{\url{https://github.com/Adcombrink/Optimal-Continuous-CBS/tree/master/Counterexample/ValidationWithSMT}\xspace}

\newcommand{\Real}{\mathbb{R}}
\newcommand{\epsmach}{\varepsilon_\mathit{mach}}

\newcommand{\MAPFR}{MAPF$_R$\xspace}

\newcommand{\Graph}{\mathcal{G}}
\newcommand{\Vertices}{\mathcal{V}}
\newcommand{\vertex}{v}
\newcommand{\Edges}{\mathcal{E}}

\newcommand{\MetricSpace}{\mathcal{M}}
\newcommand{\coord}{\mathit{coord}}

\newcommand{\Start}{S}
\newcommand{\Goal}{G}
\newcommand{\Actions}{\mathcal{A}}
\newcommand{\action}{a}
\newcommand{\motion}{\varphi}
\newcommand{\duration}{D}
\newcommand{\wait}{w}
\newcommand{\move}{m}
\newcommand{\objective}{\sigma}

\newcommand{\CTnode}{N}
\newcommand{\Constraints}{C}
\newcommand{\Plans}{\ensuremath{\Pi}}
\newcommand{\plan}{\pi}

\newcommand{\Solutions}{\mathcal{S}}
\newcommand{\AllSolutions}{\mathcal{S}}

\newcommand{\waitat}[1]{\Vertices(#1)\xspace}
\newcommand{\moveto}[1]{\mathit{to}(#1)\xspace}
\newcommand{\movefrom}[1]{\mathit{from}(#1)\xspace}

\newcommand{\Ic}{I^c}
\newcommand{\IcStart}{t_1^c}
\newcommand{\IcEnd}{t_2^c}
\newcommand{\IcBar}{{\color{cyan}I^I}}
\newcommand{\IcBarStart}{{\color{cyan}t_1^I}}
\newcommand{\IcBarEnd}{{\color{cyan}t_2^I}}

\newcommand{\branchingTheory}{TBR\xspace}
\newcommand{\branchingImplementation}{IBR\xspace}
\newcommand{\branchingOurs}{$\delta$-BR\xspace}

\newcommand{\trajectory}{\tau}
\newcommand{\Trajectories}{\mathcal{T}}
\newcommand{\TrajectoriesSet}{\mathbb{T}}

\newcommand{\CTNodeSequence}{\varsigma}

\newcommand{\Tset}{{\color{cyan}\mathcal{F}\!}}
\newcommand{\IntStart}{s}
\newcommand{\IntEnd}{e}

\newcommand{\OCCBS}{OC-CBS\xspace}

\newcommand{\New}[1]{{#1}\xspace}
\newcommand{\NewConflict}[1]{{#1}\xspace}
\newcommand{\NewPrecision}[1]{{#1}\xspace}

\journal{Artificial Intelligence}

\begin{document}

\begin{frontmatter}



\title{Optimal Multi-Agent Path Finding in Continuous Time}


\author[chalmers]{Alvin Combrink\corref{cor1}}
\ead{combrink@chalmers.se}
\author[chalmers]{Sabino Francesco Roselli}
\ead{rsabino@chalmers.se}
\author[chalmers]{Martin Fabian}
\ead{fabian@chalmers.se}

\cortext[cor1]{Corresponding author.}
\affiliation[chalmers]{organization={Division of Systems and Control, Department of Electrical Engineering, Chalmers University of Technology},
            city={G{\"o}teborg},
            country={Sweden}}

\begin{abstract}
    
Continuous-time Conflict Based Search (CCBS) has been widely used as an exact baseline for Continuous-time Multi-Agent Path Finding (\MAPFR),
and its correctness guarantees underpin a range of continuation methods built on top of it.
Recent work, however, has shown that CCBS's guarantees of exactness and solution completeness do not in fact hold:
optimal solutions can be removed from the search, causing the algorithm to return suboptimal solutions. 
This paper establishes
sufficient conditions for exactness and solution completeness in CCBS-style algorithms,
and introduces Optimal Continuous-time Conflict-Based Search (\OCCBS) which satisfies these conditions.
\OCCBS therefore guarantees an optimal solution on every solvable \MAPFR instance.
Experiments on benchmark problems 
show that \OCCBS remains competitive with CCBS in runtime while providing formal correctness guarantees.
Because \OCCBS is a drop-in replacement for CCBS, it also restores the theoretical guarantees of existing methods that relied on CCBS's now-invalidated correctness.
Finally, the framework and correctness criteria offer a general foundation for analyzing and designing future exact \MAPFR solvers.

\end{abstract}




\end{frontmatter}



\section{\New{Introduction}}
\label{sec:Introduction}

The problem of finding collision-free paths for agents in a shared environment is well-studied and found in a diverse range of domains. 
These include 
warehouses and factories~\cite{wurman2008coordinating, MAPF_buzzword, brown2020optimal, varambally2022mapf, brorsson_TUVE},
video games~\cite{snape2012reciprocal, Ma_Yang_Cohen_Kumar_Koenig_2017, harabor2022benchmarks}, 
traffic management systems~\cite{LiAAAI23, goenawan2025astmautonomoussmarttraffic}, 
airport surface management~\cite{morris2016planning, li2019departure}, 
search and rescue missions~\cite{abdulameer2025multi}, 
and many more.
Each agent is tasked with moving from its current position to a goal. However, the presence of other agents means they cannot simply follow the shortest path; they must avoid collisions and deadlocks while reaching their goals as quickly as possible.
The literature refers to this as the \emph{Multi-Agent Path Finding} problem (MAPF).

In its classical formulation~\cite{SurveyStern2019}, 
time is discretized and agents move on an undirected graph with unit-length edges. 
At every timestep, the agents can either wait at a vertex or move to an adjacent vertex. 
These agents are represented as size-less points, and collisions occur when two agents occupy the same vertex at the same timestep or traverse the same edge in opposite directions at the same time.
A solution moves each agent from its starting vertex to its goal vertex without colliding with other agents, and an objective function defines the solution's quality.
The two most common objective functions are the \emph{sum-of-costs} (SOC) and \emph{makespan}, respectively being either the sum or maximum over the time when agents reach their goals.
This classical formulation is known to be NP-hard to solve optimally~\cite{NPhardness}.

Early focus on solving the classical MAPF problem was placed on reducing computation and searching for sub-optimal solutions by, e.g., establishing a priority order among agents or restricting the search to a limited time-window~\cite{erdmann1987multiple, Silver}.
Shortly thereafter, exact methods such as $M^*$~\cite{Mstar}, \emph{Increasing Cost Tree Search}~\cite{ICTS}, and \emph{Conflict Based-Search} (CBS)~\cite{CBS} were introduced. 
To handle the solution space that grows exponentially with the number of agents, 
these methods avoid an exhaustive search by adopting a ``lazy'' approach of planning agents individually and addressing collisions only as they occur.
Of the three, CBS has had the largest adoption, leading to many continuations. 
To name a few, 
\emph{Meta-Agent CBS}~\cite{sharon2012meta} and \emph{Improved CBS}~\cite{boyarski2015icbs} for performance enhancements,
\emph{Enhanced CBS}~\cite{barer2014suboptimal} and \emph{CBS-Budget}~\cite{CBSB} for bounded sub-optimal solutions, 
\emph{Rolling Horizon Collision Resolution}~\cite{li2021lifelong} for \emph{lifelong MAPF}, 
and \emph{Multi-Goal CBS}~\cite{tang2024mgcbs} where agents are assigned multiple goal vertices.
The many solution methods to the classical MAPF problem and its variations are comprehensively detailed in, e.g.,~\cite{SurveyStern2019, SurveySalzman2020, SurveyMa2022}.

The classical MAPF problem has enjoyed widespread adoption within the MAPF community, 
however, 
the simplifying assumptions of discrete time, unit-length edges, and point-agent representations introduce drawbacks to its practical applicability and solution quality.
For instance, real-world environments do not always follow grid-like structures, 
and agents within the same space may have different sizes, shapes, and motion models. 
Ensuring that no two agents located at different vertices or on two different edges collide limits the resolution of the graph representation of the environment, and introducing artificial wait times to accommodate for different agent speeds and edge lengths while still moving in lockstep can inflate the objective value.

The \emph{Continuous-time Multi-Agent Path Finding} problem (\MAPFR) was introduced to relax these simplifying assumptions on classical MAPF~\cite{CCBS}.
In \MAPFR, 
edges represent potentially several trajectories within the real-world environment, taking any amount of time to complete;
agents can traverse edges at any real-valued time;
and collision detection is based on the true agent shapes rather than their position on the graph.
Besides yielding a model with higher fidelity to the real world, continuous time, being a relaxation of discrete time, can yield higher quality solutions~\cite{GRACE}. 

The \MAPFR problem is relatively young compared to its classical counterpart, yet important steps have been taken in the continuous-time direction.
The common \emph{prioritized planning} heuristic strategy is extended in~\cite{kasaura2022prioritized} to continuous-time for finding sub-optimal solutions.
Optimal solutions to a simplified version of \MAPFR, where agents traverse edges in straight lines at constant speeds, are addressed by the work in~\cite{kolarik2023multi}.
A similar formulation applied to the \emph{lifelong} MAPF problem~\cite{ma2017lifelongmultiagentpathfinding} is presented in~\cite{CPLP}.
\emph{ECBS-CT}~\cite{ECBS_CT} solves the \emph{Multi-Agent Motion Planning} problem (a generalization of \MAPFR) both optimally and bounded-suboptimally. Although the problem formulation permits arbitrary agent volumes, ECBS-CT discretizes space into cells for collision checking, which overapproximates agent geometry and introduces false-positive collision detection.

To our knowledge, the only known method to address \MAPFR in its entirety for optimal solutions is \emph{Continuous-time Conflict-Based Search} (CCBS)~\cite{CCBS}, which is claimed to be 
\NewPrecision{\emph{exact}} (only returns optimal solutions) and \emph{solution complete} (guaranteed to terminate on any solvable \MAPFR problem).
Numerous continuation works have since built upon CCBS,
such as~\cite{walker2024clique} for performance enhancements, \emph{$T$-Robust CCBS}~\cite{tan2024robust} for
delay-robust solutions,
\cite{yakovlev2024optimal}~for \emph{Any-Angle MAPF}, 
and~\cite{kulhan2023multi} with a modified version of CCBS for quadcopter drones.
Despite CCBS's widespread adoption, a recent critical assessment in~\cite{CCBS_revisit} has established that CCBS does not fulfill its guarantees.
The theoretical description of CCBS does not guarantee termination under the existence of an optimal solution (a condition for solution completeness) and the publicly available implementation\footnote{\GithubCCBS} does not guarantee the return of an optimal solution.
This is shown in~\cite{CCBS_revisit} theoretically and with counter-examples. 
To our knowledge, neither~\cite{CCBS_revisit} nor any work since proposes an exact and solution complete solver.
Thus, these results compromise the guarantees in CCBS continuation methods that rely on its soundness and solution completeness, and reveal a significant gap within the \MAPFR solution method portfolio.

This work introduces \emph{Optimal Continuous-time Conflict-Based Search} (\OCCBS), 
which finds an optimal solution in finite time to any solvable \MAPFR problem. 
We develop a framework that provides sufficient conditions for \NewPrecision{exactness} and solution completeness of CCBS-like algorithms. 
The publicly available CCBS implementation does not satisfy these sufficient conditions, which by itself does not mean that CCBS is not \NewPrecision{exact}, but it does complement prior evidence in~\cite{CCBS_revisit} of \NewPrecision{inexactness}.
\NewPrecision{
When proving that \OCCBS satisfies these conditions,
we explicitly address the fact that any practical implementation cannot represent continuous values exactly.
Rather than leaving this limitation implicit, our guarantees are with respect to this finite-precision model, and hold within it.
That is, \OCCBS finds solutions that are optimal up to the precision of the underlying representation.}
\New{
For the objective function which optimality is stated with respect to, our assumptions are mild: the objective function never decreases when an agent's plan takes longer, and bounding the objective value also bounds each single-agent plan duration. Both SOC and makespan satisfy them.}
\OCCBS is validated on a constructed counterexample and benchmark problems, showing that although CCBS is generally able to find solutions faster due to its aggressive search space pruning, \OCCBS finds solutions of equal and occasionally better quality.
Furthermore, our provided implementation demonstrates that \OCCBS's branching rule can be adopted as a drop-in replacement in existing CCBS implementations. 
Finally, the introduced framework and non-termination criteria provide analytical tools for reasoning about CCBS-like solvers and their extensions, thereby laying a foundation for the next generation of \NewPrecision{exact} and solution complete \MAPFR algorithms.

The remainder of this article is outlined as follows:
In Section~\ref{sec:Background}, we formally define the \MAPFR problem, soundness, \NewPrecision{exactness}, and solution completeness. CCBS is also introduced, along with details surrounding the findings in~\cite{CCBS_revisit}.
The framework for analyzing branching rules and CCBS's algorithmic properties is presented in Section~\ref{sec:Preliminaries}.
In Section~\ref{sec:ValidationIncompleteness}, we apply the framework to CCBS's implemented branching rule.
\OCCBS is presented in Section~\ref{sec:ModifyingCCBS}, along with \NewPrecision{exactness} and solution completeness proofs.
Section~\ref{sec:Experiments} presents empirical results.
Finally, conclusions are found in Section~\ref{sec:Conclusions}.

\section{Background}
\label{sec:Background}

In this section, 
we formally introduce the \MAPFR problem and define what we mean by \emph{soundness}, \NewPrecision{\emph{exactness}}, and \emph{solution completeness}. 
Thereafter, CCBS~\cite{CCBS} and the recent findings in~\cite{CCBS_revisit} on CCBS's guarantees are presented and elaborated.

\subsection{Problem Formulation}
\label{sec:Problem_Formulation}

A \MAPFR problem is defined in~\cite{CCBS} by a tuple $\langle \Graph, \MetricSpace, \Start, \Goal, \coord, \Actions \rangle$, where 
$\Graph=\langle \Vertices, \Edges \rangle$ represents a directed graph with vertices $\Vertices$ and edges $\Edges\subseteq \Vertices\times\Vertices$,
$\MetricSpace$ denotes a 2D metric space, 
$\Start$ and $\Goal$ respectively map agents to start and goal vertices,
$\coord:\Vertices\rightarrow\MetricSpace$ maps vertices to coordinates in $\MetricSpace$, 
and $\Actions$ comprises a finite set of \emph{move actions}. 

\begin{figure}[htbp]
    \begin{subfigure}[t]{0.48\textwidth}
        \centering
        \includegraphics[width=\linewidth]{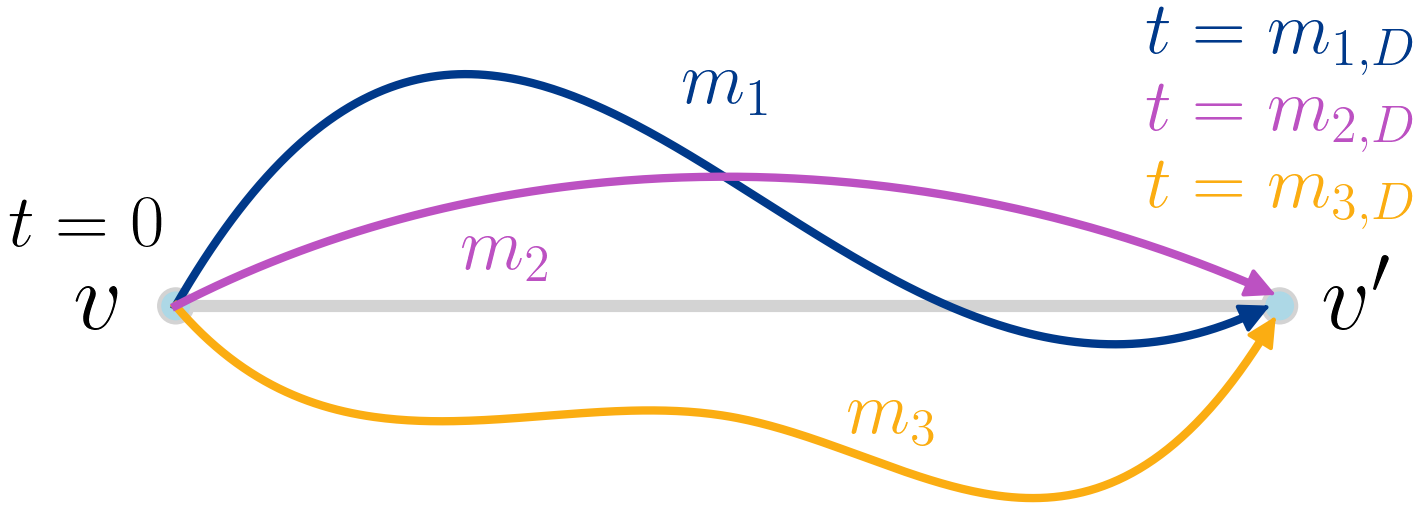}
        \caption{An edge $(\vertex,\vertex')\in\Edges$ and its associated move actions $\move_1$, $\move_2$, and $\move_3$. Each action specifies a trajectory in $\MetricSpace$ from $\vertex$ to $\vertex'$, starting at time $t=0$ and ending after potentially different durations ($\move_{1,\duration}$, $\move_{2,\duration}$, $\move_{3,\duration}$).}
        \label{fig:ExampleMotions}
    \end{subfigure}
    \hfill
    \begin{subfigure}[t]{0.48\textwidth}
        \centering
        \includegraphics[width=\linewidth]{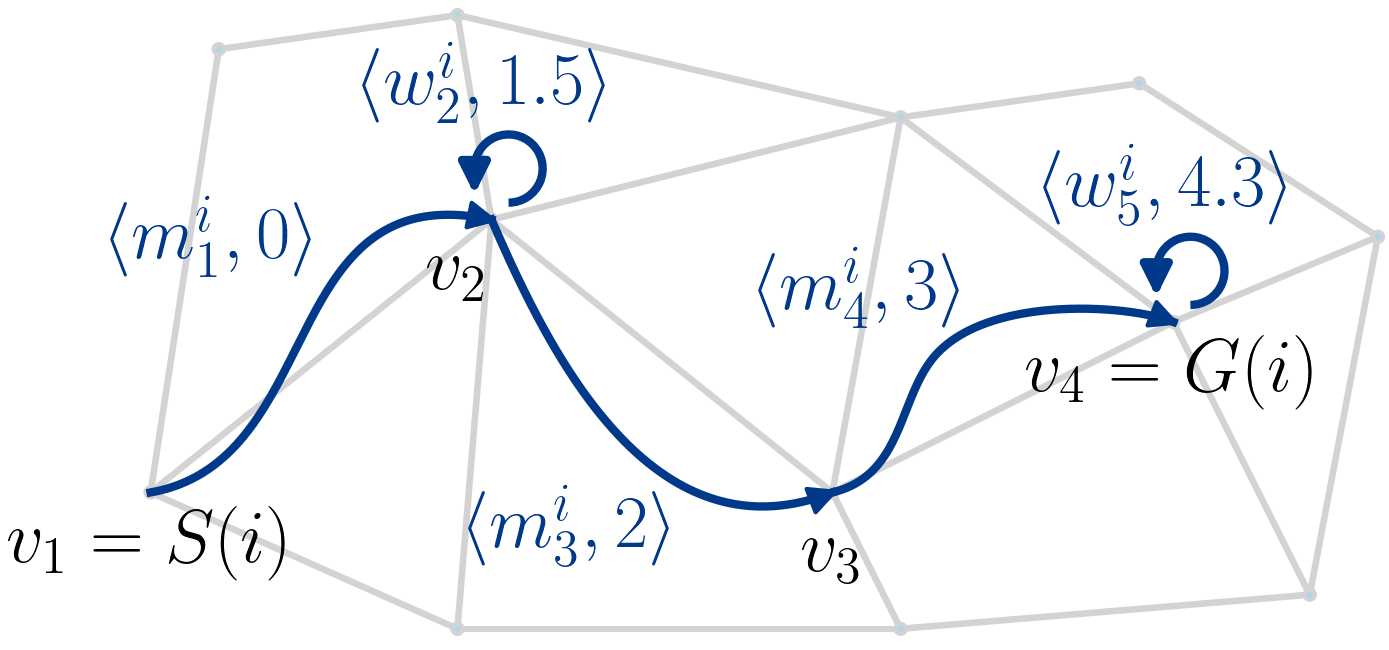}
        \caption{An illustration of a valid plan $\plan_i = \langle \langle\move_1^i, 0\rangle, \langle\wait_2^i, 1.5\rangle, \langle\move_3^i, 2\rangle, \langle\move_4^i, 3\rangle, \langle\wait_5^i, 4.3\rangle \rangle$ with action durations $\move_{1,\duration}^i=1.5$, $\wait_{2,\duration}^i=0.5$, $\move_{3,\duration}^i=1$, $\move_{4,\duration}^i=1.3$, $\wait_{5,\duration}^i=\infty$.}
        \label{fig:ExamplePlan}
    \end{subfigure}
\end{figure}

A move action $\move=(\move_\motion, \move_\duration)$ offers one of possibly several ways for an agent to traverse an edge $(\vertex, \vertex')\in\Edges$. 
The \emph{motion function} $\move_\motion: [0, \move_\duration]\rightarrow\MetricSpace$ 
\New{maps time to metric space, starting at $\vertex$ and ending at $\vertex'$;}
that is, $\move_\motion(0) = \coord(\vertex)$ and $\move_\motion(\move_\duration)=\coord(\vertex')$.
\New{We refer to $\move_\duration$ as the motion function's \emph{duration},
and require $\move_\duration \in \mathbb{R}^{>0}$ for every move action in $\Actions$.}
Let the source and target vertices be denoted $\movefrom{\move}=\vertex$ and $\moveto{\move}=\vertex'$, respectively. 
Each edge in $\Edges$ admits a finite number of move actions, providing multiple options for an agent to traverse an edge. 
Figure~\ref{fig:ExampleMotions} illustrates this concept through edge $(\vertex, \vertex')$ and three associated move actions $\move_1$, $\move_2$, and $\move_3$.
While all move actions start and end at the same vertices, their trajectories (by $\move_{1,\motion}$, $\move_{2,\motion}$, $\move_{3,\motion}$) and durations ($\move_{1,\duration}$, $\move_{2,\duration}$, $\move_{3,\duration}$) may be distinct.

In addition to move actions, there are also \emph{wait actions} not included in $\Actions$.
Wait actions allow for agents to wait at vertices, which may be necessary to e.g.\ avoid colliding with other agents.
A wait action $\wait = (\wait_\motion,\wait_\duration)$ maps to a single vertex $\vertex\in\Vertices$ such that its motion function $\wait_\motion: [0,\wait_\duration] \rightarrow \coord(\vertex)$ with
\New{$\wait_D\in\Real^{>0}\cup\{\infty\}$}.
We denote \New{$\waitat{\wait}=\movefrom{\wait}=\moveto{\wait}=\vertex$}.

A \emph{timed action} $\langle \action, t \rangle$ (with $\action$ either a move or wait action) specifies that an agent executes $\action$ starting at time 
\New{$t\in\Real^{\geq 0}$}, such that the motion function $\action_\motion$ is shifted by $t$.
Specifically, an agent executing $\langle \action, t \rangle$ is located at $\action_\motion(t'-t)$ during $t'\in[t, t+\action_\duration]$.
The difference between an action and a timed action is therefore that an action specifies how an agent moves and the duration to do so, while a timed action additionally specifies when the movement occurs.

Agents can traverse the graph toward their goals by executing multiple timed actions, one after the other.
A \emph{plan} $\plan_i$ for an agent $i$ is a finite sequence of timed actions,
\begin{equation}
    \plan_i = \langle \langle\action_1^i, t_1^i\rangle, \langle \action_2^i, t_2^i\rangle, \dots, \langle \action_n^i, t_n^i\rangle\rangle.
\end{equation}
For a plan $\plan_i$ to be \emph{valid}, 
it must start at time $0$ ($t_1^i=0$), 
start and end at $i$'s start and goal vertices ($\movefrom{\action_1^i}=\Start(i)$ and $\moveto{\action_n^i}=\Goal(i)$),
and every successive timed action must start when and where the previous ends,

\begin{equation*}
    \left(t_k^i=t_{k-1}^i + \action_{k-1,\duration}^i\right) \wedge  \left( \moveto{\action_{k-1}^i} = \movefrom{\action_k^i}\right) \quad \forall k=2,..,n.
\end{equation*}
Additionally, the final action $\action_n^i$ is 
an \emph{infinite wait action} with duration $\action_{n,D}^i = \infty$
to capture that agents always remain idle indefinitely at their last location. 
The duration of $\plan_i$ is $\plan_{i,\duration} = \sum_{k=1}^{n-1} \action_{k,\duration}^i$, naturally excluding the final infinite wait action.

Figure~\ref{fig:ExamplePlan} illustrates an example of a valid plan $\plan_i$, taking agent $i$ from $\Start(i)$ to $\Goal(i)$.
The first timed move action $\langle \move_1^i=\langle\move_{1,\motion}^i, 1.5\rangle, 0\rangle$ specifies that the agent begins at $t=0$ to move according to the motion function $\move_{1,\motion}^i$ which has duration $1.5$. 
This is followed by the timed wait action $\langle\wait_2^i=\langle\wait_{2,\motion}^i, 0.5\rangle, 1.5 \rangle$ specifying a wait at $\waitat{\wait_2^i} = \vertex_2$ from $t=1.5$ until $t=1.5+0.5=2.0$. 
Thereafter, the agent executes $\langle\move_3^i=\langle\move_{3,\motion}^i, 1\rangle, 2\rangle$ to arrive at $\vertex_3$ at $t=3$, then $\langle\move_4^i=\langle\move_{4,\motion}^i, 1.3\rangle,3\rangle$ to arrive at $\vertex_4$ at $t=4.3$. 
Finally, the timed wait action $\langle\wait_5^i=\langle\wait_{5,\motion}^i, \infty\rangle, 4.3\rangle$ specifies that $i$ waits at $\vertex_4$ indefinitely from $t=4.3$.
Naturally, the plan $\plan_i$ illustrated in the example represents merely one among countless valid plans; infinitely many alternative wait actions could be employed at any of the vertices, different move actions could be selected for the same edges, or an entirely different path through the graph could be taken.


\NewConflict{
    Typically in MAPF with volumetric agents, conflicts occur due to geometric overlaps between the agents~\cite{CCBS, li2019departure, kasaura2022prioritized};
    we adopt a more general definition that, we argue, subsumes CCBS.
    Specifically, we consider that two agents may conflict without committing to the underlying reason.
    A \emph{conflict} $\langle \langle \action^i, t^i \rangle, \langle \action^j, t^j \rangle \rangle$ means that as two agents $i$ and $j$ execute their respective timed actions $\langle \action^i, t^i \rangle$ and $\langle \action^j, t^j \rangle$,
    they will at some time jointly be in a state that is considered unacceptable---they will be \emph{in conflict}.
    Unacceptable states need not involve geometric overlap.
    In aviation, for instance, rotor wash, jet blast, and wake turbulence can cause unacceptable interactions without involving direct contact between vehicles. 
    Restricting conflicts to geometric collisions thus excludes such cases from being modeled.
    However, we limit the problem to include only \emph{pairwise} conflicts, that is, conflicts that depend on exactly two agent states.
    We therefore do not consider, for instance, the combined downwash of two drones conflicting with a third, where considering each agent pair individually would not yield a conflict.
    The CCBS and \OCCBS frameworks operate on conflicts as abstract objects; their search procedures and correctness guarantees rely only on the fact that a reported conflict is genuine and that the branching rules resolve it, not on the geometric nature of the conflict itself.
    Nonetheless, throughout this article we use geometric-based conflicts in our examples and naming conventions to match the typical MAPF setting.
    A geometric-based conflict occurs when the intersection volume $V$ between two agents' volumes is non-zero, $V>0$. Specifically, two agents tangentially touching means that $V=0$ and is therefore not a conflict.}

For a conflict $\langle \langle \action^i, t^i \rangle, \langle \action^j, t^j \rangle \rangle$,
if $\action^i$ and $\action^j$ are both move actions then it is a \emph{move-move} conflict, if instead $\action^i$ is a move action and $\action^j$ is a wait action then it is a \emph{move-wait} conflict.
By convention, the move action in a move-wait conflict is stored in the first position. 
Note that a \emph{wait-wait} conflict can only arise after a move-move or move-wait (assuming agents start in conflict-free positions).
Therefore, we sufficiently consider only move-move and move-wait conflicts.
Two plans $\plan_i$ and $\plan_j$ conflict if there exists a conflict $\langle \langle \action^i, t^i \rangle, \langle \action^j, t^j \rangle \rangle \in \plan_i\times\plan_j$.

A \emph{joint plan} $\Plans$ is a set containing one valid plan for each agent, 
a \emph{solution} is a joint plan that does not contain any two conflicting plans, and
an optimal solution $\Plans^*$ is a solution that minimizes an objective function $\objective$ over all solutions.
The \emph{sum-of-costs}~(SOC) $\objective(\Plans) = \sum_{\plan\in\Plans} \plan_\duration$ (sum over plan durations) and \emph{makespan} $\objective(\Plans) = \max_{\plan\in\Plans} \plan_\duration$ (maximum over plan durations) are the two most common objective functions in the MAPF context.  

\New{
We assume that $\objective(\Plans)$ is decomposable into a function $f$ over the individual agent plan durations, 
$\objective(\Plans) = f((\plan_{\duration})_{\plan\in\Plans})$, so that a joint plan enters the objective function only through the durations of the plans contained within it. 
We place two further assumptions on $\objective$:
\textbf{Monotonicity}: $f$ is non-decreasing in each argument. That is, for joint plans $\Plans$ and $\Plans'$, with $\plan_{i,\duration} \leq \plan'_{i,\duration}$ for every agent $i$, it holds that $\objective(\Plans) \leq \objective(\Plans')$. 
\textbf{Properness}: For every $c<\infty$ there exists $B(c)<\infty$ such that $f(x) \leq c$ implies $x_i \leq B(c)$ for every agent $i$.
In particular, $\objective(\Plans) \leq c$ implies $\plan_{i,\duration} \leq B(c)$ for every agent $i$.

Monotonicity says that no agent taking longer can improve the objective; properness says that a bounded objective value admits only bounded plan durations. 
The two are independent and are used in different parts of our analysis. 
Both SOC and makespan satisfy them, with $B(c)=c$ in either case since all plan durations are non-negative.
All proofs and guarantees in this work hold for any objective satisfying Monotonicity and Properness.
}


\subsection{Algorithmic Guarantees}
\label{sec:AlgGuarantees}

In this paper, we say that a solution algorithm to the \MAPFR problem is 
\begin{itemize}
    \item \emph{sound} if it only returns solutions;
    \item \NewPrecision{\emph{exact} if it only returns optimal solutions\footnote{We use \emph{exact} since \emph{optimal} (used in~\cite{CCBS}) typically refers to an algorithm's time and space complexity, not the quality of the solutions that it produces.}};
    \item \emph{solution complete} if it is guaranteed to terminate with a solution on any solvable \MAPFR problem.
\end{itemize}

Note that exactness implies soundness, and that
\emph{completeness} guarantees the return of a solution if one exists or the report of no solution otherwise, while the weaker \emph{solution completeness} (as defined above) only guarantees the former. 

\NewPrecision{
    We consider algorithms implemented under finite-precision arithmetic.
    Continuous values can therefore only be represented up to a certain numerical precision.
    Crucially, this does not mean that time is discretized; 
    rather, it means that an algorithm under this setting operates over a continuous domain whose values are approximated to finite precision.
    Exact continuous-time solutions are therefore unattainable in practice, and the above theoretical guarantees are stated with respect to this finite-precision setting.
    The precise consequences of this assumption are discussed in Section~\ref{sec:Preliminaries:finite-precision}.
}

\subsection{Continuous-time Conflict Based Search}
\label{sec:Background:CCBS}

CCBS is argued in~\cite{CCBS} to be an exact and solution-complete \MAPFR solver for minimal SOC.
Based on CBS, CCBS simultaneously builds and searches a binary constraint tree (CT).

To explain the CT and how it is searched, we begin by briefly introducing CCBS's underlying planner \emph{Constrained Safe Interval Path Planning} (CSIPP) which is covered in more detail in Section~\ref{sec:Proposed:termination:CSIPP}. 
CSIPP---a variant of \emph{Safe Interval Path Planning} (SIPP)~\cite{SIPP}---computes a minimal-duration valid plan for a single agent that satisfies a given set of constraints $C$.
In SIPP, an $A^*$~\cite{A_star} search is performed in the vertex-safe interval state space, with edges in $\Edges$ providing transitions from one state to the next, and safe intervals defining when an edge may be traversed or a vertex may be occupied.
In CSIPP, constraints define the safe intervals and move actions provide transitions between states.

In the CT, a node $\CTnode$ is associated with a set of constraints $\CTnode_\Constraints$ and a joint plan $\CTnode_\Plans$.
The plans in $\CTnode_\Plans$ are computed using CSIPP given the constraint set $\CTnode_\Constraints$, 
such that every plan $\plan_i\in\CTnode_\Plans$ is valid under $\CTnode_\Constraints$ and has minimal duration.
The CCBS search starts at the CT root node $\CTnode^R$ with the empty constraint set $\CTnode^R_\Constraints = \varnothing$. 
Since $\CTnode^R_\Constraints$ is empty, every plan $\plan_i\in\CTnode^R_\Plans$ is unconstrained (apart from being valid) and therefore has the shortest possible duration over all valid plans for agent $i$.
At every iteration of CCBS, a node $\CTnode$ minimizing $\objective(\CTnode_\Plans)$ over all unexpanded nodes is selected for expansion.
If $\CTnode_\Plans$ is a solution (i.e., conflict-free), then it is returned as an optimal solution.
Otherwise, a \emph{branching rule} is applied to $\CTnode$ to spawn two child nodes.
To do so, an arbitrary conflict in $\CTnode_\Plans$ is selected (say between agents $i$ and $j$) and used to construct a pair of constraints $\langle c_i, c_j \rangle$. These constraints $c_i$ and $c_j$ respectively constrain the plans of agents $i$ and $j$ to avoid this specific conflict.
The child nodes $\CTnode^i$ and $\CTnode^j$ each inherit one of the constraints in addition to all constraints at $\CTnode$: $\CTnode^i_\Constraints = \CTnode_\Constraints \cup \{c^i\}$ and $\CTnode^j_\Constraints = \CTnode_\Constraints \cup \{c^j\}$.
The plans in $\CTnode_\Plans^i$ and $\CTnode_\Plans^j$ are then computed using CSIPP to satisfy $\CTnode_\Constraints^i$ and $\CTnode_\Constraints^j$, respectively.

\begin{figure}
    \centering
    \begin{subfigure}[b]{0.45\linewidth}
        \centering
        \includegraphics[width=0.75\textwidth]{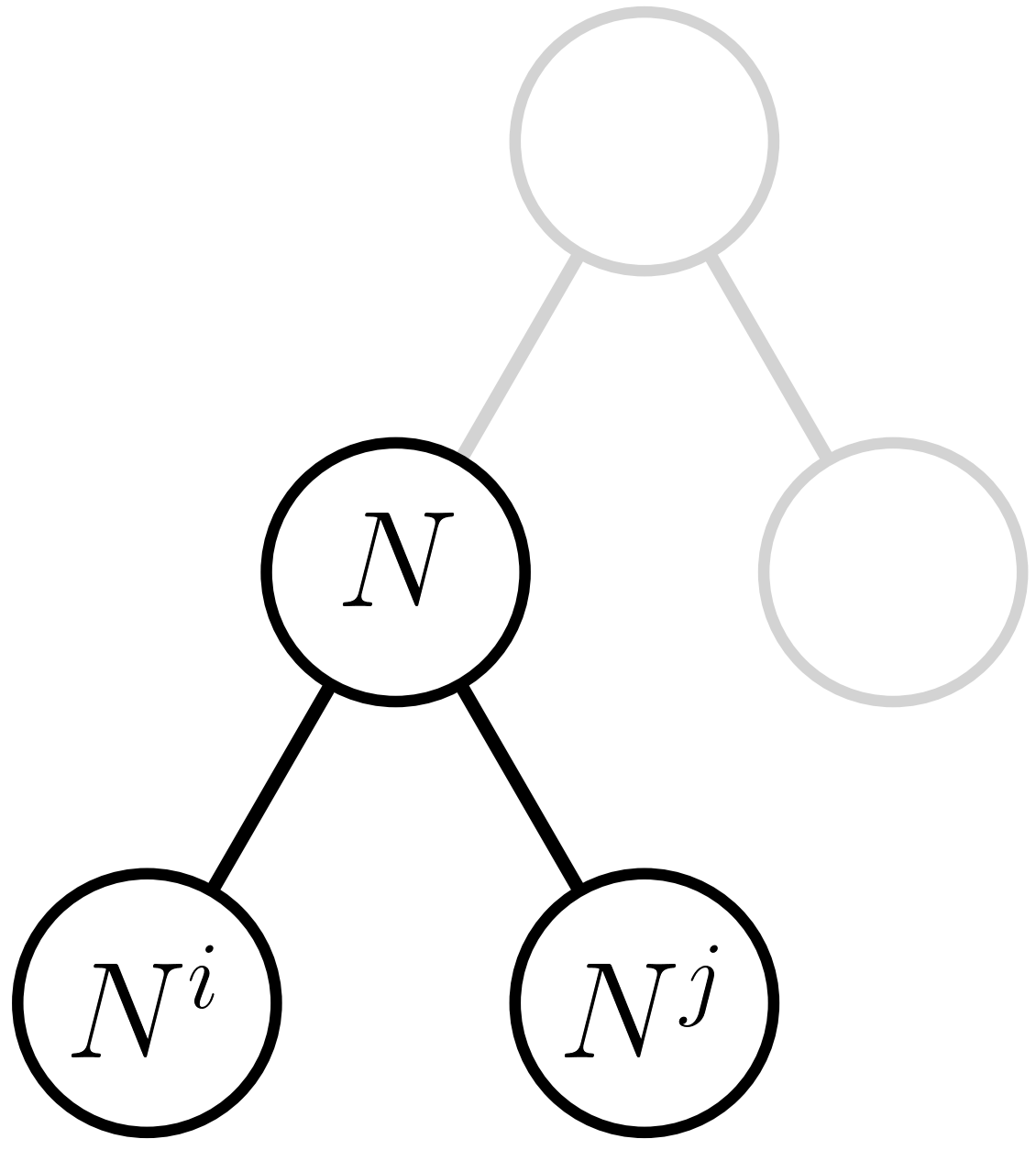}
        \caption{The CT node $\CTnode$ where a conflict is detected in $\CTnode_\Plans$, and its two children $\CTnode^i$ and $\CTnode^j$.}
        \label{fig:ExampleCT_1}
    \end{subfigure}
    \hfill
    \begin{subfigure}[b]{0.45\linewidth}
        \centering
        \includegraphics[width=1\textwidth]{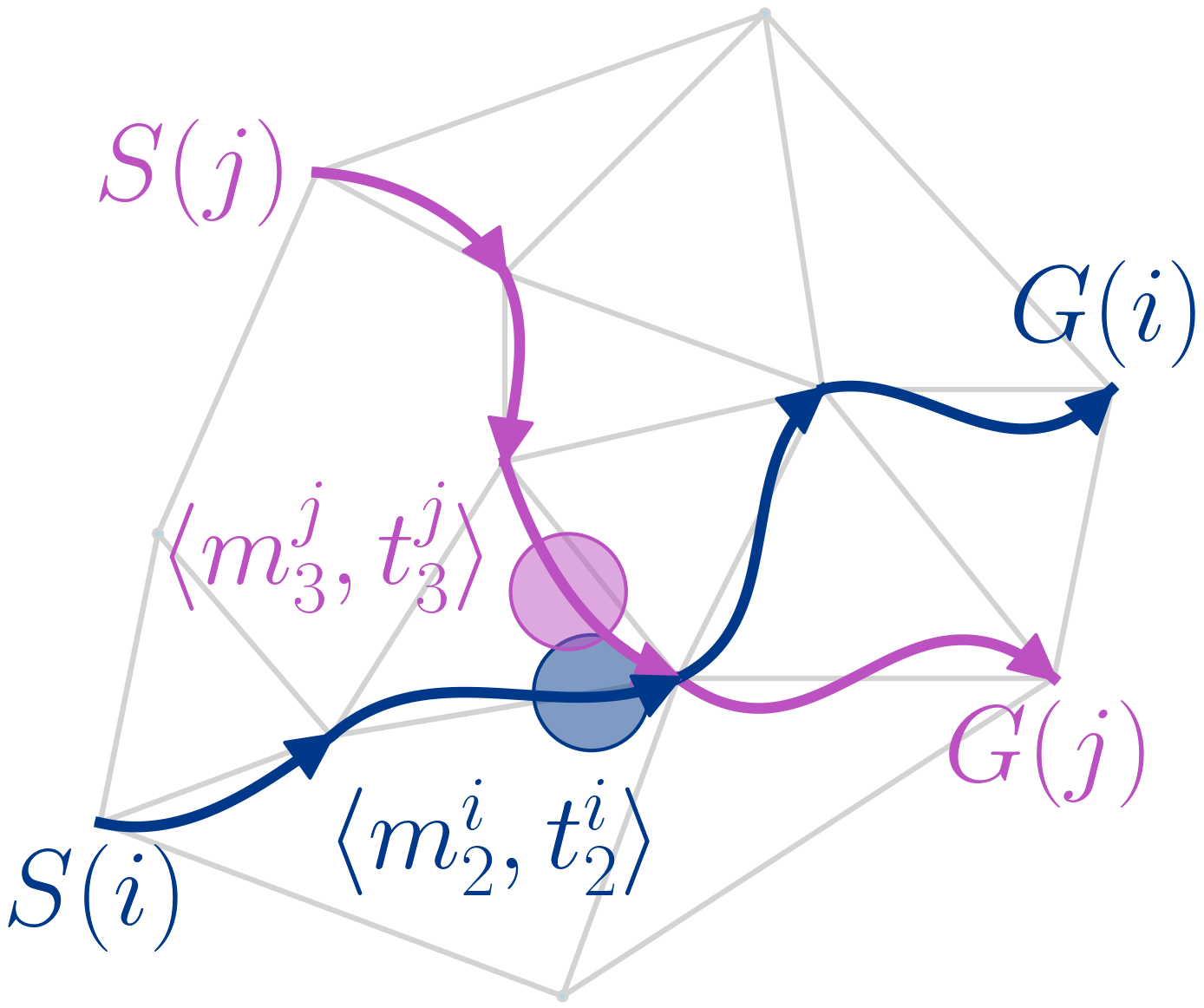}
        \caption{The joint plan $\CTnode_\Plans = \left\{\plan_i, \plan_j\right\}$ contains a conflict $\langle \langle\move^i_2, t^i_2\rangle, \langle\move^j_3, t^j_3\rangle \rangle$ between agents $i$ and $j$.}
        \label{fig:ExampleCT_2}
    \end{subfigure}
    \vfill
    \begin{subfigure}[b]{0.45\linewidth}
        \centering
        \includegraphics[width=1\textwidth]{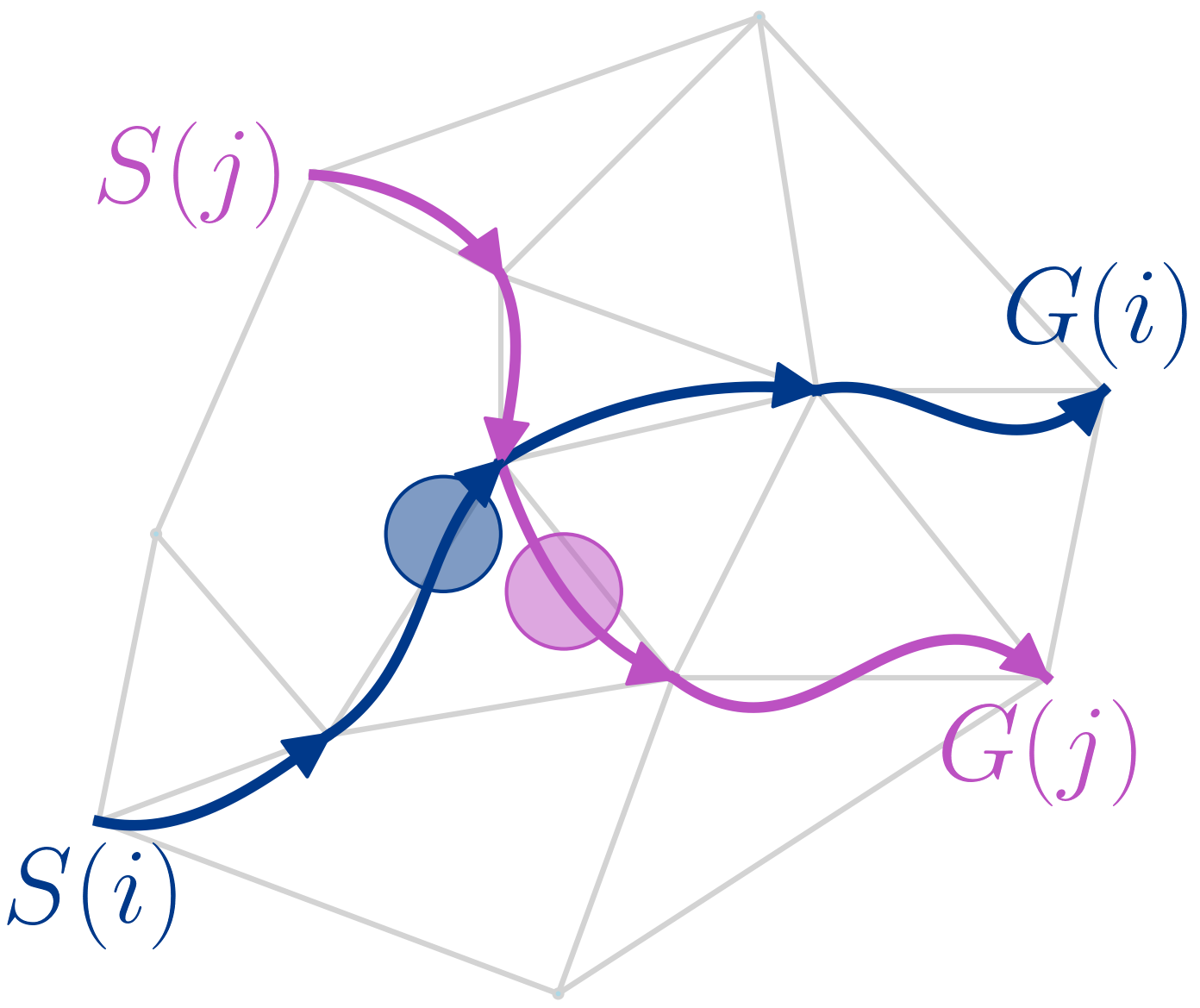}
        \caption{The joint plan $\CTnode^i_\Plans$. Additional constraints on agent $i$ are added to $\CTnode^i_\Constraints$ to specifically avoid the collision detected at $\CTnode$. A new plan $\plan_i\in\CTnode^i_\Plans$ satisfying $\CTnode^i_\Constraints$ is then computed, where $i$ traverses another path in the graph.}
        \label{fig:ExampleCT_3}
    \end{subfigure}
    \hfill
    \begin{subfigure}[b]{0.45\linewidth}
        \centering
        \includegraphics[width=1\textwidth]{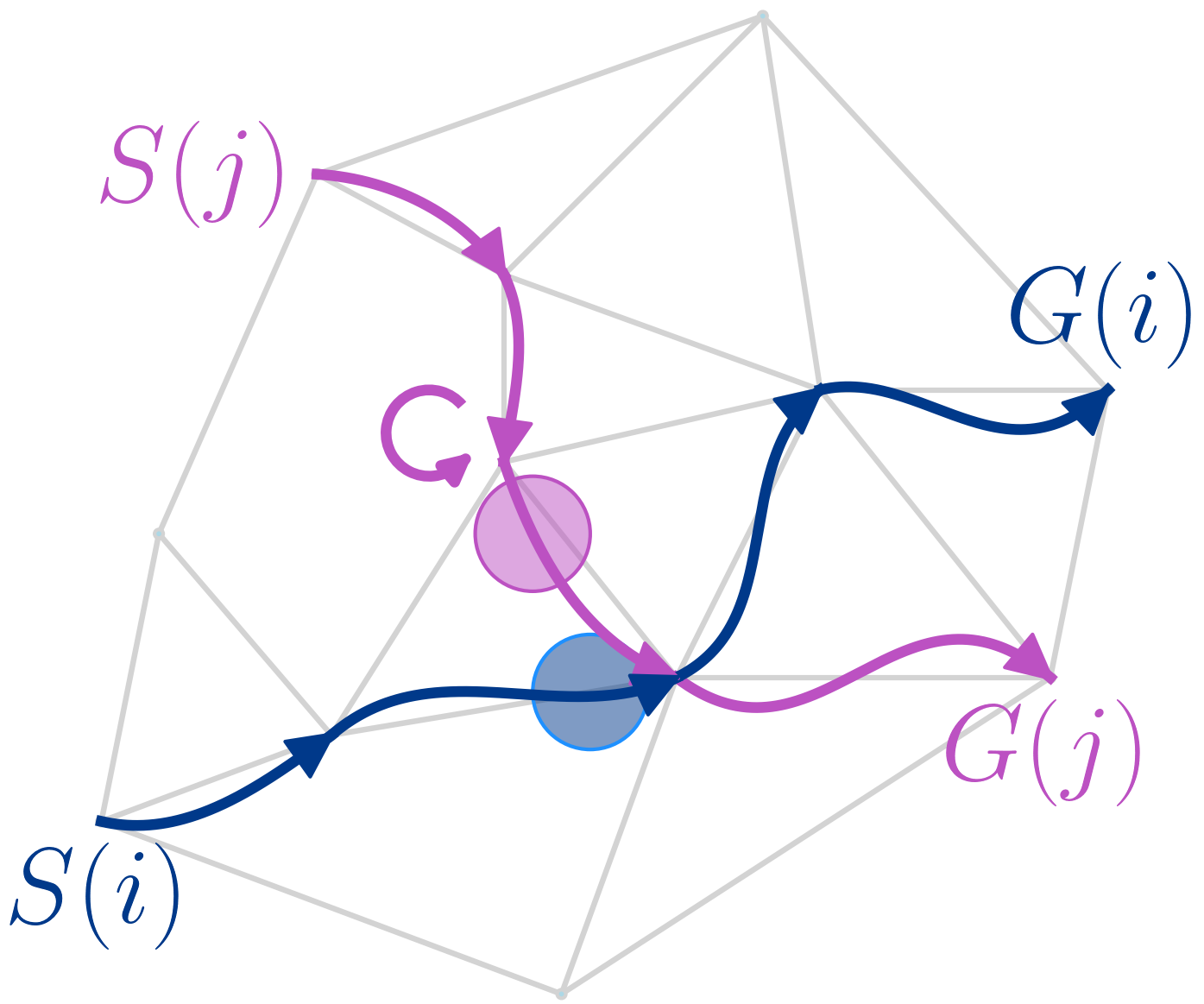}
        \caption{The joint plan $\CTnode^j_\Plans$. Additional constraints on agent $j$ are added to $\CTnode^j_\Constraints$ to specifically avoid the collision detected at $\CTnode$. A new plan $\plan_j\in\CTnode^j_\Plans$ satisfying $\CTnode^j_\Constraints$ is then computed, where $j$ waits for some time at a node.}
        \label{fig:ExampleCT_4}
    \end{subfigure}
    \caption{An example of a branching rule applied to a conflict $\langle \langle\move^i_2, t^i_2\rangle, \langle\move^j_3, t^j_3\rangle \rangle$ at CT node $\CTnode$. For agent $i$ ($j$), a child node $\CTnode^i$ ($\CTnode^j$) is spawned with additional constraints on $i$ ($j$) to avoid this specific conflict. A new plan is then computed which satisfies these constraints.}
    \label{fig:ExampleCT}
\end{figure}

We clarify the above by introducing an example in Figure~\ref{fig:ExampleCT}, which will later be revisited in more detail. 
Only two agents $i$ and $j$ are considered in this example.
At the CT node $\CTnode$ (Figure~\ref{fig:ExampleCT_1}) a conflict is detected in the joint plan $\CTnode_\Plans=\left\{\plan_i, \plan_j\right\}$, illustrated in Figure~\ref{fig:ExampleCT_2}.
Specifically, $i$ and $j$ collide while respectively executing  $\langle\move^i_2, t^i_2\rangle$ and $\langle\move^j_3, t^j_3\rangle$, making $\langle \langle\move^i_2, t^i_2\rangle, \langle\move^j_3, t^j_3\rangle \rangle$ a conflict.
Based on this conflict, a branching rule is used to create a pair of constraints $\langle c^i, c^j \rangle$. 
We will explore these constraints in more detail later; 
for now it is sufficient to know that $c^i$ constrains when $i$ may execute $\move^i_2$, and $c^j$ constrains when $j$ may execute $\move^j_3$, to avoid this specific conflict.
Consider the first child node $\CTnode^i$ (Figure~\ref{fig:ExampleCT_1}) with constraints $\CTnode^i_\Constraints = \CTnode_\Constraints \cup \{c^i\}$.
Using CSIPP, the plans in $\CTnode^i_\Plans$ are computed to satisfy $\CTnode^i_\Constraints$. 
In practice, it is sufficient to only compute $i$'s plan since only the constraints on $i$ have changed, while all other agents can inherit their plans from $\CTnode$. 
Figure~\ref{fig:ExampleCT_3} illustrates $\CTnode^i_\Plans$, where agent $i$ now takes a different path through the graph.
Similarly, the other child node $\CTnode^j$ has constraints $\CTnode^j_\Constraints = \CTnode_\Constraints \cup \{c^j\}$, and a new plan for $j$ is computed using CSIPP. The joint plan $\CTnode^j_\Plans$ is illustrated in Figure~\ref{fig:ExampleCT_4}, where $j$ now waits for some time at a vertex before continuing.
In this particular example, the resulting joint plans at the child nodes happen to be conflict-free and are therefore both solutions. 
In general, however, the child nodes' joint plans may still contain conflicts, requiring further CCBS iterations until a solution is found.

\NewPrecision{If it can be correctly determined that a selected CT node's joint plant $\CTnode_\Plans$ is indeed a solution (at which point that solution will be returned) then CCBS is sound since it only returns solutions.}
The \NewPrecision{exactness} and solution completeness proofs of CCBS in~\cite{CCBS} rely on the CT satisfying the following properties, where $\CTnode$ is a parent to $\CTnode^i$ and $\CTnode^j$ and $\Solutions(\CTnode)$ is the set of all solutions satisfying $\CTnode_\Constraints$ and therefore reachable from $\CTnode$:
\begin{align}
    \Solutions\left(\CTnode\right) &= \Solutions\left(\CTnode^i\right) \cup \Solutions\left(\CTnode^j\right), \label{eq:CCBSproof:sound} \\
    \objective\left( \CTnode_\Plans \right) &\leq \min\left( \objective( \CTnode^i_\Plans ), \objective( \CTnode^j_\Plans ) \right).\label{eq:CCBSproof:optimal}
\end{align}

Property~\eqref{eq:CCBSproof:sound} ensures that no solutions are removed when branching a node.
At the CT root $\CTnode^R$, since $\CTnode^R_\Constraints=\varnothing$ it holds that all solutions to a \MAPFR problem are contained in $\Solutions(\CTnode^R)$. 
If property~\eqref{eq:CCBSproof:sound} holds then $\Solutions(\CTnode^R) = \Solutions(\CTnode^i)\cup\Solutions(\CTnode^j)$ (with $\CTnode^i$ and $\CTnode^j$ being children of $\CTnode^R$) which recursively implies that all solutions are reachable from the CT root.
Since the search starts at $\CTnode^R$, all solutions to a \MAPFR problem (including all optimal solutions) are reachable in the search and therefore can be found.
Since it is the branching rule that determines $\Solutions(\CTnode^i)$ and $\Solutions(\CTnode^j)$ (by the constraint pair that it constructs), it is the branching rule that determines if property~\eqref{eq:CCBSproof:sound} is satisfied.

Property~\eqref{eq:CCBSproof:optimal} is automatically satisfied by CSIPP.
Since plans in $\CTnode_\Plans$ are computed using CSIPP given constraints $\CTnode_\Constraints$, every plan $\plan_i\in\CTnode_\Plans$ is the shortest-duration valid plan over all valid plans satisfying $\CTnode_\Constraints$.
For all nodes $\CTnode'$ in the sub-tree rooted at $\CTnode$ we know that $\CTnode_\Constraints\subseteq\CTnode'_\Constraints$ since constraints are only added during a branching, never removed. 
Adding constraints can never result in shorter-duration plans. 
Therefore it holds for all agents $i$ that $\plan_{i,\duration}\leq\plan_{i,\duration}'$ where $\plan_{i,\duration}\in\CTnode_\Plans$ and $\plan_{i,\duration}'\in\CTnode_\Plans'$,
\New{and the monotonicity assumption on $\objective$ applied to these component-wise inequalities yields $\objective(\CTnode_\Plans) \leq \objective(\CTnode'_\Plans)$,
thereby satisfying property~\eqref{eq:CCBSproof:optimal}}.

In this work, we focus our attention on two branching rules: 
the \emph{theoretical branching rule}~(\branchingTheory) used in the correctness proofs in~\cite{CCBS}, and the \emph{implemented branching rule} (\branchingImplementation) found in the publicly available reference CCBS implementation\footnote{\GithubCCBS.}.

\subsubsection{The Theoretical Branching Rule}
\label{sec:CCBS:TBR}
The \branchingTheory uses the abstract definition of actions, encompassing move and wait actions alike, such that move-move and move-wait conflicts are addressed using the same procedure.
Given a conflict $\langle \langle \action^i, t^i \rangle, \langle \action^j, t^j \rangle \rangle$, the \branchingTheory finds the earliest time for each agent to execute its action without colliding with the other agent.
That is, the earliest time $t^i_u > t^i$ for agent $i$ to execute $\langle \action_i, t^i_u\rangle$ without colliding with $j$ executing $\langle \action^j, t^j \rangle$ is found.
The interval $[t^i, t^i_u)$ is denoted the \emph{unsafe interval} since $\langle \action_i, t\rangle$ for any $t\in[t^i, t^i_u)$ certainly conflicts with $\langle \action^j, t^j \rangle$.
This unsafe interval is used to define the \emph{motion constraint} $c_i = \langle i, \action^i, [t^i, t^i_u) \rangle$ specifying that $i$ is forbidden from executing $\langle \action^i, t \rangle$ for any $t\in[t^i, t^i_u)$.
Likewise for agent $j$, the unsafe interval $[t^j, t^j_u)$ is found and the motion constraint $c_j = \langle j, \action_j, [t^j, t^j_u) \rangle$ is defined.
Letting the conflict be detected at CT node $\CTnode$, such that $\langle \langle \action^i, t^i \rangle, \langle \action^j, t^j \rangle \rangle \in \plan_i\times\plan_j$ with $\plan_i, \plan_j \in \CTnode_\Plans$, the \branchingTheory uses the constraint pair $\langle c_i, c_j \rangle$ to branch on $\CTnode$.
The CT using the \branchingTheory is shown in~\cite{CCBS} to satisfy properties~\eqref{eq:CCBSproof:sound} and~\eqref{eq:CCBSproof:optimal}, from which it is concluded that CCBS is sound, \NewPrecision{exact}, and solution complete.

Let us now revisit the example in Figure~\ref{fig:ExampleCT} where the conflict $\langle\langle\move_2^i,t_2^i\rangle, \langle\move_3^j, t_3^j\rangle\rangle$ was detected. 
From this conflict, the motion constraints $c^i = \langle i, \move_2^i, [t_2^i, t^i_u) \rangle$ and $c^j = \langle j, \move_3^j, [t_3^j, t^j_u) \rangle$ are formed. 
At node $\CTnode^i$ with $\CTnode^i_\Constraints = \CTnode_\Constraints \cup \{c^i\}$, agent $i$ is no longer permitted to execute $\langle \move_2^i, t\rangle$ for $t\in[t_2^i, t^i_u)$. Therefore, an alternative plan is found where $i$ avoids executing $\move_2^i$ entirely.
Similarly at node $\CTnode^j$ where $\CTnode^j_\Constraints = \CTnode_\Constraints \cup \{c^j\}$, agent $j$ is no longer permitted to execute $\langle \move_3^j, t\rangle$ for $t\in[t_3^j, t^j_u)$. Therefore, an alternative plan is found where $j$ waits long enough to eventually execute $\move^j_3$ at $t_u^j$.

\subsubsection{The Implemented Branching Rule}

The \branchingImplementation differs from the \branchingTheory by how move-wait conflicts $\langle \langle \move^i, t^i \rangle, \langle \wait^j, t^j \rangle \rangle$ are branched on.
To avoid confusion between the \branchingImplementation and the \emph{described} implementation in~\cite{CCBS}, we first explain the latter and then how the \branchingImplementation differs.
With the \branchingTheory, the motion constraint $\langle j, \wait^j, [t^j, t^j_u) \rangle$ is created for the move-wait conflict above.
Such a motion constraint forbids $j$ from starting the execution of $\wait^j$ at any time in $[t^j, t^j_u)$. That is, $\langle \wait^j, t \rangle$ is forbidden for all $t\in[t^j, t^j_u)$.
However, in~\cite{CCBS}, CSIPP is described to implement such a motion constraint as a \emph{vertex constraint} $\langle j, \waitat{\wait^j}, [t^j, t^j_u) \rangle$, forbidding $j$ from occupying the vertex $\waitat{\wait^j}$ during $[t^j, t^j_u)$.
Observe that such a vertex constraint is more constraining than the motion constraint;
the motion constraint forbids \emph{specifically} $\wait^j$ from being executed \emph{starting} at any time in $[t^j, t^j_u)$, 
while the vertex constraint forbids 
\emph{every wait action $\wait$ at the same vertex} ($\waitat{\wait}=\waitat{\wait^j}$) from being executed if it \emph{overlaps with} $[t^j, t^j_u)$, i.e., if $[t, t+\wait_\duration]\cap[t^j, t^j_u) \neq \varnothing$.
Therefore, already here we see a dissonance between the \branchingTheory and the described implementation.

Similar to the described implementation, the \branchingImplementation also implements vertex constraints. 
However, the \branchingImplementation uses the \emph{intersection interval} $\IcBar$ instead of the unsafe interval $[t^j, t^j_u)$.
\New{
$\IcBar$ is formally defined in Section~\ref{sec:Preliminaries:intervals} below, but informally, $\IcBar$}
is the time interval during which agent $i$ while executing $\langle \move^i, t^i \rangle$ would collide with agent $j$ waiting for all time at $\waitat{\wait^j}$.
Thus, the \branchingImplementation creates a vertex constraint $\langle j,\waitat{\wait^j}, \IcBar \rangle$.
The intersection interval $\IcBar$ is not necessarily the same as the unsafe interval $[t^j, t^j_u)$ since it does not take into account when $j$ begins executing its wait action.
\New{
For example, if $\IcBar = [\IcBarStart, \IcBarEnd)$ and agent $j$ begins waiting at $\IcBarStart < t^j < \IcBarEnd$, 
then the unsafe interval is $[t^j, \IcBarEnd) \neq \IcBar$.}
Besides this, timed move actions in both move-move and move-wait conflicts are handled in the same way as in the \branchingTheory, resulting in the motion constraint $\langle i, \action^i, [t^i, t^i_u) \rangle$.

Therefore, the \branchingImplementation uses the constraint pair $\langle \langle i,\move^i,[t^i,t^i_u)\rangle,\langle j,\move^j,[t^j,t^j_u)\rangle\rangle$ for move-move conflicts and $\langle \langle i,\move^i,[t^i,t^i_u)\rangle,\langle j,\waitat{\wait^j}, \IcBar \rangle\rangle$ for move-wait conflicts.
No proofs of \NewPrecision{exactness} and solution completeness are provided for CCBS under this branching rule.

\subsection{A Critique on CCBS}
The work in~\cite{CCBS_revisit} questions the correctness of CCBS. 
The claims are two-fold: 
\begin{enumerate}
    \item Under the \branchingTheory, CCBS does not guarantee termination;\label{claim:noterm}
    \item Under the \branchingImplementation, CCBS violates property~\eqref{eq:CCBSproof:sound}.\label{claim:violation}
\end{enumerate}
At its core, the identified issue 
concerns 
how wait actions are handled when resolving move-wait conflicts.
For Claim~\ref{claim:noterm}, it is highlighted in~\cite{CCBS_revisit} that under the \branchingTheory on a wait action $\langle\wait^i, t^i\rangle$ (giving rise to a motion constraint $\langle i, \wait^i, [t^i, t^i_u)\rangle$), only one wait duration $\wait^i_\duration$ out of infinitely many real-valued durations is removed from the search.
For instance, a wait action $\wait = (\wait^i_\motion, \wait^i_\duration+\epsilon)$ with $\epsilon$ arbitrarily small is still permitted during $[t^i, t^i_u)$.
Therefore, despite satisfying property~\eqref{eq:CCBSproof:sound} by ensuring that any optimal solution is reachable from the CT root, there may exist infinitely many nodes between the root and such a solution.
Hence, CCBS under the \branchingTheory is not proven to always terminate after a finite number of iterations despite the existence of a solution and therefore lacks a proof of solution completeness.
Essentially,~\cite{CCBS_revisit} identifies that properties~\eqref{eq:CCBSproof:sound} and~\eqref{eq:CCBSproof:optimal} are not sufficient to guarantee solution completeness.

For Claim~\ref{claim:violation}, counter examples are provided in~\cite{CCBS_revisit} 
demonstrating situations where the \branchingImplementation on move-wait conflicts removes valid solutions from the search (violating property~\eqref{eq:CCBSproof:sound}), and consequently causing CCBS to return sub-optimal solutions. 
Thus, CCBS under the \branchingImplementation does not always return optimal solutions and is therefore not \NewPrecision{exact}.


\section{Preliminaries}
\label{sec:Preliminaries}

This section establishes the theoretical foundations for analyzing branching rules in CCBS.
We first introduce a decomposed form of CCBS constraints and extend to this the notion of \emph{sound constraint pairs}. 
We then define a \emph{sound branching rule} as one that preserves all solutions during node expansion and prove that a branching rule is sound if and only if its underlying constraint pair is sound.
\New{Using these definitions, we show that a sound branching rule, together with a lower-bound property on CT node objective values that CSIPP guarantees, is sufficient to ensure that a CCBS-like algorithm is \NewPrecision{exact}.}
If such a sound algorithm is also guaranteed to terminate, then it is solution complete.
Finally, we formally introduce the collision interval and intersection interval.


\subsection{Sound Branching}

A definition for a \emph{sound} pair of constraints $\langle c_1, c_2 \rangle$ is provided in~\cite{CCBS, atzmon2018robust}. 
Let the set $\AllSolutions$ contain all solutions to a given \MAPFR problem, and let $c(\Plans)$ be true if a joint plan $\Plans$ satisfies a constraint $c$, false otherwise.
\begin{definition}[Sound pair of constraints~\cite{CCBS, atzmon2018robust}]
    \label{def:sound_constraints}
    For a given \MAPFR problem with solutions $\AllSolutions$, a pair of constraints $\langle c_1, c_2 \rangle$ is sound if
    \begin{equation}
        c_1(\Plans) \vee c_2(\Plans), \quad \forall \Plans\in\AllSolutions.
    \end{equation}
\end{definition}
Throughout this article, we shall consider a decomposed form of the motion and vertex constraints by describing them as \emph{constraint sets} comprised of \emph{forbidding constraints}.
These constraint forms are equivalent, however, the decomposed form allows us to analyze constraint pairs used in branching rules more explicitly than when using motion and vertex constraints.
A forbidding constraint $\langle i, \action, t\rangle$ forbids $i$ from executing the specific timed action $\langle \action, t \rangle$.
A motion constraint $c = \langle i, \action^i, [t^i, t^i_u) \rangle$,
which forbids $i$ from executing action $\action_i$ starting at any time $t\in[t^i, t^i_u)$,
is equivalent to the dense constraint set
\begin{equation*}
    C=\left\{ \langle i, \action^i, t \rangle \mid t\in[t^i, t^i_u) \right\}
\end{equation*}
such that $c(\Plans) \Leftrightarrow C(\Plans)$ where $C(\Plans) = \bigwedge_{c'\in C} c'(\Plans)$. 
We denote this equivalence by $c \Leftrightarrow C$.
Similarly, a vertex constraint $c=\langle i, \vertex, I \rangle$ forbidding $i$ from occupying $\vertex$ at any time $t\in I$ is equivalent to $C=\left\{\langle i, \wait, t \rangle \mid \waitat{\wait} = \vertex, [t, t+\wait_\duration]\cap I \neq \varnothing \right\}$.

We now provide a corresponding soundness definition for a pair of constraint sets, and then define soundness of a branching rule. 
Then, in Lemma~\ref{lemma:sound_branching}, we show that a branching rule using a pair of constraint sets is sound if and only if the constraint set pair is sound. 
\begin{definition}[Sound pair of constraint sets]
    \label{def:sound_constraint_sets}
    For a given \MAPFR problem, a pair of constraint sets $\langle C_1, C_2 \rangle$ is sound if $\forall\langle c_1, c_2 \rangle \in C_1 \times C_2$, $\langle c_1, c_2 \rangle$ is sound.
\end{definition}

\begin{definition}[Sound branching rule]
    \label{def:sound_branching_mechanism}
    A branching rule applied to a CT node $\CTnode$ to spawn children $\CTnode^1$ and $\CTnode^2$ is sound if property~\eqref{eq:CCBSproof:sound} always holds: $S(\CTnode) = S(\CTnode^1)\cup S(\CTnode^2)$.
\end{definition}
\begin{lemma}[Equivalence of sound branching and sound constraint set pair]
    \label{lemma:sound_branching}
    Let a branching rule applied to a CT node $\CTnode$ use a constraint set pair $\langle C_1, C_2 \rangle$ to spawn two children $\CTnode^1$ and $\CTnode^2$, such that $\CTnode^1_\Constraints = \CTnode_\Constraints \cup \left\{C_1\right\}$ and $\CTnode^2_\Constraints = \CTnode_\Constraints \cup \left\{C_2\right\}$.
    The branching rule is sound iff $\langle \Constraints_1, \Constraints_2 \rangle$ is sound.
\end{lemma}
\begin{proof}
    To prove the branching rule's soundness, we must show that property~\eqref{eq:CCBSproof:sound}, $\Solutions(\CTnode) = \Solutions(\CTnode^1) \cup \Solutions(\CTnode^2)$, is always satisfied.
    For each child $N^i$, $i=1,2$, the solutions permitted under the constraints $\CTnode^i_\Constraints$ are
    \begin{align*}
        \Solutions(\CTnode^i) &= \left\{ \Plans\in\AllSolutions \mid c(\Plans), c\in\CTnode^i_\Constraints \right\} \\
        &= \left\{ \Plans\in\AllSolutions \mid c(\Plans),  c\in\CTnode_\Constraints \cup \left\{C_i\right\} \right\} \\
        &= \left\{ \Plans\in\Solutions(\CTnode) \mid C_i(\Plans) \right\}.
    \end{align*}
    Therefore, for the branching rule to be sound it must hold for every solution $\Plans\in\Solutions(\CTnode)$ that $C_1(\Plans) \vee C_2(\Plans)$ since then $\Solutions(\CTnode) = \Solutions(\CTnode^1) \cup \Solutions(\CTnode^2)$.
    For every $\Plans\in\Solutions(\CTnode)$, consider two cases: 
    \begin{enumerate}
        \item $C_1(\Plans) \vee C_2(\Plans)$ holds, then we immediately know that  $\Plans\in\Solutions(\CTnode^1)\cup\Solutions(\CTnode^2)$;
        \item otherwise, assume without loss of generality that $\exists c_1\in C_1: \neg c_1(\Plans)$. 
    \end{enumerate}
    For the latter case, we start by assuming that $\langle C_1, C_2 \rangle$ is sound. 
    By Definition~\ref{def:sound_constraint_sets}, every pair $\langle c_1, c_2 \rangle \in C_1 \times C_2$ is sound, meaning $c_1(\Plans)\vee c_2(\Plans)$ holds.
    Since we assumed $\neg c_1(\Plans)$, it follows that $c_2(\Plans)$ must be true.
    This reasoning applies to every pair $\langle c_1, c_2 \rangle \in C_1 \times C_2$, such that $\forall c_2\in C_2: c_2(\Plans)$ holds.
    This means that $C_2(\Plans)$ is true and therefore that $\Plans\in\Solutions(\CTnode^2)$.
    Therefore, if $\langle C_1, C_2 \rangle$ is sound then it holds for all $\Plans\in\Solutions(\CTnode)$ that $\Plans\in\Solutions(\CTnode^1)\cup\Solutions(\CTnode^2)$. Thus, if $\langle C_1, C_2 \rangle$ is sound then the branching rule is sound.

    
    Assume instead that $\langle C_1, C_2 \rangle$ is unsound,
    then by Definition~\ref{def:sound_constraint_sets} there exists an unsound pair $\langle c_1, c_2 \rangle\in C_1 \times C_2$. 
    Since $\langle c_1, c_2 \rangle$ is unsound, there exists a valid solution $\Plans$ such that $\neg(c_1(\Plans) \vee c_2(\Plans))$, implying $\neg C_1(\Plans)$ and $\neg C_2(\Plans)$.
    Therefore, $\Plans\not\in\Solutions(\CTnode^1)\cup\Solutions(\CTnode^2)$.
    Since $\Solutions(\CTnode)$ can contain such a solution $\Plans$, it does not always hold for all $\Plans\in\Solutions(\CTnode)$ that $\Plans\in\Solutions(\CTnode^1)\cup\Solutions(\CTnode^2)$. 
    Thus, if $\langle C_1, C_2 \rangle$ is unsound then the branching rule is unsound.
\end{proof}


Finally, we put branching rules back into the context of the CCBS algorithm by 
showing in Theorem~\ref{theorem:soundness} that CCBS is \NewPrecision{exact} when using a sound branching rule.
Additionally, we show in Theorem~\ref{theorem:solution-completeness} that CCBS is solution complete if it is both sound and guaranteed to terminate.

\New{

Property~\eqref{eq:CCBSproof:optimal} relates a CT node to the joint plans of its descendants, and therefore establishes that the minimum objective value over all unexpanded CT nodes is non-decreasing over the search. 
It does not, however, relate a node's objective value to the \emph{solutions} reachable from that node, which is what a best-first argument requires:
without such a bound, a node's objective value could exceed that of a solution in its own sub-tree, and best-first search would then expand a sub-optimal solution first.
We therefore make explicit the admissibility property 
\begin{equation}
    \objective(\CTnode_\Plans) \leq \objective(\Plans),\quad \forall \Plans \in \Solutions(\CTnode). \label{eq:OCCBS_property:admissibility}
\end{equation}
Property~\eqref{eq:OCCBS_property:admissibility} corresponds to the lower-bound lemma underlying the optimality proof of CBS~\cite{CBS}, where the cost of a CT node is shown to lower-bound the cost of every solution the node permits. 
Like property~\eqref{eq:CCBSproof:optimal}, this is satisfied by CSIPP under $\objective$'s monotonicity:
every plan $\plan_i$ of a solution $\Plans\in\Solutions(\CTnode)$ is valid and satisfies $\CTnode_\Constraints$, and is therefore no shorter than the corresponding CSIPP plan in $\CTnode_\Plans$;
applying $\objective$'s monotonicity assumption to these component-wise inequalities gives $\objective(\CTnode_\Plans) \leq \objective(\Plans)$.
Properties~\eqref{eq:CCBSproof:optimal} and~\eqref{eq:OCCBS_property:admissibility} are independent, since the joint plans compared in~\eqref{eq:CCBSproof:optimal} need not be solutions.
Property~\eqref{eq:OCCBS_property:admissibility} is what carries exactness in Theorem~\ref{theorem:soundness}.
}

\begin{theorem}[\NewPrecision{Exactness}]
    \label{theorem:soundness}
    CCBS is \NewPrecision{exact} when using a sound branching rule.
\end{theorem}
\begin{proof}
    \New{
    Let $\Plans^*$ be an optimal solution with $\objective(\Plans^*) = \objective^*$. 
    Since the branching rule is sound,
    $\Solutions(\CTnode) = \Solutions(\CTnode^1) \cup \Solutions(\CTnode^2)$ at every branching,
    and $\Solutions(\CTnode^R)$ contains all solutions.
    By induction over iterations, there is therefore at every iteration an unexpanded CT node $\CTnode$ with $\Plans^* \in \Solutions(\CTnode)$,
    and by property~\eqref{eq:OCCBS_property:admissibility} it satisfies $\objective(\CTnode_\Plans) \leq \objective(\Plans^*) = \objective^*$.
    Since CCBS expands a node minimizing $\objective(\CTnode_\Plans)$ over all unexpanded nodes, the node selected at that iteration also has objective value at most $\objective^*$.
    CCBS terminates only when the selected node's joint plan is conflict-free,
    that is, when it is a solution.
    The returned solution therefore has objective value at most $\objective^*$, and since $\objective^*$ is the optimal objective value, the two are equal. 
    Hence CCBS returns an optimal solution.
    }

\end{proof}

\begin{theorem}[Solution completeness]
    \label{theorem:solution-completeness}
    CCBS is solution complete if it is sound and guaranteed to terminate.
\end{theorem}
\begin{proof}
    CCBS's soundness ensures that all solutions are reachable from the CT root.
    CCBS has exactly one termination condition: when it selects a CT node $\CTnode$ such that $\CTnode_\Plans$ is a solution, then $\CTnode_\Plans$ is returned. 
    CCBS continues expanding nodes until this condition is met.
    Since all solutions are reachable from the CT root where CCBS begins its search, and CCBS is guaranteed to terminate, it must eventually encounter and return a solution.
    Therefore, CCBS is solution complete if it is sound and guaranteed to terminate.
\end{proof}

Note that Theorem~\ref{theorem:soundness} shows that CCBS is \NewPrecision{exact} if the branching rule is sound. 
However, it does not show that CCBS is \NewPrecision{not exact} if the branching rule is unsound.
Thus, the branching rule being sound is a \emph{sufficient but not necessary} condition for CCBS to be \NewPrecision{exact}.

\subsection{Collision Interval and Intersection Interval}
\label{sec:Preliminaries:intervals}

We now formally define the \emph{collision interval} and \emph{intersection interval}, discuss the consequences of continuous-time MAPF algorithms being run on machines with \NewPrecision{finite-precision}, and end with comments on practical implementations for computing these intervals for \NewConflict{geometric-based conflict detection.}

Given a conflict $\langle\langle \action^i, t^i\rangle,\langle \action^j, t^j\rangle\rangle$, the collision interval $\Ic$ is the \emph{first maximally connected} time interval during which agents $i$ and $j$ are \NewConflict{in conflict} while executing their respective timed actions.
\New{
    By \emph{maximally connected} we mean that $\Ic=[\IcStart, \IcEnd)$ cannot be extended in either direction without including times when $i$ and $j$ do not conflict (i.e., they do not conflict at time $t<\IcEnd$ arbitrarily close to $\IcEnd$ or at exactly $\IcEnd$), 
    and by \emph{first} maximally connected we mean that $i$ and $j$ are not in conflict for all time $t<\IcStart$ or at time $t=\IcEnd$.
}
\NewConflict{With geometric-based conflicts, $\Ic$ would then be the first maximally connected time interval during which the agents' volumes intersect.}
Such a collision interval must exist for any move-move or move-wait conflict, otherwise 
there would be no
conflict.

\New{
An intersection interval $\IcBar$ exists only for a move-wait conflict $\langle\langle\move^i, t^i\rangle, \langle\wait^j, t^j\rangle\rangle$.
It is a time interval during which the agents are in conflict when $i$ executes $\langle\move^i, t^i\rangle$ and $j$ remains stationary at $\waitat{\wait^j}$ \emph{for all time}. 
That is, the intersection interval is like the collision interval, except that instead of the waiting agent occupying its vertex during only a limited time window, it instead waits for all time $t\in(-\infty, \infty)$.
Put another way, if $j$ was a stationary object at its vertex, $\IcBar$ records the time interval during which $i$ conflict with this object while executing its action.
If multiple such intervals exist, then $\IcBar$ is the first such interval to intersect the collision interval $\Ic$.}


\New{
    Restricting attention to a single intersection interval at a time is sufficient, even when multiple such intervals exist.
    Intersection intervals can be handled independently, one at a time. Provided that the branching rule is sound (Definition~\ref{def:sound_branching_mechanism}), i.e., $S(\CTnode) = S(\CTnode^1) \cup S(\CTnode^2)$, resolving a conflict for one interval partitions the solution space without discarding any valid solutions. Repeating this process ensures that all intersection intervals are eventually addressed, and thus all conflicts are resolved.  
}

\begin{figure}[H]
    \centering
    \begin{subfigure}[t]{0.35\linewidth}
        \centering
        \includegraphics[width=1\textwidth]{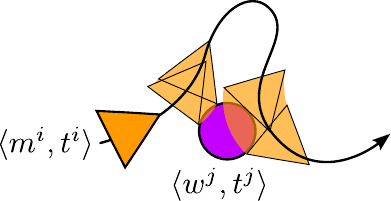}
        \caption{\New{The volumes of agent $i$ (triangle) and agent $j$ (circle) intersect two times while $i$ executes $\langle \move^i, t^i \rangle$ and $j$ waits at $\waitat{\wait^j}$.}}
        \label{fig:IntersectionExample_geometry}
    \end{subfigure}
    \hfill
    \begin{subfigure}[t]{0.6\linewidth}
        \centering
        \includegraphics[width=1\textwidth]{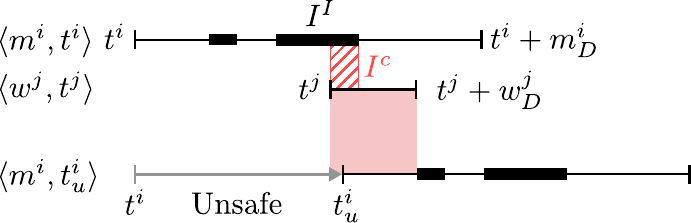}
        \caption{The timeline shows each timed action, 
        the geometric intersection intervals (black bars along the move action's timeline), 
        the collision interval $\Ic$ (red), 
        the intersection interval $\IcBar$,
        and agent $i$'s unsafe interval $[t^i, t^i_u)$.}
        \label{fig:IntersectionExample_Timeline}
    \end{subfigure}
    \caption{A geometric and timeline representation of a conflict $\langle \langle \move^i, t^i \rangle, \langle \wait^j, t^j \rangle \rangle$.}
    \label{fig:IntersectionExample}
\end{figure}

\New{
Figure~\ref{fig:IntersectionExample} provides a concrete example of a move-wait conflict and its unsafe, collision, and intersection intervals. 
The geometric representation in Figure~\ref{fig:IntersectionExample_geometry} shows how agent $i$'s volume while executing $\langle \move^i,t^i\rangle$ intersects the volume of agent $j$ located at $\waitat{\wait^j}$ during two maximally connected intervals.
Figure~\ref{fig:IntersectionExample_Timeline} shows the timeline representation of each agent's timed actions. 
Unlike the geometric representation, the timeline shows \emph{when} $j$ actually occupies $\waitat{\wait^j}$. 
Considering that $j$ must be located at $\waitat{\wait^j}$ for the agents to collide, the collision interval $\Ic$ is shown as the first maximally connected time interval during which $j$ is located at $\waitat{\wait^j}$ and $i$ intersects $j$'s volume.
From the definition of an intersection interval, the intersection interval $\IcBar$ is in this case the second of the two geometric intervals. 
Finally, $t^i_u$ is the first time after $t^i$ when $i$ executing $\langle \move^i, t^i_u\rangle$ does not collide with $j$'s, forming $i$'s unsafe interval $[t^i, t^i_u)$. A collision will occur if $i$ executes $m^i$ at any time in $[t^i, t^i_u)$.
}

It will be useful to make explicit that $\Ic$ is the intersection of the time interval during which $j$ occupies $\waitat{\wait^j}$ with the intersection interval $\IcBar$,
\begin{equation}
    \label{eq:ic_icbar_intervals}
    \Ic = [t^j, t^j + \wait^j_\duration] \cap \IcBar.
\end{equation}

\subsubsection{\NewPrecision{Finite-precision Considerations}}
\label{sec:Preliminaries:finite-precision}


\NewPrecision{

    From Section~\ref{sec:AlgGuarantees}, we assume that \OCCBS is implemented under finite-precision arithmetic.
    All continuous values are thus represented up to a fixed numerical precision $\epsmach$, as is standard in any practical implementation.
    This setting introduces a fundamental trade-off.
    A conservative treatment of conflict avoidance, biased toward safety, may reject solutions that rely on distinctions finer than the available precision.
    An aggressive treatment, biased toward permissiveness, may admit conflicts whose duration falls within numerical uncertainty, leading to unsoundness.

    Intervals of extremely short duration (on the order of a few $\epsmach$) cannot be reliably distinguished from rounding artifacts. 
    If such an interval represents the time of a potential conflict,
    classifying it as conflict or non-conflict becomes numerically ambiguous.
    We resolve this ambiguity by treating such intervals as non-conflicts; temporal overlaps that are too small to be resolved under finite precision are not regarded as conflicts.
    
    Accordingly, we restrict the method to intervals that are resolvable under finite precision. 
    All unsafe intervals, collision intervals, and intersection intervals are assumed to be sufficiently large relative to $\epsmach$. 
    Specifically for \OCCBS, we assume that
    \begin{equation}
        \epsmach \ll \min\left(\gamma |I|, (1-\gamma) |I| \right)
        \label{eq:finite_precision_intervals}
    \end{equation}
    where $\gamma\in(0,1)$ is a user-defined parameter introduced in Section~\ref{sec:proposal:delta_BR}, and $\ll$ denotes \emph{larger by several orders of magnitude}.
    This condition ensures that both subintervals induced by $\gamma$ remain well above the numerical precision, so that their evaluation is not affected by rounding effects.
    As a result, a pair $\langle \langle \action^i, t^i\rangle,\langle\action^j, t^j\rangle\rangle$ is not considered a conflict if its associated collision or unsafe interval fails to satisfy~\eqref{eq:finite_precision_intervals}, since such overlaps are indistinguishable from numerical noise or can be eliminated by perturbations below machine precision. 

    This requirement couples the choice of $\gamma$ to the minimum interval length that the algorithm is expected to handle: $\gamma$ cannot be selected independently of the smallest meaningful interval.
    However, to give an example showing that this limitation has little practical relevance, consider the following case:
    under double-precision floating-point arithmetic ($\epsmach \approx 2.22 \times 10^{-16}$), a time-horizon of $1000$~seconds yields an absolute precision of approximately $2.22 \times 10^{-13}$.
    Choosing $\gamma=0.9$ and requiring the smaller portion $(1-\gamma)|I|$ to exceed this precision by six orders of magnitude (thereby limiting numerical imprecision) implies $|I| \gtrsim 2.22 \times 10^{-6}$ seconds, or $2.22$~microseconds.
    This lower bound depends on available numerical precision rather than the absolute time scale. 

    Under the above assumption, a property that will be important for the proofs in Section~\ref{sec:ModifyingCCBS} is that all relevant intervals $I$ are bounded away from zero by an amount larger than zero. 
    That is, $\exists c$ such that $0<\epsmach \ll c<|I|$. 
    
}

\subsubsection{\New{Practical Geometric-Based Conflict Detection}}

\New{
\NewConflict{With geometric-based conflict detection},
unsafe intervals, collision intervals, and intersection intervals arise from the geometry of agent volumes and motions in $\MetricSpace$, and represent time intervals during which agent volumes intersect.
For simple cases, such as circular agents moving along straight-line trajectories with constant velocity or acceleration, these intervals can be computed analytically~\cite{walker2019collision}.
More general motions and agent geometries can be handled using standard collision detection methods from computational geometry~\cite{RealtimeCollision, quinlan1994efficient, jimenez2004efficient}, which are typically numerical but accurate up to a specified precision.

We make two assumptions regarding conflict detection.
First, we assume the existence of a procedure that, given two actions, determines their intersection intervals up to a precision consistent with the finite-precision model described in Section~\ref{sec:Preliminaries:finite-precision}.
In particular, intervals shorter than the precision induced by $\epsmach$ are not considered meaningful, and conflict-free guarantees are inherently limited by this numerical precision as well as modeling and sensing errors.
Second, we assume consistency: for a given pair of actions, the conflict detection procedure always returns the same result up to this precision. This ensures that conflicts are handled consistently across the search.

Geometric collision detection for general motions and arbitrary agent volumes can be computationally expensive~\cite{walker2019collision}.
To mitigate this, intersection intervals (and corresponding intervals for move-move conflicts) can be precomputed and stored, as done in~\cite{kasaura2022prioritized}.
This also ensures consistency, since stored values remain fixed. 
Given an intersection interval, unsafe intervals can be derived directly.
Specifically, for a conflict $\langle\langle \action^i, t^i\rangle,\langle \action^j, t^j\rangle\rangle$, there exists an intersection interval $\IcBar = [\IcBarStart, \IcBarEnd)$ such that $\IcBarStart \leq t^i \leq \IcBarEnd$. 
The earliest time $t^i_u>t^i$ at which $\langle \action^i, t^i_u\rangle$ no longer conflicts with $\langle \action^j, t^j\rangle$ is then given by $t^i_u = \IcBarEnd$.

The publicly available implementation of CCBS assumes circular agents and straight line traversals at constant speed. 
Since our implementation builds upon theirs, the experiments in Section~\ref{sec:Experiments} also make this same assumption. However, instead of a binary search with a specified precision, we use the analytical interval computation without precomputations. 
}



\section{Analyzing CCBS's Implemented Branching Rule}
\label{sec:ValidationIncompleteness}

\New{
Although~\cite{CCBS_revisit} already shows that the \branchingImplementation is not a sound branching rule and therefore that CCBS does not satisfy the sufficient (but not necessary) conditions for \NewPrecision{exactness} established in Theorem~\ref{theorem:soundness}, we apply here the framework from Section~\ref{sec:Preliminaries} as an illustrative example and as complementary evidence.
}


Lemma~\ref{lemma:epsilon_collision_free} provides intermediate results that are used in Theorem~\ref{theorem:CCBS_unsound} to show that the \branchingImplementation is not sound.

\begin{lemma}[Conflict timing for time-shifted move actions]
    \label{lemma:epsilon_collision_free}
    Let $\langle \langle \move^i, t^i \rangle, \langle \wait^j, t^j \rangle\rangle$ not be a conflict. 
    If $i$ instead executes $\langle \move^i, t^i - \epsilon \rangle$ for some $\epsilon>0$, then a conflict with $j$ executing $\langle \wait^j, t^j \rangle$ can only occur within the time interval $[t^j + \wait^j_\duration - \epsilon, t^j + \wait^j_\duration]$.
\end{lemma}
\begin{proof}
    Since $j$ is stationary at $\waitat{\wait_j}$ throughout $[t^j, t^j + \wait^j_\duration]$ and the move action $\move^i$ is identical in both $\langle \move^i, t^i\rangle$ and $\langle \move^i, t^i - \epsilon\rangle$, we can analyze when a conflict occurs by comparing agent $i$'s motion in both cases.
    When agent $i$ executes $\langle \move^i, t^i - \epsilon\rangle$, its motion during the interval $[t^j, t^j + \wait^j_\duration - \epsilon]$ corresponds exactly to its motion during $[t^j + \epsilon, t^j + \wait^j_\duration]$ when executing $\langle \move^i, t^i\rangle$. 
    Since no conflict occurred during $[t^j + \epsilon, t^j + \wait^j_\duration]$ when executing $\langle \move^i, t^i\rangle$, and agent $j$ remains stationary at the same location $\waitat{\wait^j}$, no conflict occurs during $[t^j, t^j + \wait^j_\duration - \epsilon]$ in the shifted case.
    Therefore, any conflict between $i$ and $j$ executing $\langle \move^i, t^i - \epsilon\rangle$ and $\langle \wait^j, t^j \rangle$ must occur during the remaining time interval $[t^j + \wait^j_\duration - \epsilon, t^j + \wait^j_\duration]$. 
\end{proof}
\begin{figure}
    \centering
    \includegraphics[width=0.75\linewidth]{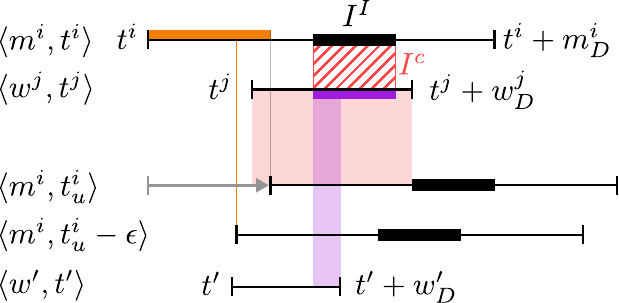}
    \caption{Timeline of a conflict $\langle \langle \move^i, t^i \rangle, \langle \wait^j, t^j \rangle \rangle$ with intersection interval $\IcBar$ and the collision interval $\Ic$. The resulting constraints forbid $i$ from executing $\move^i$ starting at $[t^i, t^i_u)$ (orange) and $j$ from occupying $\waitat{\wait^j}$ during $\IcBar$ (purple). 
    The timed action $\langle\move^i,t^i_u\rangle$ and two non-conflicting but forbidden timed actions $\langle\move^i,t^i_u-\epsilon\rangle$ and $\langle\wait',t'\rangle$ are shown.} 
    \label{fig:CCBSUnsound}
\end{figure}
\begin{theorem}[Unsoundness of the \branchingImplementation]
    \label{theorem:CCBS_unsound}
    CCBS's \branchingImplementation is not sound.
\end{theorem}
\begin{proof}
    Consider a move-wait conflict $\langle \langle \move^i, t^i \rangle, \langle \wait^j, t^j \rangle \rangle$ with intersection interval $\IcBar = [\IcBarStart, \IcBarEnd)$. 
    From this conflict, the \branchingImplementation creates the constraints $\langle i, \move^i, [t^i, t^i_u) \rangle$ and $\langle j, \waitat{\wait^j}, \IcBar \rangle$.
    Letting $C_i \Leftrightarrow  \langle i, \move^i, [t^i, t^i_u) \rangle$ and $C_j \Leftrightarrow \langle j, \waitat{\wait^j}, \IcBar \rangle$, the \branchingImplementation branches using the constraint set pair $\langle C_i, C_j \rangle$.
    The move-wait conflict and the constraint intervals $[t^i, t^i_u)$ and $\IcBar$ are illustrated in the upper part of Figure~\ref{fig:CCBSUnsound}.

    By construction of $C_i$, for $0 < \epsilon < t^i_u - t^i$ we know that $\langle i, \move^i, t^i_u - \epsilon \rangle\in C_i$.
    By definition of $t^i_u$, $\langle \move^i, t^i_u\rangle$ and $\langle \wait^j, t^j \rangle$ do not conflict.
    It then follows from Lemma~\ref{lemma:epsilon_collision_free} that any conflict between agent $i$ executing $\langle \move^i, t^i_u - \epsilon\rangle$ and $j$ executing $\langle \wait^j, t^j \rangle$ must occur during the time interval $[t^j + \wait^j_\duration - \epsilon, t^j + \wait^j_\duration]$. 
    The timelines of $\langle \move^i, t^i_u\rangle$ and $\langle \move^i, t^i_u-\epsilon\rangle$ are illustrated in the lower part of Figure~\ref{fig:CCBSUnsound}.

    The constraint set $C_j$ contains one forbidding constraint for each timed action that requires occupying $\waitat{\wait^j}$ during $\IcBar$.
    We now aim to find a forbidden timed wait action $\langle\wait',t'\rangle$ (i.e., $\langle j, \wait', t'\rangle\in C_j$) that does not conflict with $\langle \move^i, t^i_u-\epsilon\rangle$.
    Let
    \begin{equation*}
        [t', t'+\wait'_{\duration}] \subseteq [t_j, t_j+\wait^j_{\duration}],
    \end{equation*}
    that is, we restrict our search to timed wait actions that execute during a subset time interval of when $\langle \wait^j, t^j \rangle$ is executed.
    Recall that a conflict between $i$ and $j$ executing $\langle \move^i, t^i_u-\epsilon\rangle$ and $\langle \wait^j, t^j \rangle$ can only occur during $[t^j + \wait^j_\duration - \epsilon, t^j + \wait^j_\duration]$.
    The same also holds for $\langle\wait',t'\rangle$ since it entails $j$ waiting stationary at the same vertex; a conflict between $i$ and $j$ executing $\langle \move^i, t^i_u-\epsilon\rangle$ and $\langle\wait',t'\rangle$ can only occur during $[t^j + \wait^j_\duration - \epsilon, t^j + \wait^j_\duration]$.
    We can use this by constraining $\langle\wait',t'\rangle$ to end before the start of this interval, that is $t'+\wait'_{\duration} < t^j + \wait^j_\duration - \epsilon$, and by doing so ensuring that $\langle\wait',t'\rangle$ and $\langle \move^i, t^i_u -\epsilon\rangle$ do not collide:
    \begin{equation}
        \label{eq:theorem:CCBS_unsound:forbidden_wait}
        [t', t'+\wait'_{\duration}] \subseteq [t^j, t^j + \wait^j_{\duration} - \epsilon).
    \end{equation}
    Finally, to ensure that $\langle j, \wait', t'\rangle$ is indeed a member of $C_j$, we require $[t', t'+\wait'_{\duration}]$ to intersect with $\IcBar$.
    Since $\epsilon$ can be arbitrarily small, let $\epsilon<|\IcBar|$.
    It follows that the interval $[t^j+\wait^j_{\duration}-\epsilon, t^j+\wait^j_\duration]$ of length $\epsilon$ can only span at most a subset of $\IcBar$.
    Thus, a \New{non-zero} subset of $[t^j, t^j + \wait^j_{\duration} - \epsilon)$ is also included in $\IcBar$.
    Therefore, there exists a $\langle \wait', t'\rangle$ such that~\eqref{eq:theorem:CCBS_unsound:forbidden_wait} is satisfied and $[t', t' + \wait_\duration')\cap \IcBar \neq \varnothing$.
    These properties ensure that $\langle \wait', t'\rangle$ does not conflict with $\langle \move^i, t^i_u -\epsilon\rangle$ and that $\langle j, \wait', t'\rangle\in C_j$. 
    An example of such a timed wait action is illustrated at the bottom of Figure~\ref{fig:CCBSUnsound}.

    In conclusion, the timed actions $\langle \move^i, t^i_u - \epsilon\rangle$ and $\langle \wait', t'\rangle$ do not conflict.
    Therefore, a solution $\Plans$ can contain both of these timed actions.
    However, letting $c_i = \langle i, \move^i, t^i_u - \epsilon \rangle$ and $c_j = \langle j, \wait', t'\rangle$, $c_i(\Plans) \vee c_j (\Plans)$ is not satisfied.
    Thus, $\langle c_i, c_j \rangle$ is not sound, and since $c_i\in C_i$ and $c_j\in C_j$, $\langle C_i, C_j \rangle$ is not sound either.
    By Lemma~\ref{lemma:sound_branching}, the \branchingImplementation is therefore not sound.
\end{proof}


\section{\New{Optimal Continuous-Time Conflict-Based Search}}
\label{sec:ModifyingCCBS}

\New{
This section details \OCCBS, and is divided into five parts.
    Section~\ref{sec:proposal:delta_BR} introduces \OCCBS's branching rule, \branchingOurs, and discusses its similarities and differences with shifting constraints from~\cite{CCBS_revisit}.
    Thereafter, Section~\ref{sec:proposal:soundness} shows that \branchingOurs is a sound branching rule.
    Sections~\ref{sec:Proposed:non-termination_requirement} and~\ref{sec:Proposed:termination} together provide a proof by contradiction that \OCCBS terminates within a finite number of iterations.
    Informally, Section~\ref{sec:Proposed:non-termination_requirement} assumes for the sake of contradiction that \OCCBS, using a sound branching rule, never terminates on a solvable \MAPFR instance. 
    It is shown that the CT must then contain an infinite path of nodes whose objective value never exceeds the optimal objective value $\objective^*$. 
    Intuitively, this means that the reachable set of joint plans along this path is progressively restricted due to the accumulating set of constraints, however, without ever removing \emph{all} joint plans at or below $\objective^*$.
    Section~\ref{sec:Proposed:termination} then shows that, under \branchingOurs, it is impossible for joint plans at or below $\objective^*$ to persist after an infinite accumulation of constraints, contradicting the assumption that every node on the path permits such a joint plan.
    Therefore, no such infinite path can exist, and \OCCBS must terminate after a finite number of iterations.
    Finally, Section~\ref{sec:proposed:final} ties these results together and concludes that \OCCBS is \NewPrecision{exact} and solution complete.
}

\subsection{\New{Branching rule \branchingOurs}}
\label{sec:proposal:delta_BR}

\New{
The branching rule \branchingOurs handles move-move conflicts in the same way as the \branchingImplementation. However, the novelty lies in the constraints generated for move-wait conflicts.
We begin by explaining \emph{shifting constraints}~\cite{CCBS_revisit} on which \branchingOurs's handling of move-wait conflicts is based, then \branchingOurs, and finally how they differ.}

\begin{figure}
    \centering
    \includegraphics[width=0.75\linewidth]{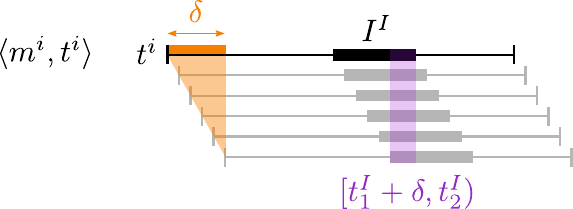}
    \caption{Shifting the start of the timed move action $\langle \move^i, t^i\rangle$ also shifts the intersection interval $\IcBar$. Executing $\move^i$ at any time in the (orange) interval $[t^i, t^i+\delta)$ will result in a shifted intersection interval which necessarily includes the (purple) interval $[\IcBarStart+\delta, \IcBarEnd)$.} 
    \label{fig:ShiftingConstraint}
\end{figure}

\New{
    Shifting constraints are a sound (does not remove solutions from the search) pair of motion and vertex constraints  that are generated from a move-wait conflict. 
    Recall from Section~\ref{sec:Preliminaries:intervals} that,
    given a move-wait conflict $\langle\langle\move^i,t^i\rangle,\langle\wait^j,t^j\rangle\rangle$ with collision interval $\Ic$,
    the intersection interval $\IcBar=[\IcBarStart, \IcBarEnd)$ is a maximally connected time interval during which $i$ and $j$ are in conflict when $i$ executes $\langle\move^i,t^i\rangle$ and $j$ occupies $\waitat{\wait^j}$ for all time.
    There may exists several such intervals; the intersection interval is the chronologically first such interval that intersects $\Ic$.
    If $j$ occupies $\waitat{\wait^j}$ at any time during $\IcBar$ a conflict occurs between the two agents. 
    If instead $i$ executes $\langle\move^i,t^i + \Delta t\rangle$ (the same move action but time-shifted by $\Delta t \in \Real$) then the intersection interval $\IcBar' = [\IcBarStart + \Delta t, \IcBarEnd + \Delta t)$ is also shifted by $\Delta t$. Then a conflict occurs if $j$ occupies $\waitat{\wait^j}$ at any time during $\IcBar'$.
    Figure~\ref{fig:ShiftingConstraint} illustrates the timeline of $\move^i$ starting at $t^i$ and incrementally later times.
    Now, let some $\delta\in[0, |\IcBar|)$ be given.
    If $i$ executes $\langle\move^i,t^i+\Delta t\rangle$ for any $\Delta t\in[0,\delta]$, then the resulting time-shifted intersection interval $[\IcBarStart + \Delta t, \IcBarEnd + \Delta t)$ will at least contain the interval $[\IcBarStart + \delta, \IcBarEnd)$, as illustrated in Figure~\ref{fig:ShiftingConstraint}. 
    For any choice of $\Delta t \in [0, \delta]$, a conflict occurs if $j$ occupies $\waitat{\wait^j}$ at any time during $[\IcBarStart + \delta, \IcBarEnd)$.
    By this reasoning,~\cite{CCBS_revisit} shows that the motion constraint $\langle i, \move^i, [t^i, t^i + \delta) \rangle$ and vertex constraint $\langle j, \waitat{\wait^j}, [\IcBarStart + \delta, \IcBarEnd) \rangle$ is a sound constraint  pair since any solution must satisfy at least one of them.
}

\New{
    Furthermore, \cite{CCBS_revisit} identifies that the choice of $\delta$ is non-trivial.
    If $\delta=0$, then the motion constraint's interval $[t^i, t^i + \delta)$ becomes singular. 
    Consequently, agent $i$ is still permitted to execute $\langle \move^i, t^i+\epsilon \rangle$ for any arbitrarily small $\epsilon$ (positive or negative), thus requiring infinitely many CT branchings in order to avoid conflicts between $i$ and $j$.
    If instead $\delta>0$, then the constraint pair is unable to eliminate the conflict where $j$ waits until exactly $\IcBarStart$ since the vertex constraint interval $[\IcBarStart + \delta, \IcBarEnd)$ cannot contain $\IcBarStart$. The child node of such a branching would encounter the same conflict, resulting in non-termination.
}

\begin{figure}[H]
    \centering
    \begin{subfigure}[b]{0.40\linewidth}
        \centering
        \includegraphics[width=1\textwidth]{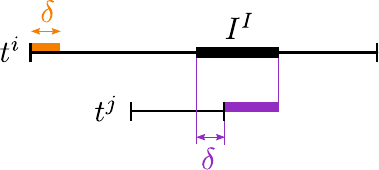}
        \caption{\normalsize\ $\delta = t^j + \wait^j_\duration - \IcBarStart$}
        \label{fig:deltaBR_example1}
    \end{subfigure}
    \hfill
    \begin{subfigure}[b]{0.40\linewidth}
        \centering
        \includegraphics[width=1\textwidth]{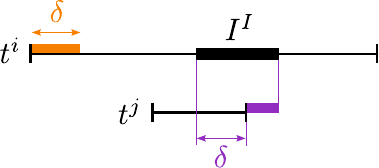}
        \caption{\normalsize\ $\delta = t^j + \wait^j_\duration - \IcBarStart$}
        \label{fig:deltaBR_example2}
    \end{subfigure}
    \vfill
    \begin{subfigure}[b]{0.40\linewidth}
        \centering
        \includegraphics[width=1\textwidth]{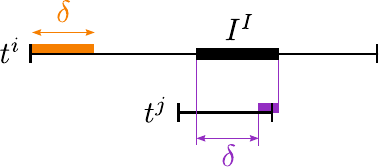}
        \caption{\normalsize\ $\delta = \gamma|\IcBar|$}
        \label{fig:deltaBR_example3}
    \end{subfigure}
    \hfill
    \begin{subfigure}[b]{0.40\linewidth}
        \centering
        \includegraphics[width=1\textwidth]{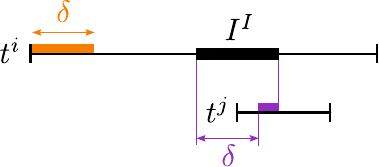}
        \caption{\normalsize\ $\delta = \gamma|\IcBar|$}
        \label{fig:deltaBR_example4}
    \end{subfigure}
    \caption{An illustration of \branchingOurs applied to four different move-wait conflicts, $\langle\langle\move^i,t^i\rangle,\langle\wait^j,t^j\rangle\rangle$, where the start time $t^j$ of wait action $\wait^j$ progressively increases for each sub-figure. $\delta = \min( \gamma|\IcBar|,\; t^j + \wait^j_\duration - \IcBarStart)$ with $\gamma=0.7$ is shown; $\delta = t^j + \wait^j_\duration - \IcBarStart$ in figures~\ref{fig:deltaBR_example1} and~\ref{fig:deltaBR_example2}, and $\delta = \gamma|\IcBar|$ in figures~\ref{fig:deltaBR_example3} and~\ref{fig:deltaBR_example4}.
    Agent $i$'s constrained interval $[t^i, t^i + \delta)$ (orange) and agent $j$'s constrained interval $[\IcBarStart + \delta, \IcBarEnd)$ (purple) are shown.}
    \label{fig:deltaBR_examples}
\end{figure}

\New{
We now introduce \branchingOurs, and thereafter compare it with shifting constraints.
}
\begin{definition}[\branchingOurs]
    \label{def:proposed_branching}
    \branchingOurs handles move-move and move-wait conflicts differently:
    \begin{itemize}
        \item Given a move-move conflict $\langle \langle \move^i, t^i \rangle, \langle \move^j, t^j \rangle \rangle$, the \branchingTheory/\branchingImplementation is applied: 
        the unsafe intervals $[t^i, t^i_u)$ and $[t^j, t^j_u)$ are found and the pair of motion constraints $\langle i, \move^i,  [t^i, t^i_u)\rangle$ and $\langle j, \move^j,  [t^j, t^j_u)\rangle$ are used to branch on.
        That is,
        agent $i$ is forbidden from executing $\langle \move^i, t\rangle$ for any $t\in[t^i, t^i_u)$ and 
        agent $j$ is forbidden from executing $\langle \move^j, t \rangle$ for any $t\in[t^j, t^j_u)$.
        \item Given a move-wait conflict $\langle \langle \move^i, t^i \rangle, \langle \wait^j, t^j \rangle \rangle$ with 
        intersection interval $\IcBar = [\IcBarStart, \IcBarEnd)$, let 
        \begin{equation}
            \label{eq:delta}
            \delta = \min\left( \gamma|\IcBar|,\; t^j + \wait^j_\duration - \IcBarStart \right)
        \end{equation}
        where $0<\gamma<1$ is a fixed constant.
        For agent $i$, the motion constraint $\langle i, \move^i,  [t^i, t^i + \delta)\rangle$ is created.
        For agent $j$, the vertex constraint $\langle j, \waitat{\wait^j}, [\IcBarStart + \delta, \IcBarEnd) \rangle$ and motion constraints $\langle j, \move, [\IcBarStart + \delta, \IcBarEnd) \rangle$ for every $\move\in\Actions: \movefrom{\move}=\waitat{\wait^j}$ is created.
        That is, agent $i$ is forbidden from executing $\move^i$ at any time in $[t^i, t^i + \delta)$, and agent $j$ is forbidden from occupying $\waitat{\wait_j}$ during $[\IcBarStart + \delta, \IcBarEnd)$ or executing any move action from $\waitat{\wait_j}$ at any time in $[\IcBarStart + \delta, \IcBarEnd)$.
    \end{itemize}
\end{definition}
The fixed constant $\gamma\in(0,1)$ controls which agent in a move-wait conflict is more constrained: setting $\gamma$ near $1$ increases the constrained interval on the moving agent while setting $\gamma$ closer to $0$ increases the constrained interval on waiting agent.
Figure~\ref{fig:deltaBR_examples} illustrates four different cases where \branchingOurs (with $\gamma=0.7$) is applied to a move-wait conflict $\langle\langle\move^i,t^i\rangle,\langle\wait^j,t^j\rangle\rangle$.
In the first two cases (Figures~\ref{fig:deltaBR_example1} and~\ref{fig:deltaBR_example2}) we have $\delta = t^j + \wait^j_\duration - \IcBarStart$ since $t^j + \wait^j_\duration - \IcBarStart < \gamma|\IcBar|$.
Thus, agent $j$ is forbidden from occupying $\waitat{\wait^j}$ from $t^j + \wait^j_\duration$ until the end of $\IcBar$.
On the other hand, in the second two cases (Figures~\ref{fig:deltaBR_example3} and~\ref{fig:deltaBR_example4}) we have $\delta = \gamma|\IcBar|$.
Thus, a non-singular part of agent $j$'s occupation of $\waitat{\wait^j}$ is included in the constrained interval $[\IcBarStart + \delta, \IcBarEnd)$. 
Furthermore, it is also visible that agent $j$'s constrained interval $[\IcBarStart + \delta, \IcBarEnd)$ is smaller than agent $i$'s constrained interval $[t^i, t^i + \delta)$ as a result of $\gamma>0.5$.
\New{Specifically for the case in Figure~\ref{fig:deltaBR_example4}, although $j$'s action preceding $\langle \wait^j, t^j\rangle$ must also conflict with $\langle \move^i, t^i\rangle$ since it ends at $\waitat{\wait^j}$, this case must still be handled. Since CCBS (and \OCCBS) selects an \emph{arbitrary} conflict to branch on, it may select $\langle\langle\move^i,t^i\rangle,\langle\wait^j,t^j\rangle\rangle$ rather than the conflict with $j$'s preceding action.}

\New{
    Unlike the definition of shifting constraints in~\cite{CCBS_revisit} which leaves $\delta$ arbitrary,
    for a move-wait conflict $\langle\langle\move^i,t^i\rangle,\langle\wait^j,t^j\rangle\rangle$, 
    \branchingOurs creates shifting constraints with a specifically selected $\delta$-value, and additionally motion constraints on $j$ for every move action from the vertex where $j$ waits.
    The conclusion in~\cite{CCBS_revisit} regarding the case of $\delta>0$ is to our knowledge correct; 
    the vertex constraint $\langle j, \waitat{\wait^j}, [\IcBarStart + \delta, \IcBarEnd) \rangle$ with any $\delta>0$ is not sufficient to avoid this specific conflict in the next CT node if $j$'s wait action ends at exactly $\IcBarStart$.
    \branchingOurs's selection of a value for $\delta$ does not always avoid this effect either; for the case of $\delta = t^j + \wait^j_\duration - \IcBarStart$ (see Figures~\ref{fig:deltaBR_example1} and~\ref{fig:deltaBR_example2}), the wait action ends at exactly when the vertex constraint interval $[\IcBarStart + \delta, \IcBarEnd)$ begins. 
    Consequently, a wait action that ends infinitesimally earlier is not avoided by the vertex constraint, potentially resulting in a similar infinite branching issue.
    This is the reason behind \branchingOurs's additional motion constraints on $j$; if the vertex constraint does not constrain the wait action by an amount bounded away from zero, then the motion constraints do on the subsequent move action.  
    The shifting constraints are shown in~\cite{CCBS_revisit} to be sound; we show in Theorem~\ref{theorem:new_procedure_sound} that the addition of motion constraints on $j$ preserves soundness.
    We show in Section~\ref{sec:Proposed:termination:wait_reductions} that the vertex constraint and the motion constraints together ensure that progress toward an optimal solution is bounded away from zero at every branching on a move-wait conflict, and show in Section~\ref{sec:Proposed:termination:short_plans} how this leads to \OCCBS terminating in finite time.
}

\New{
In \branchingOurs, the value of $\delta$ is selected by~\eqref{eq:delta} as the minimum of $\gamma|\IcBar|$ and $t^j + \wait^j_\duration - \IcBarStart$. Bounding $\delta$ to $t^j + \wait^j_\duration - \IcBarStart$ ensures that $j$'s wait action and the constrained interval $[\IcBarStart + \delta, \IcBarEnd)$ overlap by at least a singular value $t^j + \wait^j_\duration$ (specifically so that the additional motion constraints on $j$ invalidate $j$'s current plan), visualized in Figures~\ref{fig:deltaBR_example1} and~\ref{fig:deltaBR_example2}. 
Otherwise, $\delta=\gamma|\IcBar|$ provides control via $\gamma$ over the balance between how much each agent is constrained.
}

\subsection{Soundness}
\label{sec:proposal:soundness}

We now show that \branchingOurs is sound.
\branchingOurs applies \branchingTheory/\branchingImplementation on move-move conflicts, which is shown in Section~4.2 of~\cite{CCBS} to satisfy property~\eqref{eq:CCBSproof:sound} and is therefore sound. Hence, we refer the reader to there for more details.
What remains is to prove \branchingOurs's soundness on move-wait conflicts.
\begin{theorem}[\branchingOurs's move-wait soundness]
    \label{theorem:new_procedure_sound}
    \branchingOurs is sound when applied to a move-wait conflict. 
\end{theorem}
\begin{proof}
    Consider a move-wait conflict $\langle \langle \move^i, t^i \rangle, \langle \wait^j, t^j \rangle \rangle$ with intersection interval $\IcBar = [\IcBarStart, \IcBarEnd)$, on which we apply \branchingOurs.
    For agent $i$, \branchingOurs creates the motion constraint $\langle i, \move^i, [t^i, t^i + \delta)\rangle$.
    Since this motion constraint is in the form of a shifting constraint, and by~\eqref{eq:delta} we know that $\delta < |\IcBar|$, we know that a conflict occurs if agent $j$ occupies vertex $\waitat{\wait^j}$ at any time during $[\IcBarStart+\delta, \IcBarEnd)$. 
    The motion constraint $\langle i, \move^i, [t^i, t^i + \delta)\rangle$ can equivalently be expressed in its decomposed form: a set of forbidding constraints $\Constraints_i = \left\{ \langle i, \move^i, t \rangle \mid t \in [t^i, t^i+\delta) \right\}$.

    On the other hand, \branchingOurs creates for agent $j$ a vertex constraint $\langle j, \waitat{\wait^j}, [\IcBarStart+\delta, \IcBarEnd)\rangle$ and the motion constraints $\langle j, \move, [\IcBarStart+\delta, \IcBarEnd)\rangle$ for every $\move\in\Actions: \movefrom{m} = \waitat{\wait^j}$.
    These can collectively be represented by the set of forbidding constraints 
    \begin{align*}
        C_j &= \left\{ \langle j, \wait, t\rangle \mid \waitat{\wait} = \waitat{\wait^j}, [t, t+\wait_\duration] \cap [\IcBarStart+\delta, \IcBarEnd) \neq\varnothing\right\} \\
        &\cup \left\{ \langle j, \move, t \rangle \mid \movefrom{\move} = \waitat{\wait^j}, \move\in\Actions, t\in[\IcBarStart+\delta, \IcBarEnd)\right\}
    \end{align*}

    \New{
    For every forbidding constraint $\langle j,\action,t\rangle\in C_j$, the timed action $\langle \action, t\rangle$ involves $j$ being located at $\waitat{\wait^j}$ some time in $[\IcBarStart+\delta, \IcBarEnd)$; the wait actions overlap with $[\IcBarStart+\delta, \IcBarEnd)$ and the move actions start from $\waitat{\wait^j}$ at some time in $[\IcBarStart+\delta, \IcBarEnd)$.}
    Thus, if agent $i$ executes any of the actions forbidden by $C_i$ and agent $j$ executes any of the actions forbidden by $C_j$, then a conflict will occur.

    The above implies that for every $\langle c_i, c_j\rangle = \langle \langle i, \action^i, t^i\rangle, \langle j, \action^j, t^j\rangle\rangle \in C_i\times C_j$ the pair of timed actions $\langle \langle \action^i, t^i\rangle, \langle \action^j, t^j\rangle\rangle$ is a conflict.
    Thus, any solution must satisfy $c_i \vee c_j$, meaning that $\langle c_i, c_j\rangle$ is sound for every $\langle c_i, c_j\rangle \in C_i\times C_j$ and therefore that $\langle C_i, C_j \rangle$ is sound. 
    By Lemma~\ref{lemma:sound_branching}, this means that \branchingOurs is sound on any move-wait conflict.
\end{proof}

To conclude, \branchingOurs is shown to be sound when applied to move-move conflicts (in~\cite{CCBS}) and move-wait conflicts (in Theorem~\ref{theorem:new_procedure_sound}).
As these are the only two types of conflict considered, \branchingOurs is therefore sound.

\subsection{A Requirement for Non-termination}
\label{sec:Proposed:non-termination_requirement}

To prove that \New{\OCCBS} is solution complete and therefore terminates on a solvable \MAPFR problem, we employ a proof by contradiction. We begin by assuming the opposite: that \New{\OCCBS} never terminates on a solvable \MAPFR problem despite the existence of a solution. This will result in a condition which must hold if \New{\OCCBS} never terminates.

We will establish in this section that if \New{\OCCBS} runs indefinitely, then the CT must contain an infinite sequence of nodes that are continuously selected and branched upon. By analyzing this infinite sequence, we derive a specific requirement that any non-terminating execution must satisfy. It will later be shown in Section~\ref{sec:Proposed:termination} that \New{\OCCBS} does not satisfy this requirement, thus proving termination.
Specifically, what we establish here is:
\begin{enumerate}
    \item[] \ref{sec:Proposed:non-termination_requirement:infinite_sequence}) \nameref{sec:Proposed:non-termination_requirement:infinite_sequence}: There exists an infinite descending sequence in the CT with each node containing conflicts (i.e., is not a solution) and having an objective value $\leq\objective^*$, with $\objective^*$ being the optimal.
    \item[] \ref{sec:Proposed:non-termination_requirement:trajectory_concentration}) \nameref{sec:Proposed:non-termination_requirement:trajectory_concentration}: We define a \emph{trajectory} in Definition~\ref{def:trajectory} and a mapping from a CT node to a trajectory in Definition~\ref{def:proposed:node_trajectory_map}, and find that infinitely many nodes in the sequence map to the same trajectory $\trajectory^\infty$. 
    \item[] \ref{sec:Proposed:non-termination_requirement:constraint_accumulation}) \nameref{sec:Proposed:non-termination_requirement:constraint_accumulation}: Constraints accumulate unboundedly along the infinite node sequence, yet there must always exist another node mapping to $\trajectory^\infty$ which still maintains an objective value lower than the optimal value.
\end{enumerate}
The final point formulates the requirement on the branching rule for non-termination.

We begin by defining trajectories and how plans map to them.
\New{
    Informally, a trajectory $\trajectory$ describes the sequence of move actions an agent takes when following a plan $\plan$ that maps to $\trajectory$, denoted $\plan\sim\trajectory$, but without timing details.
    Multiple plans can therefore map to the same trajectory, as plans can contain various wait actions.
    For example, the following two plans 
    \begin{itemize}
        \item $\plan_i = \langle \langle\move_1^i, 0\rangle, \langle\wait_2^i, 1.5\rangle, \langle\move_3^i, 2\rangle, \langle\move_4^i, 3\rangle, \langle\wait_5^i, 4.3\rangle \rangle$ introduced in Figure~\ref{fig:ExamplePlan}, and
        \item $\plan_i' = \langle \langle\move_1^i, 0\rangle, \langle\move_3^i, 1.5\rangle, \langle\move_4^i, 2.5\rangle\rangle$
    \end{itemize}
    both map to the same trajectory $\trajectory_i = \langle \move_1^i, \move_3^i, \move_4^i \rangle$.
}
\begin{definition}[Trajectory]
    A trajectory $\trajectory=\langle\move_1,\move_2\dots,\move_n\rangle\in\Actions^n$ is a finite sequence of move actions.
    \label{def:trajectory}
\end{definition}
Each move action $\move_i\in\trajectory$ has a duration $\move_{i,\duration}$. Thus, the duration of $\trajectory$ is $\trajectory_\duration = \sum_{k=1}^n \move_{k,\duration}$.
\begin{definition}[Joint Trajectory]
    A joint trajectory $\Trajectories$ is a set containing one trajectory $\trajectory_i$ for each agent $i$.
\end{definition}
\begin{definition}[Plan-Trajectory Mapping]
    A plan $\plan$ maps to a trajectory $\trajectory$, denoted $\plan\sim\trajectory$, if $\trajectory$ contains exactly every move action in $\plan$ in the order that they appear.
    Similarly, a joint plan $\Plans$ maps to a joint trajectory $\Trajectories$,
    $\Plans\sim\Trajectories$, if it holds for every agent $i$ that $\plan_i\sim\trajectory_i$ with $\plan_i\in\Plans,\trajectory_i\in\Trajectories$.
\end{definition}


Observe that $\trajectory_\duration\leq\plan_\duration$ if $\plan\sim\trajectory$ since $\plan$ contains at least all the move actions in $\trajectory$ but possibly additional wait actions.
\New{
    Since each trajectory $\trajectory$ has a duration $\trajectory_\duration$, a joint trajectory $\Trajectories$ can be evaluated on the objective function, $\objective(\Trajectories)$.
    Since $\trajectory_{i,\duration} \leq \plan_{i,\duration}$ holds for every agent $i$, the monotonicity of $\objective$ gives that if $\Plans \sim \Trajectories$ then    
}   
\begin{equation}
    \label{eq:trajectorys_leq_plans}
    \objective\left(\Trajectories\right)\leq\objective\left(\Plans\right).
\end{equation}

Finally, we let $\Trajectories$ denote the set of all joint trajectories, define the set $\TrajectoriesSet^c \subseteq \Trajectories$ to contain those joint trajectories with an objective value at or below some fixed constant $c\in\Real$, and show that
\New{
    $\TrajectoriesSet^c$ is finite for every $c<\infty$.
    To see this, let $\Trajectories\in\TrajectoriesSet^c$.
    By the properness of $\objective$, $\trajectory_{i,\duration} \leq B(c)$ for every $\trajectory_i \in \Trajectories$.
    Every move action has strictly positive duration and $\Actions$ is finite,
    so the shortest duration over all move actions in $\Actions$ is strictly positive, and each $\trajectory_i \in \Trajectories$ therefore consists of a bounded number of move actions.
    As $\Actions$ is finite there are finitely many such trajectories, and a joint trajectory comprises exactly one trajectory per agent over the finite set of agents. 
    Hence $\TrajectoriesSet^c$ is finite.
}

\subsubsection{Infinite CT node sequence}
\label{sec:Proposed:non-termination_requirement:infinite_sequence}

Since \branchingOurs is sound (Section~\ref{sec:proposal:soundness}), all solutions (including all optimal solutions) are reachable from the CT root.
Non-termination of \New{\OCCBS} implies that at every iteration, a non-solution CT node $\CTnode$ is expanded instead of a node representing an optimal solution with objective value $\objective^*$. 
\New{
By soundness and property~\eqref{eq:OCCBS_property:admissibility},
the set of all unexpanded nodes always contains a node whose objective value is at most $\objective^*$.
Since the search is best-first, the expanded node $\CTnode$ must therefore satisfy $\objective(\CTnode_\Plans) \leq \objective^*$.
}
Additionally, $\CTnode_\Plans$ must contain a conflict, otherwise $\CTnode_\Plans$ would constitute a valid solution.
For non-termination to persist indefinitely, such nodes must be available at every iteration. 
This leads to the following recursive requirement: at every iteration there must exist at least one node $\CTnode$ satisfying the above two conditions, with at least one child node also satisfying the above two conditions, otherwise eventually all nodes satisfying these properties will be depleted and the algorithm will terminate with an optimal solution.
In summary, non-termination implies \New{the following necessary condition:} at every iteration there exists at least one \New{unexpanded} CT node $\CTnode$ where
\begin{enumerate}[label=\textnormal{(\Roman*)}]
    \item $\objective\left( \CTnode_\Plans \right) \leq \objective^*$, \label{proposal:super_optimality}
    \item $\CTnode_\Plans$ contains conflicts and is therefore not a solution, and \label{proposal:always_conflicting}
    \item at least one child node of $\CTnode$ satisfies~\ref{proposal:super_optimality} and~\ref{proposal:always_conflicting}.\label{proposal:recursive_expansion}
\end{enumerate}
\New{The above is an implication from non-termination, from which} it follows the existence of an infinite descending path
\begin{equation}
    \CTNodeSequence = \langle \CTnode^1, \CTnode^2, \dots \rangle
\end{equation}
in the CT, where for every $k=1,2,\dots$ it holds that $\CTnode^k$ satisfies~\ref{proposal:super_optimality}-\ref{proposal:recursive_expansion} and $\CTnode^{k+1}$ is a child of $\CTnode^k$.

\subsubsection{Concentration on $\trajectory^\infty$}
\label{sec:Proposed:non-termination_requirement:trajectory_concentration}

Consider a node $\CTnode^k\in\CTNodeSequence$. 
When $\CTnode^k$ is expanded, \New{\OCCBS} detects a conflict between two plans in $\CTnode^k_\Plans$ and applies \New{\branchingOurs} to create two child nodes. 
By definition of sequence $\CTNodeSequence$, exactly one of these children continues the sequence as $\CTnode^{k+1}$.
For one of the plans involved in the conflict, say $\plan_i\in\CTnode^k_\Plans$, a constraint is created for agent $i$ and added to the constraint set $\CTnode^{k+1}_\Constraints$.
In the context of our analysis, we say that $\plan_i$ is ``branched on'' at $\CTnode^k$.
\begin{definition}[Node-trajectory mapping]
    \label{def:proposed:node_trajectory_map}
    A node $\CTnode\in\CTNodeSequence$, where $\plan$ is branched on, maps to the trajectory $\trajectory$ for which $\plan\sim\trajectory$.
\end{definition}

With Definition~\ref{def:proposed:node_trajectory_map}, every one of the infinitely many nodes in $\CTNodeSequence$ map to a trajectory. 
However, we can show that infinitely many nodes in $\CTNodeSequence$ map to one specific trajectory, denoted $\trajectory^\infty$.
Consider $\CTnode\in\CTNodeSequence$.
There exists some $\Trajectories\in\TrajectoriesSet$ for which $\CTnode_\Plans\sim\Trajectories$, and $\CTnode$ maps to a trajectory $\trajectory\in\Trajectories$.
By inequality~\eqref{eq:trajectorys_leq_plans}, $\objective\left(\Trajectories\right)\leq\objective\left(\CTnode_\Plans\right)$, and by condition~\ref{proposal:super_optimality}, $\objective\left( \CTnode_\Plans \right) \leq \objective^*$.
This means that $\objective\left( \Trajectories \right) \leq \objective^*$ and therefore that $\Trajectories\in\TrajectoriesSet^{\objective^*}$.
Since $\TrajectoriesSet^{\objective^*}$ is finite and each $\Trajectories\in\TrajectoriesSet^{\objective^*}$ contains a finite number of trajectories,
the set of all possible trajectories that nodes in $\CTNodeSequence$ map to is finite.
However, $\CTNodeSequence$ contains infinitely many nodes.
By the Pigeonhole principle, there exists at least one trajectory $\trajectory^\infty$ which infinitely many nodes in $\CTNodeSequence$ map to.

\subsubsection{Unbounded Constraint Accumulation}
\label{sec:Proposed:non-termination_requirement:constraint_accumulation}

\New{
At every branching of a node $\CTnode^{k}\in\CTNodeSequence$, a new constraint $c$ is created and added to the next node $\CTnode^{k+1}\in\CTNodeSequence$ and included in all subsequent nodes $\CTnode^h\in\CTNodeSequence$ for $h>k$, meaning that $c\in\CTnode^h_\Constraints$.
}
Since infinitely many nodes in $\CTNodeSequence$ map to $\trajectory^\infty$, constraints from branching on plans $\plan\sim\trajectory^\infty$ accumulate unboundedly along $\CTNodeSequence$,
yet there must always exist some $\plan'\sim\trajectory^\infty$ permitted under these constraints.
On the other hand, by condition~\ref{proposal:super_optimality}, every $\CTnode^k\in\CTNodeSequence$ must satisfy $\objective\left(\CTnode^k_\Plans\right)\leq\objective^*$.
\New{
By the properness of $\objective$, this bounds every individual plan duration.
Setting $c=B(\objective^*)<\infty$, it holds for all $\CTnode^k \in \CTNodeSequence$ that $\forall\plan\in\CTnode^k_\Plans: \plan_\duration \leq c$.
}

Finally, we have arrived at our requirement for non-termination:
for some fixed $c<\infty$ and trajectory $\trajectory^\infty$, there must exists some plan $\plan\sim\trajectory^\infty$ with duration $\plan_\duration\leq c$ that is permitted under the constraints accumulated from infinitely many branchings on plans also mapping to $\trajectory^\infty$.




\subsection{Solution Completeness}
\label{sec:Proposed:termination}

In this section, we prove that \New{\OCCBS} terminates within finite iterations by showing that the non-termination requirement established in Section~\ref{sec:Proposed:non-termination_requirement} cannot be satisfied, thereby completing our proof by contradiction.
This is done by showing that each branching operation reduces the \New{bounded} set of feasible \New{move execution and vertex occupancy} times for $\trajectory^\infty$, or increases the duration of all plans mapping to $\trajectory^\infty$, in each case by \NewPrecision{an amount bounded away from zero}, eventually eventually making it impossible to construct any plan $\plan\sim\trajectory^\infty$ with a bounded duration.

The proof is structured in five stages:
\begin{enumerate}
    \item[] \ref{sec:Proposed:termination:permitted_times}) \nameref{sec:Proposed:termination:permitted_times}: We formalize how constraints produced under \branchingOurs restrict move execution and vertex occupancy times, establishing the framework for analyzing constraint accumulation.
    \item[] \ref{sec:Proposed:termination:CSIPP}) \nameref{sec:Proposed:termination:CSIPP}: We explain how plans are constructed by the underlying path planner, CSIPP, focusing on how it selects execution times from safe intervals. 
    \item[] \ref{sec:Proposed:termination:move_reductions}) \nameref{sec:Proposed:termination:move_reductions}: We show how \branchingOurs applied to a timed move action (in a move-move or move-wait conflict) reduces feasible move execution times, either by an \NewPrecision{amount bounded away from zero} or by an entire safe interval.
    \item[] \ref{sec:Proposed:termination:wait_reductions}) \nameref{sec:Proposed:termination:wait_reductions}: We show how \branchingOurs applied to a timed wait action reduces feasible move execution or feasible vertex occupancy times by \NewPrecision{amounts bounded away from zero}.
    \item[] \ref{sec:Proposed:termination:short_plans}) \nameref{sec:Proposed:termination:short_plans}: Using the results from Sections~\ref{sec:Proposed:termination:move_reductions} and~\ref{sec:Proposed:termination:wait_reductions}, and the fact that all timed actions in a bounded-duration plan must end before a fixed time, we show that the feasible move execution and vertex occupancy times are exhausted after finite iterations of the search.
    Therefore, we show that \New{\OCCBS} does not satisfy the non-termination requirement and is therefore guaranteed to terminate.
\end{enumerate}

\subsubsection{Permitted Execution Times}
\label{sec:Proposed:termination:permitted_times}

To prove that the non-termination requirement cannot be satisfied, we establish a framework to analyze how constraints accumulate and restrict agent movements.
At the CT root $\CTnode^R$ where the constraint set is empty, $\CTnode^R_\Constraints=\varnothing$, all agents are permitted to execute all actions at any time $t\geq0$.
However, at nodes further down in the CT with non-empty constraint sets, any number of constraints may restrict when agents can execute move actions and occupy vertices.
Thus, under the constraints $\CTnode_\Constraints$ at CT node $\CTnode$, we define for a specific agent $i$
\begin{itemize}
    \item $\Tset_\move\subseteq[0,\infty)$ as the set of all times when \New{it is feasible for $i$} to execute a move action $\move\in\Actions$, and
    \item $\Tset_\vertex\subseteq[0,\infty)$ as the set of all times when \New{it is feasible for $i$} to occupy a vertex $\vertex\in\Vertices$.
\end{itemize}

Since constraints under the various branching rules restrict move execution and vertex occupancy over time intervals (Definition~\ref{def:proposed_branching}), every set $\Tset_x$,
with $x$ ranging over all moves $\move\in\Actions$ and vertices $\vertex\in\Vertices$,
can be described as the union of a finite number of maximally connected time intervals:
\begin{equation}
    \label{eq:permitted_execution_time}
    \Tset_x = \bigcup_{h=1}^{N_x} S_x^h,\quad S_x^h = [\IntStart_x^h, \IntEnd_x^h).
\end{equation}
We refer to $S_x^h$ as a \emph{safe interval} for $i$ to execute a move (when $x=\move$) or occupy a vertex (when $x=\vertex$). 


\subsubsection{CSIPP Planning Behavior}
\label{sec:Proposed:termination:CSIPP}

Understanding how CSIPP selects execution times is crucial for proving that constraint additions cause reductions \NewPrecision{whose length is bounded away from zero} in feasible time intervals.
The key insight is that CSIPP, like SIPP~\cite{SIPP} which it is based on, systematically chooses the earliest possible time within a safe interval for each action. 
This property prevents CSIPP from making arbitrarily small timing adjustments when constraints are added and thereby postponing termination indefinitely.

\begin{figure}
    \centering
    \includegraphics[width=0.5\linewidth]{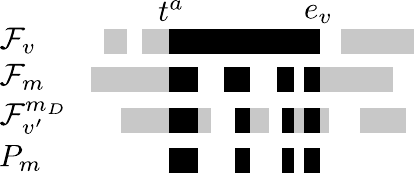}
    \caption{Illustration of the construction of $P_\move$ as the intersection between $[t^a, e_\vertex)\subseteq\Tset_\vertex$, $\Tset_\move$, and $\Tset_{\vertex'}^{\move_\duration}$.} 
    \label{fig:CSIPP_safe_intervals}
\end{figure}

CSIPP performs an A$^*$~\cite{A_star} search in the vertex-safe interval space. 
Each search state is represented by a tuple $(\vertex, S_\vertex)$ where agent $i$ is located at $\vertex\in\Vertices$ during the safe interval $S_\vertex = [s_\vertex, e_\vertex)\subseteq\Tset_\vertex$.
Let $t^a\in S_\vertex$ be the time when $i$ arrives at $\vertex$.
Unless state $(\vertex, S_\vertex)$ is a goal state and $e_\vertex=\infty$, agent $i$ must execute a move action to leave $\vertex$ before the end of the safe interval, $e_\vertex$.
For each move action $\move$ with $\movefrom{\move}=\vertex$, agent $i$ can execute $\move$ at any time $t$ satisfying
\begin{itemize}
    \item $t\in[t^a, e_\vertex)$ (executing $\move$ after arriving at $\vertex$ and before it is unsafe to remain there),
    \item $t\in\Tset_\move$ (executing $\move$ when it safe to do so), and
    \item $t+\move_\duration \in\Tset_{\vertex'}$ (executing $\move$ to arrive at the next node $\vertex' = \moveto{\move}$ when it is safe to do so),
\end{itemize}
In other words, it is safe to execute $\move$ at some time $t\in P_\move$ where
\begin{equation}
    \label{eq:CSIPP}
    P_\move = [t^a, e_\vertex)\cap\Tset_\move\cap\Tset_{\vertex'}^{\move_\duration}
\end{equation}
and $\Tset_{\vertex'}^{\move_\duration} = \{t-\move_\duration \mid t\in\Tset_{\vertex'}\}$ to compensate for the arrival at $\vertex'$ being $\move_\duration$ after $t$.
By~\eqref{eq:permitted_execution_time}, $\Tset_\move$ and $\Tset_{\vertex'}^{\move_\duration}$ are unions of maximally connected intervals, such that $P_\move$ is also a union of maximally connected intervals.
Figure~\ref{fig:CSIPP_safe_intervals} illustrates how $P_\move$ is formed from the intersection between $[t^a, e_\vertex)$, $\Tset_\move$ and $\Tset_{\vertex'}^{\move_\duration}$.
For every maximally connected interval $[s_\move, e_\move)\subseteq P_\move$, CSIPP creates an edge representing that agent $i$ executes $\langle \move, s_\move\rangle$, leading to the next state $(\vertex', S_{\vertex'})$ where $S_{\vertex'}\subseteq\Tset_{\vertex'}$ is the safe interval at $\vertex'$ such that $s_\move+\move_\duration\in S_{\vertex'}$.

CSIPP always selects the earliest time $s_\move$ in a safe interval $[s_\move, e_\move)\subseteq P_\move$ for $i$ to execute $\move$.
This earliest-time selection property is essential for our proof:
since the earliest time $s_\move$ in $[s_\move, e_\move)$ is selected, it is not possible to select an earlier time $t'<s_\move$ arbitrarily close to $s_\move$ for the move action $\move$ to be executed, since $t'\not\in P_\move$.

\subsubsection{\NewPrecision{Meaningful} Reductions For Move Actions}
\label{sec:Proposed:termination:move_reductions}

When \branchingOurs is applied to a move action $\langle \move^i, t^i \rangle\in\plan_i$ in a move-move or move-wait conflict, the motion constraint $\langle i, \move^i, [t^i, t')\rangle$ is created. 
Assuming that $\plan_i$ was returned from CSIPP, a safe interval $[s_{\move^i}, e_{\move^i})\subseteq P_{\move^i}$ with $s_{\move^i}=t_i$ was identified during the search to construct $\plan_i$.
We will first show that the constrained interval $[t^i, t')$ is \NewPrecision{bounded away from zero in length}, and then consequently that the new motion constraint removes either a prefix or the entirety of $[s_{\move^i}, e_{\move^i})$ from $P_{\move^i}$.
\NewPrecision{Throughout this and the next section, we make use of the assumptions in Section~\ref{sec:Preliminaries:finite-precision} that all unsafe intervals, collision intervals, and intersection intervals are sufficiently larger than the machine precision $\epsmach$ and are bounded away from zero in length.}

When $\langle \move^i, t^i \rangle$ is contained in a move-move conflict, \branchingOurs forbids the execution of $\move^i$ during the unsafe interval $[t^i, t^i_u)$. 
\New{
Recall that $t^i_u$ is the earliest time after $t^i$ when executing $\langle \move^i, t^i_u \rangle$ does not cause a conflict, with $[t^i, t^i_u)$ being the \emph{unsafe interval}.} 
Therefore, the interval $[t^i, t')=[t^i, t^i_u)$ is \New{bounded away from zero in length}.

Consider instead that $\langle \move^i, t^i \rangle$ is contained in a move-wait conflict $\langle\langle\move^i,t^i\rangle,\langle\wait^j,t^j\rangle\rangle$ with collision interval $\Ic = [\IcStart, \IcEnd)$ and intersection interval $\IcBar = [\IcBarStart, \IcBarEnd)$.
\branchingOurs creates the motion constraint $\langle i, \move^i, [t^i, t^i + \delta)\rangle$.
By~\eqref{eq:delta}, $\delta$ is the minimum of two positive quantities, $\gamma|\IcBar|$ and $t^j + \wait^j_\duration - \IcBarStart$:
\begin{itemize}
    \item
    \NewPrecision{
        For $\gamma|\IcBar|$, from Section~\ref{sec:Preliminaries:finite-precision} we 
        assume
        that this quantity is sufficiently larger than $\epsmach$ and bounded away from zero.}
    
    \item From~\eqref{eq:ic_icbar_intervals} we know that $\Ic\subseteq\IcBar$ and $\Ic\subseteq [t^j, t^j + \wait^j_\duration]$.
    \NewPrecision{
        The former implies that $\IcBarStart \leq \IcStart$ and the latter implies that $\IcEnd < {\IcEnd}' \leq t^j + t^j_\duration$ where ${\IcEnd}'$ is arbitrarily close to $\IcEnd$ to correct for $\Ic=[\IcStart, \IcEnd)$ not including $\IcEnd$ while $[t^j, t^j + t^j_\duration]$ does include $t^j + t^j_\duration$.
        These together give
        \begin{align*}
            \IcBarStart &\leq \IcStart < \IcEnd < {\IcEnd}' \leq t^j+\wait^j_\duration, \\
            \IcBarStart &< t^j+\wait^j_\duration, \\
            0 &< t^j+\wait^j_\duration - \IcBarStart.
        \end{align*}
        where the inequality difference is bounded away from zero due to the assumption in Section~\ref{sec:Preliminaries:finite-precision} that the length of $\Ic$ is bounded away from zero.}
\end{itemize}
\NewPrecision{Since $\delta$ is the minimum of two positive quantities which are both bounded away from zero, the length of $[t^i, t') = [t^i, t^i + \delta)$ is also bounded away from zero.
}

In both cases---when $\langle \move^i, t^i \rangle$ is contained in a move-move or move-wait conflict---a constraint $\langle i, \move^i, [t^i, t')\rangle$ is created for which we have shown above that $[t^i, t')$ is \NewPrecision{bounded away from zero in length}. 
The constraint effectively removes $[t^i, t')$ from $\Tset_{\move^i}$. 
Since $t^i=s_{\move^i}$ for some $[s_{\move^i}, e_{\move^i})\subseteq P_{\move^i}$, by~\eqref{eq:CSIPP}, either the entire safe interval $[s_{\move^i}, e_{\move^i})$ is removed (if $t' \geq e_{\move^i}$) or a prefix $[s_{\move^i}, t')$ \NewPrecision{with a length bounded away from zero} is removed from $P_{\move^i}$ (if $t' < e_{\move^i}$).
This is a useful result as it shows that \branchingOurs applied to any timed move action in a conflict necessarily \New{removes from $P_{\move^i}$ either a time interval \NewPrecision{whose length is bounded away from zero}, or an entire safe interval. Both forms of reduction are used in Section~\ref{sec:Proposed:termination:short_plans}.}

\subsubsection{\NewPrecision{Meaningful} Reductions For Wait Actions}
\label{sec:Proposed:termination:wait_reductions}

We now consider the constraints created by \branchingOurs when applied to a wait action $\langle \wait^j, t^j \rangle\in\plan_j$ in a move-wait conflict $\langle \langle \move^i, t^i \rangle, \langle \wait^j, t^j \rangle \rangle$ with conflict interval $\Ic = [\IcStart, \IcEnd)$ and intersection interval $\IcBar = [\IcBarStart, \IcBarEnd)$.
\New{We refer back to Figure~\ref{fig:deltaBR_examples} for a visual representation.}
\branchingOurs creates the vertex constraint $\langle j, \waitat{\wait^j}, [\IcBarStart + \delta, \IcBarEnd) \rangle$ and motion constraints $\langle j, \move, [\IcBarStart + \delta, \IcBarEnd) \rangle$ for each $\move\in\Actions: \movefrom{\move} = \waitat{\wait^j}$.


\NewPrecision{
    We will show here that branching on a finite wait action removes a time interval whose length is bounded away from zero from the feasible move execution times, from the feasible vertex occupancy times, or partly from each;
    and that branching on an infinite wait action leaves the vertex constraint violated by an amount bounded away from zero (a property that will have significance in the next section).
    
    We begin by showing that the length of the constrained interval $[\IcBarStart + \delta, \IcBarEnd)$ is bounded away from zero.
    From~\eqref{eq:delta} we see that $\delta \leq \gamma|\IcBar|$. 
    Recall the assumption in Section~\ref{sec:Preliminaries:finite-precision} that $(1-\gamma)|\IcBar|$ is bounded away from zero, which is equivalent to $\gamma|\IcBar|$ being bounded away from $|\IcBar|$.
    Therefore, $\delta \leq \gamma|\IcBar|$ is bounded away from $|\IcBar|$ by some fixed amount, meaning that $\IcBarStart + \delta$ is bounded away from $\IcBarStart + |\IcBar| = \IcBarEnd$ by some fixed amount. 
    Hence, the length of the constrained interval $[\IcBarStart + \delta, \IcBarEnd)$ is bounded away from zero.
}


Next, we show that $[\IcBarStart + \delta, \IcBarEnd)$ necessarily overlaps with the wait interval $[t^j, t^j+\wait^j_\duration]$.
To do so, we combine two conditions.
\NewPrecision{
    First, from Section~\ref{sec:Preliminaries:finite-precision} we know that the length of $\Ic$ is bounded away from zero by some fixed amount, and from~\eqref{eq:ic_icbar_intervals} that $\Ic= [t^j, t^j + \wait^j_\duration] \cap [\IcBarStart, \IcBarEnd)$. For the length of this intersection to be bounded away from zero, it must hold that
    \begin{equation}
        t^j<\IcBarEnd
        \label{eq:wait_start_before_ICbar_end}
    \end{equation}
    by some amount bounded away from zero.
    Second, from~\eqref{eq:delta} we know that $\delta\leq t^j + \wait^j_\duration - \IcBarStart$ and therefore that $\IcBarStart + \delta \leq t^j + \wait^j_\duration$.
    These two conditions, $t^j<\IcBarEnd$ and $\IcBarStart + \delta \leq t^j + \wait^j_\duration$, together are sufficient to guarantee overlap between $[\IcBarStart + \delta, \IcBarEnd)$ and $[t^j, t^j+\wait^j_\duration]$.
    This overlap is singular when $\IcBarStart + \delta = t^j + \wait^j_\duration$, otherwise non-zero when $\IcBarStart + \delta < t^j + \wait^j_\duration$.
}

\NewPrecision{
    We consider two cases: $\wait^j$ is either a finite wait action, or an infinite one.    
    \begin{itemize}
        \item If $\wait^j$ is a finite wait action, 
        let $t^\mathit{dep} = t^j + \wait^j_\duration$ denote the time when $j$ leaves $\waitat{\wait^j}$, and recall from above that $\IcBarStart + \delta \leq t^\mathit{dep}$.
        The constrained interval $[\IcBarStart + \delta, \IcBarEnd)$ therefore divides at $t^\mathit{dep}$ into two parts, both of which restrict $\plan_j$:
        \begin{itemize}
            \item The part before $t^\mathit{dep}$, that is $[\max(t^j, \IcBarStart+\delta), \min(t^\mathit{dep}, \IcBarEnd))$, lies within $[t^j, t^\mathit{dep})$ during which $j$ occupies $\waitat{\wait^j}$. The vertex constraint $\langle j, \waitat{\wait^j}, [\IcBarStart + \delta, \IcBarEnd) \rangle$ removes it from $\Tset_{\waitat{\wait^j}}$.
            \item The part at or after $t^\mathit{dep}$, that is, $[t^\mathit{dep}, \IcBarEnd)$, is removed from $\Tset_{\move}$ for the subsequent move action $\langle \move, t^\mathit{dep} \rangle \in\plan_j$ with $\movefrom{\move} = \waitat{\wait^j}$, by the motion constraint $\langle j, \move, [\IcBarStart + \delta, \IcBarEnd) \rangle$.
            Since $t^\mathit{dep} = s_\move$ for the safe interval $[s_\move, e_\move)\subseteq P_\move$ identified during the CSIPP search, this removal is a prefix of that safe interval, exactly as in Section~\ref{sec:Proposed:termination:move_reductions}.
        \end{itemize}
        Either part may be empty, but their union $[\max(t^j, \IcBarStart+\delta), \IcBarEnd)$ is not.
        Its length is at least $\min((1-\gamma)|\IcBar|, |\Ic|)$.
        To see this, consider two cases.
        If $t^j \leq \IcBarStart + \delta$, then the union is $[\IcBarStart+\delta, \IcBarEnd)$ with length $|\IcBar| - \delta \geq (1-\gamma)|\IcBar|$ by~\eqref{eq:delta}.
        Otherwise $t^j > \IcBarStart + \delta$, the union is $[t^j, \IcBarEnd)$, and since $\Ic = [t^j, \min(t^\mathit{dep}, \IcBarEnd))$ in this case, the length is at least $|\Ic|$.
        Both quantities are bounded away from zero by the assumptions in Section~\ref{sec:Preliminaries:finite-precision}.
        \item If $\wait^j$ is an infinite wait action then $[t^j, t^j + \wait_\duration^j] = [t^j, \infty)$. 
        From earlier we know that $t^j < \IcBarEnd$ by an amount bounded away from zero.
        This implies that the overlap $[\max(t^j, \IcBarStart), \IcBarEnd)$ between $[t^j, \infty)$ and $\IcBar$ has a length bounded away from zero.
        Recalling that the constrained interval $[\IcBarStart + \delta, \IcBarEnd)$ also ends at $\IcBarEnd$ and is also bounded away from zero in length, it then holds that $[t^j, \infty)$ and $[\IcBarStart + \delta, \IcBarEnd)$ overlap by an amount bounded away from zero.
        Therefore, $\langle \wait^j, t^j \rangle$ does not satisfy the vertex constraint $\langle j, \waitat{\wait^j}, [\IcBarStart + \delta, \IcBarEnd) \rangle$ created by \branchingOurs and the length of the overlap between the wait action and the constraint interval is bounded away from zero.
        The meaning of this in the context of our termination proof will be made clear in the next section.
    \end{itemize}
}

\subsubsection{Bounded Plan Impossibility}
\label{sec:Proposed:termination:short_plans}

Consider a CT node $\CTnode\in\CTNodeSequence$ mapping to $\trajectory^\infty$, meaning that a plan $\plan\in\CTnode_\Plans$ with $\plan\sim\trajectory^\infty$ is branched on.
To satisfy the non-termination condition~\ref{proposal:super_optimality}, 
\New{
$\plan_\duration \leq c$ with $c=B(\objective^*)$ as established in Section~\ref{sec:Proposed:non-termination_requirement:constraint_accumulation}.}
For simplicity, we suffice with the less restrictive requirement that every timed move action must \emph{begin} before $c$.
In other words, under the constraints $\CTnode_\Constraints$ and $\plan_\duration\leq c$ it holds that 
\begin{equation}
    \forall\move\in\trajectory^\infty: P_\move\subseteq[0,c). \label{eq:inftraj_move_times}
\end{equation}
\New{
    Likewise, for every vertex $\vertex$ visited in $\trajectory^\infty$ that is followed by a move action $\move\in\trajectory^\infty$, the departure time from $\vertex$ lies in $P_\move \subseteq [0, c)$. The reductions of $\Tset_\vertex$ identified in Section~\ref{sec:Proposed:termination:wait_reductions} therefore also lie within $[0, c)$. 
}

\NewPrecision{
    In the previous section, we concluded that applying \branchingOurs to an infinite wait action $\langle\wait^j_n, t^j_n\rangle\in\plan_j$ in a move-wait conflict yields a vertex constraint with an interval $[\IcBarStart + \delta, \IcBarEnd)$ that overlaps with $[t^j_n, t^j_n + \wait_{n,\duration}^j] = [t^j_n, \infty)$ by an amount bounded away from zero.
    To satisfy this constraint while still waiting indefinitely at $\waitat{\wait_n}$, agent $j$ must arrive at $\waitat{\wait_n}$ no earlier than $\IcBarEnd$.
    This invalidates the preceding move action $\langle \move^j_{n-1}, t^j_{n-1} \rangle\in\plan_j$, which would have delivered the agent at time $t^j_n < \IcBarEnd$.
    Instead, $\move^j_{n-1}$ must start no earlier than $\IcBarEnd - \move^j_{n-1,\duration}$.
    Since $\move^j_{n-1}$ is the last move action in $\trajectory^\infty$, this delay results in an increase in the duration of all plans $\plan'\sim\trajectory^\infty$ by an amount bounded away from zero.
}


\NewPrecision{
    In Section~\ref{sec:Proposed:termination:move_reductions} it was shown that applying \branchingOurs to a move action $\langle \move,t\rangle\in\plan$ in a move-move or move-wait conflict removes from $P_\move$ either a prefix whose length is bounded away from zero, or an entire maximally connected interval.
    Section~\ref{sec:Proposed:termination:wait_reductions} showed that applying \branchingOurs to a finite wait action removes a time interval whose length is bounded away from zero from $\Tset_{\waitat{\wait}}$, from $P_\move$ for the subsequent move action $\move$, or partly from each. 
    Finally, applying \branchingOurs to an infinite wait action increases the duration of all plans $\plan'\sim\trajectory^\infty$ by an amount bounded away from zero.

    \New{
    By~\eqref{eq:CSIPP}, $P_\move$ depends on the arrival time $t^a$ and the safe interval end $e_\vertex$ of the plan at the node under consideration, and is therefore not the same set at every node of $\CTNodeSequence$. 
    The plans branched on at the nodes mapping to $\trajectory^\infty$ reach $\movefrom{\move}$ through the same sequence of move actions, and since constraints only accumulate along $\CTNodeSequence$, this arrival time cannot decrease and $e_\vertex$ cannot increase.
    The left endpoints of the maximally connected intervals of $P_\move$ therefore do not move earlier along $\CTNodeSequence$, so the reductions established in Sections~\ref{sec:Proposed:termination:move_reductions} and~\ref{sec:Proposed:termination:wait_reductions} accumulate rather than being undone.  
    }
    
    Collect the finitely many sets $P_\move$ for $\move \in \trajectory^\infty$ together with the sets $\Tset_\vertex \cap [0,c)$ for the vertices $\vertex$ visited in $\trajectory^\infty$.
    By~\eqref{eq:inftraj_move_times} and the observation above, 
    all the reductions identified take place within this collection, whose total length is finite.
    Every branching of \branchingOurs therefore does one of three things:
    (i) removes from the collection a time interval whose length is bounded away from zero,
    (ii) removes an entire maximally connected interval from some $P_\move$, or
    (iii) increases the duration of all plans $\plan' \sim \trajectory^\infty$ by an amount bounded away from zero.
    Case (i) can occur only finitely often, since the total length is finite. Each such removal splits at most one maximally connected interval into two, so only finitely many maximally connected intervals are ever created; 
    case (ii) removes one and creates none, and can therefore also occur only finitely often. 
    Case (iii) can occur only finitely often before the duration of every plan $\plan'\sim\trajectory^\infty$ exceeds $c=B(\objective^*)$.
    Hence, after a finite number of branchings, no plan $\plan\sim\trajectory^\infty$ can exist that satisfies both the accumulated constraints from these branchings and the plan duration bound $\plan_\duration \leq c$.
}

To conclude, the requirement of non-termination that was established in Section~\ref{sec:Proposed:non-termination_requirement} cannot be satisfied by \New{\OCCBS}.
Therefore, termination is guaranteed after a finite number of iterations.

\subsection{\NewPrecision{Exactness} and Solution Completeness}
\label{sec:proposed:final}

Based on the results from the previous sections, we finalize our \New{\OCCBS} correctness proofs.
\begin{theorem}[\New{\OCCBS} correctness]
\New{\OCCBS} is \NewPrecision{exact} and solution complete.
\end{theorem}
\begin{proof}
    \branchingOurs was shown in Section~\ref{sec:proposal:soundness} to be sound, which means by Theorem~\ref{theorem:soundness} that \New{\OCCBS} is \NewPrecision{exact, and therefore also sound.}
    Additionally, Section~\ref{sec:Proposed:termination} showed that \New{\OCCBS} terminates within finite iterations. 
    With these soundness and termination guarantees, Theorem~\ref{theorem:solution-completeness} establishes that \New{\OCCBS} is solution complete.
\end{proof}

\section{Experimental Evaluation}
\label{sec:Experiments}


We perform experiments with the publicly available CCBS \New{and our \OCCBS, the difference being that CCBS uses the \branchingImplementation while \OCCBS uses \branchingOurs}.
Both versions are available in our repository\footnote{\GithubOurs} along with details in \ref{sec:ImplementationDetails} regarding necessary modifications to the original code base.
A counterexample is introduced in Section~\ref{sec:Experiments:counter example} for which \New{\OCCBS} produces an optimal solution and \New{CCBS} returns a sub-optimal solution, along with explanations as to why this occurs.
Thereafter, we compare the solution quality and runtime for the two methods on benchmark problems in Section~\ref{sec:Experiments:Comparison}.
Finally, in Section~\ref{sec:Experiments:Ablation} we perform an ablation study on the parameter $\gamma$ used in \branchingOurs, motivating the use of $\gamma=0.9$ unless stated otherwise.
We use circular agents with radius~$\sqrt{2}/4$ that traverse edges in straight lines at constant speed~$1$\New{, and optimize over SOC}.
All experiments are run on a 2025 Mac Studio, 16-core M4~Max CPU, 64~GB RAM, macOS Sequoia~(15.3).

\subsection{A Counterexample}
\label{sec:Experiments:counter example}

\begin{figure}
    \centering
    \includegraphics[width=0.75\linewidth]{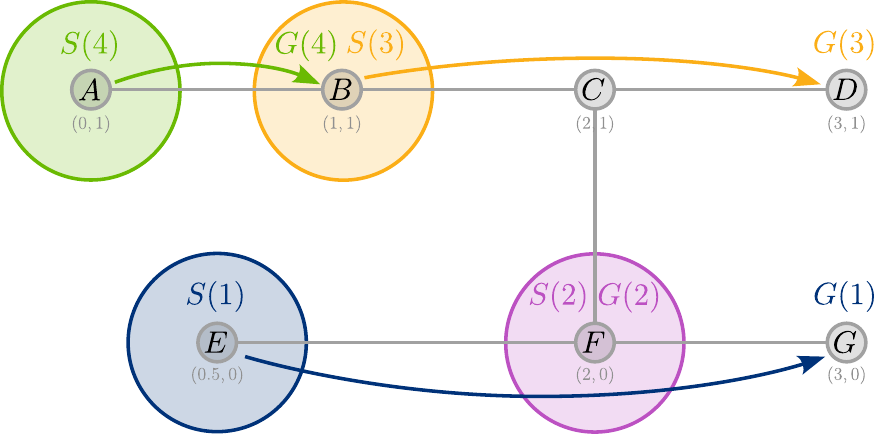}
    \caption{A \MAPFR problem with four agents with radius $r=\sqrt{2}/4$. Each agent $i\in\{1,2,3,4\}$ is shown at its start vertex $\Start(i)$ and must reach its goal vertex $\Goal(i)$ (also depicted by an arrow) without colliding with any other agent.}
    \label{fig:counterexample}
\end{figure}

The example in Figure~\ref{fig:counterexample} contains four agents, $i\in\{1,2,3,4\}$, each with a start vertex $\Start(i)$ and goal vertex $\Goal(i)$.
In this example we use $\gamma=0.5$.
Agent $1$ starts at vertex $\Start(1)=E$ and must pass vertex $F$ to reach its goal $\Goal(1)=G$, therefore, agent~$2$ with $\Start(2)=\Goal(2)=F$ must move out of the way. 
The solution found by both \New{CCBS} and \New{\OCCBS} involve agent~$2$ moving to $C$ to allow agent~$1$ to pass. 
However, agent~$2$ must avoid a conflict with agent~$3$ which passes vertex $C$ on its way to its goal vertex $D$. 
So, there are two options for the interaction between agents~$2$ and~$3$ at $C$:
\begin{enumerate}
    \item agent~$2$ waits for agent~$3$, in turn meaning that agent~$1$ must wait for agent~$2$, or
    \item agent~$3$ waits for agent~$2$, meaning that agent~$4$ must wait for agent~$3$.
\end{enumerate}
Due to the difference in length of edges $EF$ and $AB$, the time that agent~$4$ must wait for agent~$3$ is larger than the time that agent~$1$ must wait for agent~$2$.
Therefore, these two possible solutions have different SOC, and only the one where agent~$2$ waits for agent~$3$ is optimal.

We now examine how \New{CCBS} handles this problem. 
At the CT root $\CTnode^R$, with $\CTnode^R_\Constraints=\varnothing$, each agent's initial plan is generated without conflict avoidance:
\begin{equation*}
    \CTnode^R_\Plans =  \left\{
    \begin{split}
        \plan_1 &= \langle \langle EF, 0 \rangle\; \langle FG, 1.5 \rangle\; \langle G, 2.5, \infty\rangle \rangle \\
        \plan_2 &= \langle \langle F, 0, \infty \rangle\rangle \\
        \plan_3 &= \langle \langle BC, 0\rangle\; \langle CD, 1 \rangle\; \langle D, 2, \infty \rangle\rangle\\
        \plan_4 &= \langle \langle AB, 0 \rangle\; \langle B, 1, \infty \rangle\rangle
    \end{split}
    \right\}
\end{equation*}
where $\langle VU, t \rangle$ denotes a timed move action from $V$ to $U$ starting at time $t$, and $\langle V, t_1, t_2\rangle$ is a timed wait action at vertex $V$ from $t_1$ to $t_2$. 
Two conflicts appear in $\CTnode^R_\Plans$: $\langle \langle EF, 0 \rangle,  \langle F, 0, \infty \rangle\rangle$ and $ \langle \langle FG, 1.5 \rangle, \langle F, 0, \infty \rangle \rangle$, both between agents $1$ and $2$.
\New{CCBS} branches on the first conflict, with 
intersection interval $\IcBar=[0.793, 1.5)$.
This yields the constraints $\langle 1, EF, [0, \infty) \rangle$ (forbidding agent $1$ from ever traversing $EF$) and $\langle 2, F, \IcBar \rangle$ (forbidding agent $2$ from occupying $F$ during $\IcBar$). 
The first constraint is infeasible, as agent $1$ must traverse $EF$ to reach its goal $G$.
Thus, agent $2$ must vacate $F$ before $t=0.793$. 
However, the earliest time that agent $2$ can traverse $FC$ while allowing agent $3$ to pass first is at $t=1$, which is not possible due to the constraint.
Therefore, all valid solutions where agent $2$ waits for agent $3$ are removed from the search.
Consequently, agent $3$ must wait for agent $2$ to complete its detour via $C$.
The resulting solution is
\begin{equation*}
    \Plans =  \left\{
    \begin{split}
        \plan_1 &= \langle \langle EF, 0 \rangle\; \langle FG, 1.5 \rangle\; \langle G, 2.5, \infty\rangle \rangle \\
        \plan_2 &= \langle \langle FC, 0 \rangle\; \langle C, 1, 1.5\rangle\; \langle CF, 1.5 \rangle\; \langle F, 2.5, \infty \rangle\rangle \\
        \plan_3 &= \langle \langle B, 0, 1.5 \rangle\; \langle BC, 1.5\rangle\; \langle CD, 2.5 \rangle\; \langle D, 3.5, \infty \rangle\rangle\\
        \plan_4 &= \langle\langle A, 0, 1.2 \rangle\; \langle AB, 1.2 \rangle\; \langle B, 2.2, \infty \rangle\rangle
    \end{split}
    \right\}
\end{equation*}

We now examine how \New{\OCCBS} handles the same conflict. 
The conflict $\langle \langle EF, 0 \rangle,  \langle F, 0, \infty \rangle\rangle$ 
with intersection interval $\IcBar = [0.793, 1.5)$) yields
\begin{equation*}
    \delta = \min\left( 0.5\cdot\,|[0.793, 1.5)|, \infty \right) = 0.354
\end{equation*}
and the following set of constraints:
\begin{itemize}
    \item Agent $1$: motion constraint $\langle 1, EF, [0, 0.354) \rangle$,
    \item Agent $2$: vertex constraint $\langle 2, F, [1.147, 1.5) \rangle$ and motion constraints $\langle 2, FE, [1.147, 1.5)\rangle$, $\langle 2, FC, [1.147, 1.5)\rangle$, $\langle 2, FG, [1.147, 1.5)\rangle$.
\end{itemize}
Unlike the constraints produced by \New{CCBS}, these constraints exclude only infeasible behaviors while preserving feasible solutions, such as a solution where agent $2$ waits for agent $3$ to pass vertex $C$. 
The resulting solution is
\begin{equation*}
    \Plans =  \left\{
    \begin{split}
        \plan_1 &= \langle \langle E, 0, 0.5 \rangle\langle EF, 0.5 \rangle \langle FG, 2 \rangle \langle G, 3, \infty\rangle \rangle \\
        \plan_2 &= \langle \langle F, 0, 1 \rangle\langle FC, 1 \rangle \langle CF, 2\rangle \langle F, 3, \infty \rangle \rangle \\
        \plan_3 &= \langle \langle BC, 0\rangle \langle CD, 1 \rangle \langle D, 2, \infty \rangle\rangle\\
        \plan_4 &= \langle \langle AB, 0 \rangle\langle B, 1, \infty \rangle\rangle
    \end{split}
    \right\}
\end{equation*}
where agent $2$ waits at $F$ until agent $3$ has traversed past $C$.

The performance using each method is reported in Table~\ref{tab:performance_comparison},
showing that \New{\OCCBS} yields a solution with lower SOC, makespan, and computation time than \New{CCBS}.
This provides additional evidence to that in~\cite{CCBS_revisit} and Section~\ref{sec:ValidationIncompleteness} that \New{CCBS} is not sound.
Moreover, the theoretical results in Section~\ref{sec:ModifyingCCBS} ensure that the solution found by \New{\OCCBS} has optimal SOC. 
However, we further strengthen this claim by validating the optimality of \New{\OCCBS}'s solution by constructing a tailor-made Satisfiability Modulo Theory (SMT) model for this specific problem and solving it using the SMT solver Z3~\cite{de2008z3}, confirming after roughly $26$~hours of computation that the solution returned by \New{\OCCBS} is indeed optimal.
Details regarding this validation can be found in \ref{sec:SMTValidation}, and animations of \New{CCBS} and \New{\OCCBS}'s solutions can be found in the provided repository.
\begin{table}[htbp]
    \centering
    \caption{Performance comparison between the \New{CCBS} and \New{\OCCBS} on the example in Figure~\ref{fig:counterexample}.}
    \label{tab:performance_comparison}
    \begin{tabular}{@{}lccc@{}}
        \toprule
        \textbf{Method} & \textbf{Sum-of-Costs} & \textbf{Makespan} & \textbf{Computation Time} \\
        \midrule
        \New{CCBS} & $10.707$ & $3.500$ & $7.1$ ms \\
        \New{\OCCBS} & $9.000$ & $3.000$ & $1.0$ ms \\
        \bottomrule
    \end{tabular}
\end{table}

\subsection{Benchmark Comparison}
\label{sec:Experiments:Comparison}

We compare \New{CCBS} and \New{\OCCBS} on the same \emph{$2^k$-neighborhood gridmaps} from~\cite{SturtevantBenchmarks} and \emph{roadmaps} benchmark sets from~\cite{CCBS}, and additionally on our own randomly generated \emph{gridlike roadmaps}.
The same benchmarking scheme as suggested in~\cite{SurveyStern2019} is used here:
for each map, a \emph{scenario} defines a list of start and goal vertex pairs. 
We then create a problem with $n$ agents using the first $n$ start and goal vertex pairs. 
For each map and scenario, we begin by solving for $n=2$, and repeatedly incrementing $n$ until a problem cannot be solved within a time limit of $30$ seconds. 

\begin{figure}
    \centering
    \begin{subfigure}[b]{0.24\linewidth}
        \centering
        \includegraphics[width=1\textwidth]{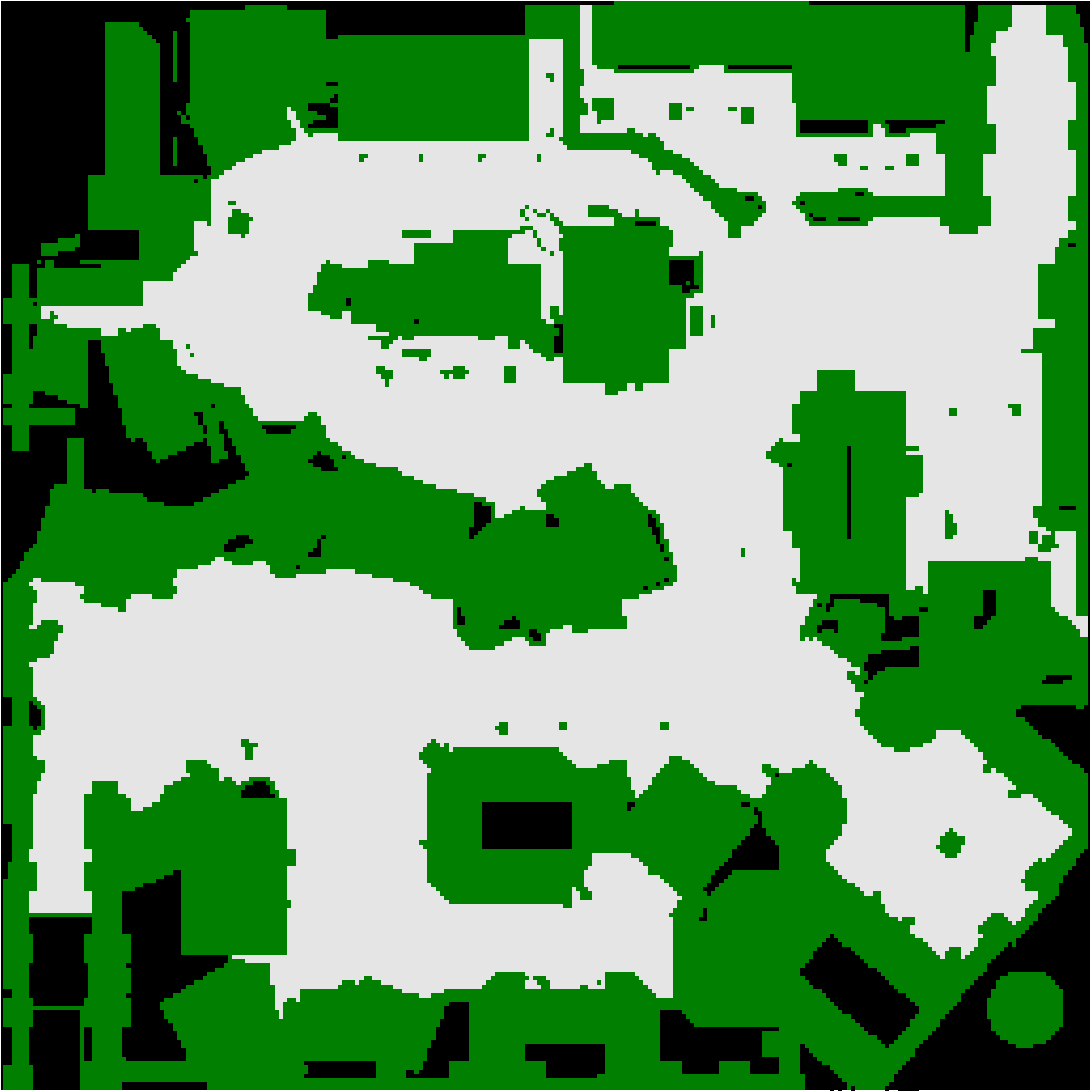}
        \caption{Den520d: $|\Vertices|=28\,178$.}
        \label{fig:gridmaps_1}
    \end{subfigure}
    \hfill
    \begin{subfigure}[b]{0.24\linewidth}
        \centering
        \includegraphics[width=1\textwidth]{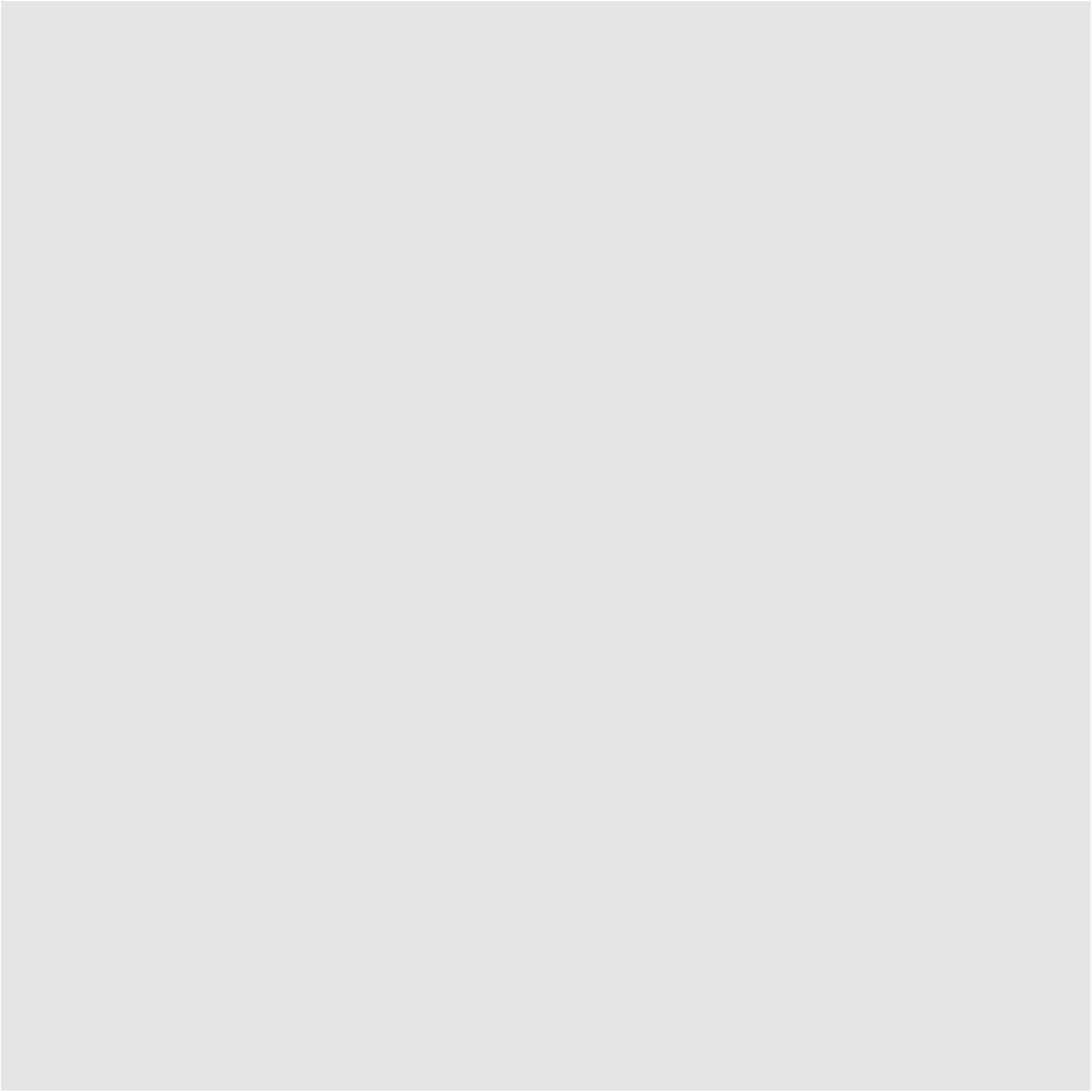}
        \caption{Empty: $|\Vertices|=256$}
        \label{fig:gridmaps_4}
    \end{subfigure}
    \hfill
    \begin{subfigure}[b]{0.24\linewidth}
        \centering
        \includegraphics[width=1\textwidth]{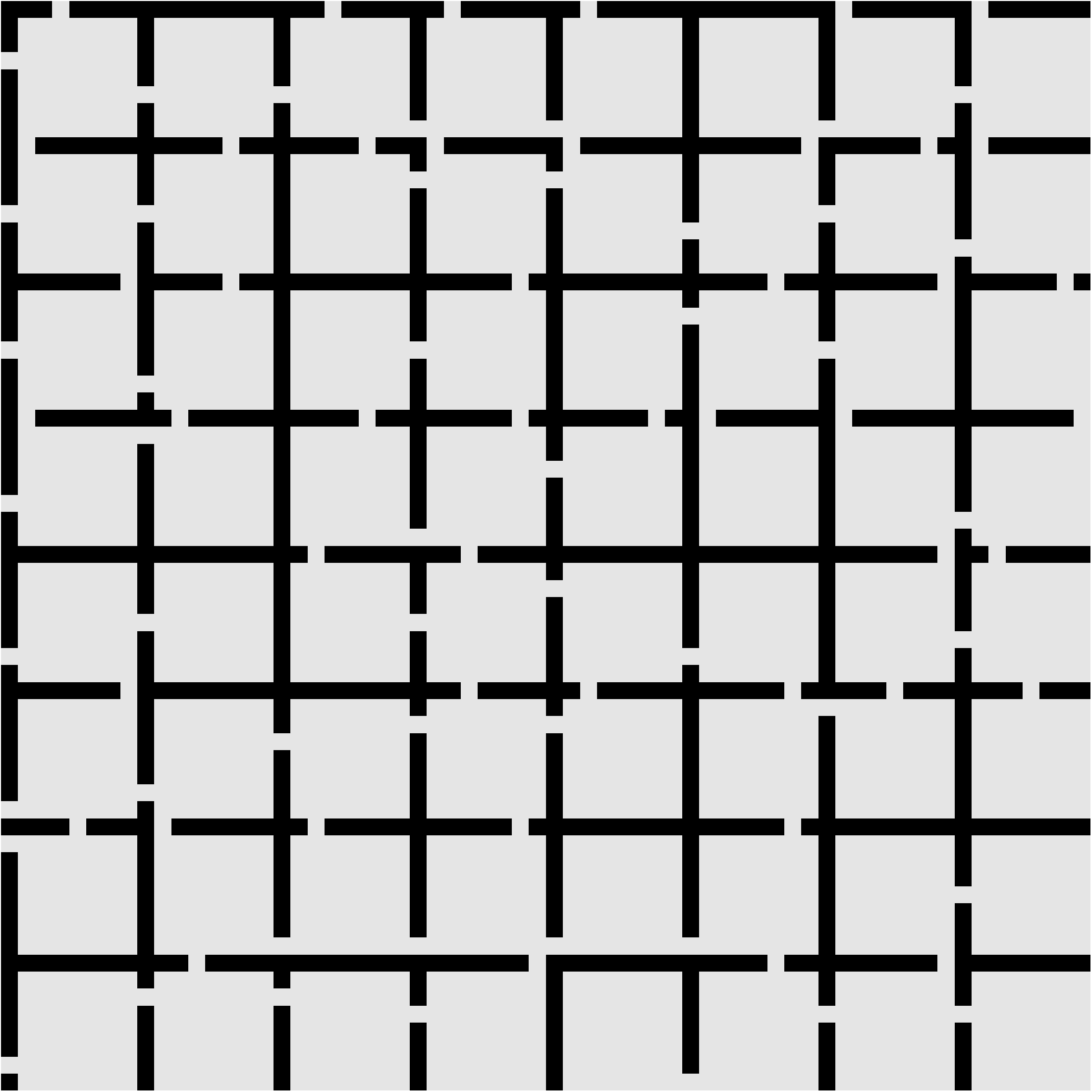}
        \caption{Room: $|\Vertices|=3\,232$}
        \label{fig:gridmaps_3}
    \end{subfigure}
    \hfill
    \begin{subfigure}[b]{0.24\linewidth}
        \centering
        \includegraphics[width=1\textwidth]{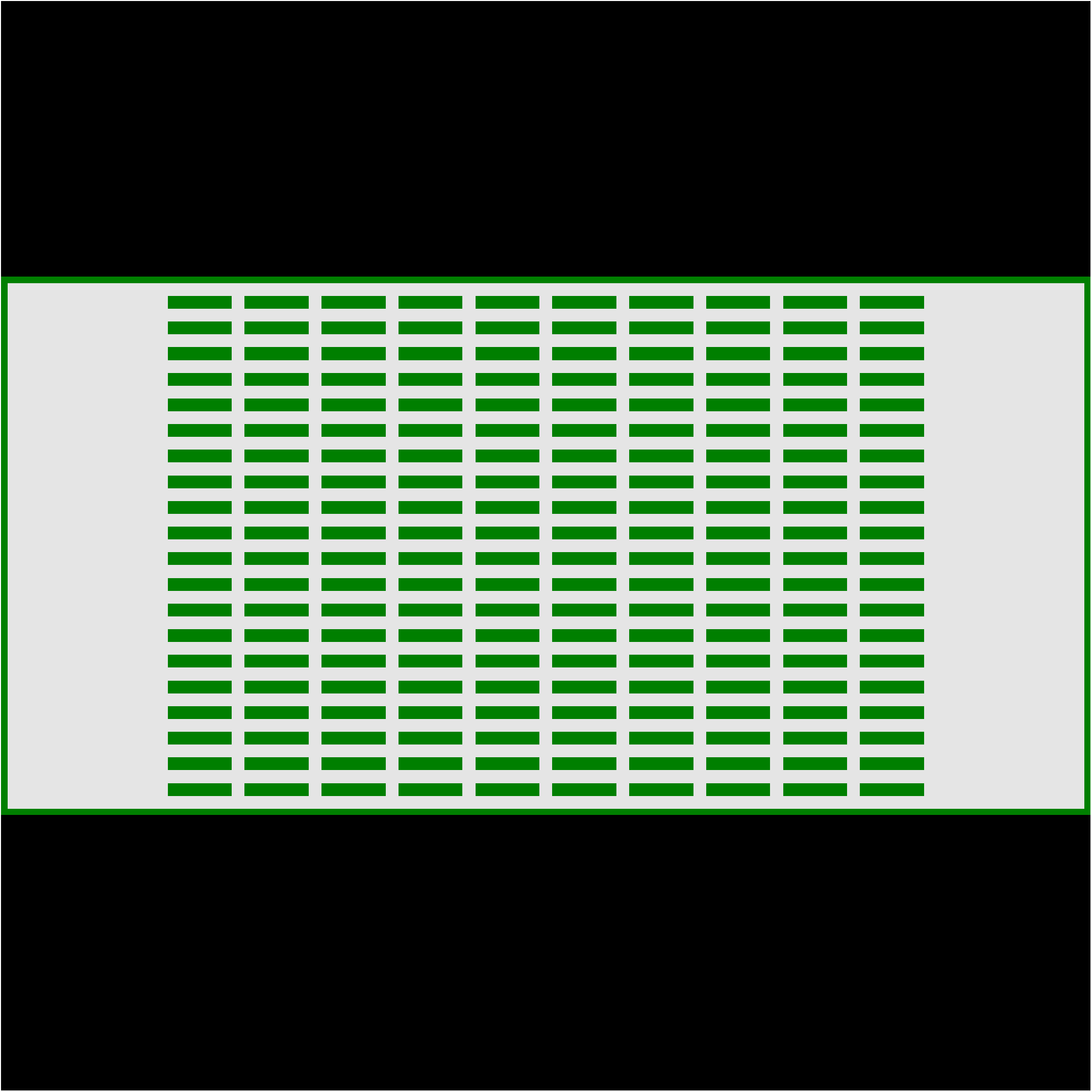}
        \caption{Warehouse: $|\Vertices|=9\,776$}
        \label{fig:gridmaps_2}
    \end{subfigure}
    \caption{\New{The four gridmaps on which \OCCBS and CCBS are benchmarked~\cite{SurveyStern2019}. The roadmaps from~\cite{CCBS} are based on \emph{Den520d}.}}
    \label{fig:gridmaps}
\end{figure}

The $2^k$-neighborhood gridmaps define a grid of traversable and blocked cells. Each traversable cell is connected to its $2^k$ neighborhood, provided that a traversal can be done without colliding with a blocked cell. 
We test with the same parameters as in~\cite{CCBS}: 
$k=2,\dots,5$ with each of the maps \emph{Den520d}, \emph{Empty-16-16}, \emph{Room-64-64-8}, and \emph{Warehouse-10-20-10-2-2} (Figure~\ref{fig:gridmaps}).
The roadmaps are based on the overall shape of Den520d with three levels of density: \New{\emph{Sparse} ($|\Vertices|=169$, $|\Edges|=349$), \emph{Dense} ($|\Vertices|=878$, $|\Edges|=7\,341$), and \emph{Mega-dense} ($|\Vertices|=11\,342$, $|\Edges|=263\,533$)}. 
We refer to~\cite{CCBS} for additional details regarding these maps.

\begin{figure}[H]
    \centering
    \begin{subfigure}[t]{0.48\linewidth}
        \centering
        \includegraphics[width=\textwidth]{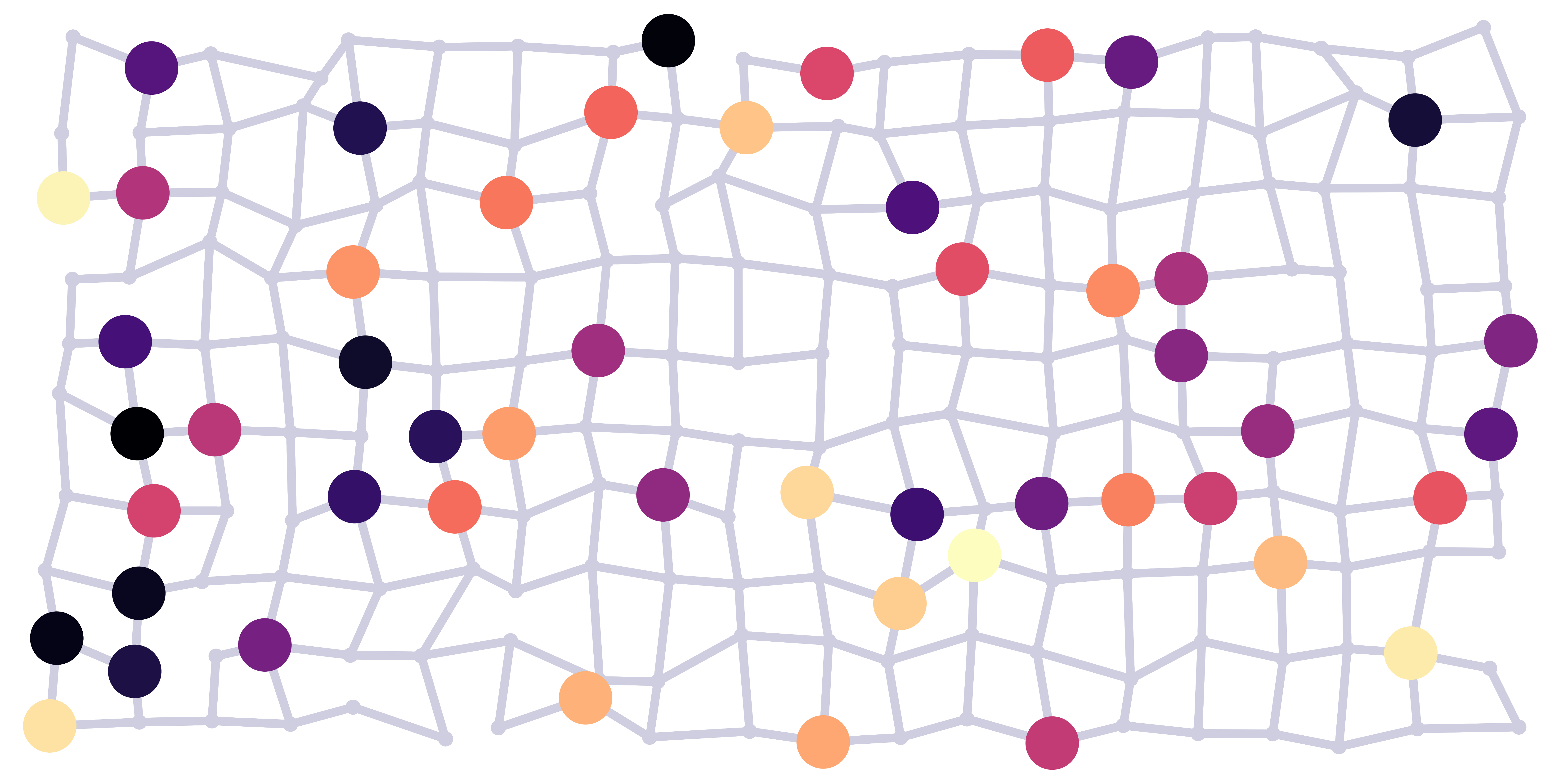}
        \caption{Average vertex degree $3.5$.}
        \label{fig:gridlike_roadmaps_350}
    \end{subfigure}
    \hfill
    \begin{subfigure}[t]{0.48\linewidth}
        \centering
        \includegraphics[width=\textwidth]{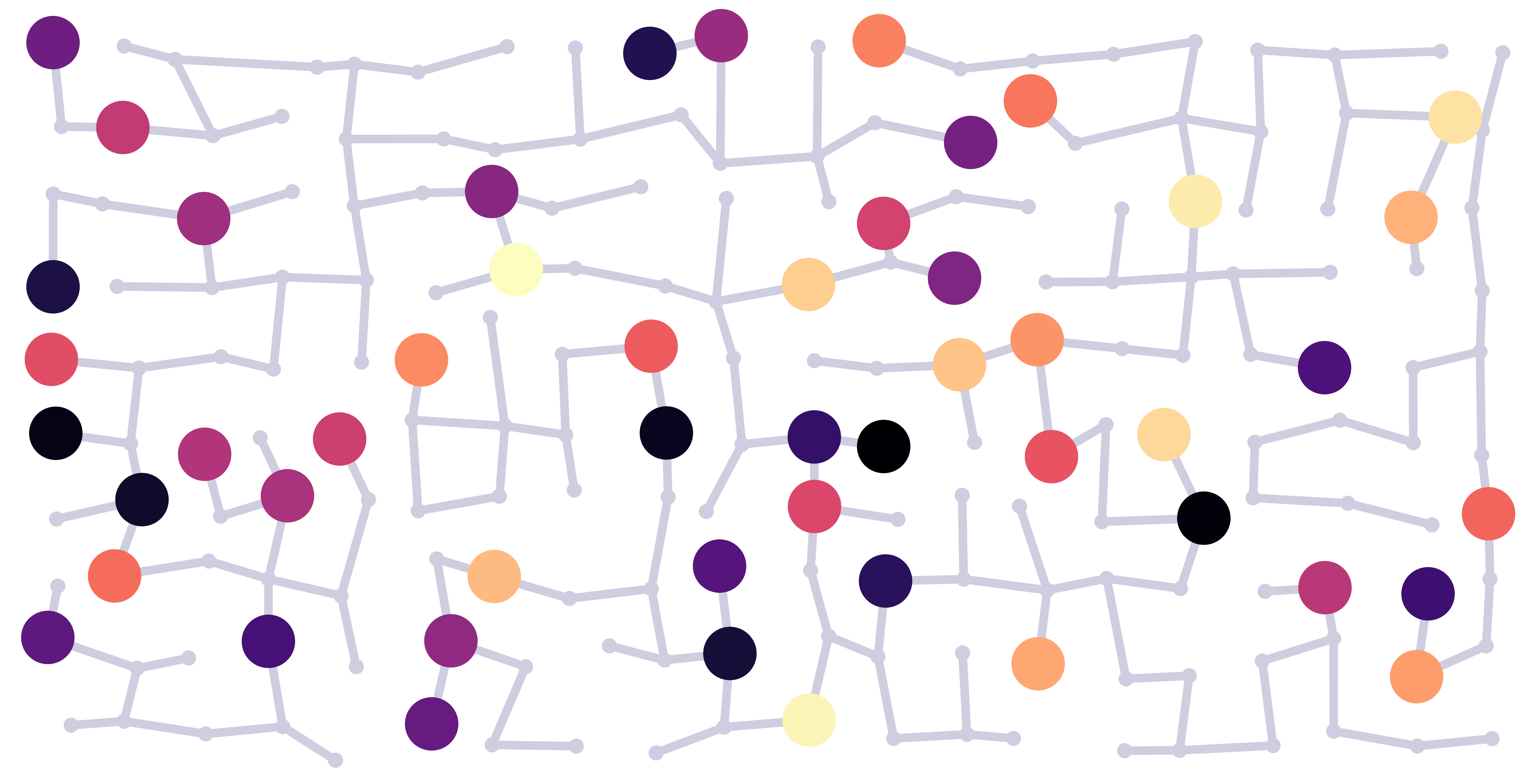}
        \caption{Average vertex degree $2.0$.}
        \label{fig:gridlike_roadmaps_200}
    \end{subfigure}
    \caption{Two gridlike roadmaps with different average degrees and agent start positions.}
    \label{fig:gridlike_roadmaps}
\end{figure}

In addition to the gridmaps and roadmaps, we introduce gridlike roadmaps that are generated with a desired average degree over the map vertices.
Two example maps with average degrees of $2.0$ and $3.5$ are shown in Figure~\ref{fig:gridlike_roadmaps}.
By lowering the average degree, the map becomes more constrained in the number of paths between two vertices. The likelihood of agent interactions consequently increases. 
These maps are generated from a $20$--by--$10$ vertex grid with spacing $1$ by applying zero-mean gaussian noise with standard deviation $0.15$ to the vertex positions, and randomly removing edges until a desired average vertex degree is achieved. Edges are only removed if the resulting graph remains connected. 

\subsubsection{Solution Quality}

On the gridmap and roadmap benchmark sets, we find no difference in solution quality between \New{CCBS} and \New{\OCCBS}; both methods produce equivalent solutions. 
However, on the gridlike roadmap benchmark set we find on $11$ problems that \New{\OCCBS} produces solutions with a lower SOC than \New{CCBS}, and equivalent solutions on all remaining problems.
Although these $11$ problems represent only a fraction of the $15\,462$~problems that both methods could solve within the time limit, it shows that \New{CCBS}'s sub-optimality issues do not only present themselves in tailor-made problems.
These results also hint at the difficulty of producing benchmark problems that fully test the limits of these algorithms. 
Additionally, the $11$~problems that trigger \New{CCBS} to return sub-optimal solutions are on maps with an average node degree distribution toward the lower end of the tested range: $1$ with average degree $2.1$, $3$ with average degree $2.2$, $4$ with average degree $2.4$, and $3$ with average degree $2.8$.
This is perhaps unsurprising, as \New{CCBS}'s unsoundness is due to how it handles move-wait conflicts; with fewer path options due to a lower average degree, the likelihood that the available paths include a vertex with a stationary agent increases.
\New{
    Finally, we note that because CCBS's IBR is unsound, it may repeatedly prune solution-containing regions that would otherwise be explored next by the CT search. More generally, the removed region may contain not only the optimal solution but also other low-cost solutions that would otherwise have been explored next, leaving the child nodes with only non-solution candidates. In such cases, the search may fail to terminate. Such cases (if they occur) are not reflected in our results as we only compare instances on which both methods return a solution within the time limit.}

\subsubsection{Runtime}
\label{sec:experiments:runtime}
\begin{figure}[htbp]
    \centering
    \begin{minipage}[t]{0.48\linewidth}
        \centering
        \includegraphics[width=\textwidth]{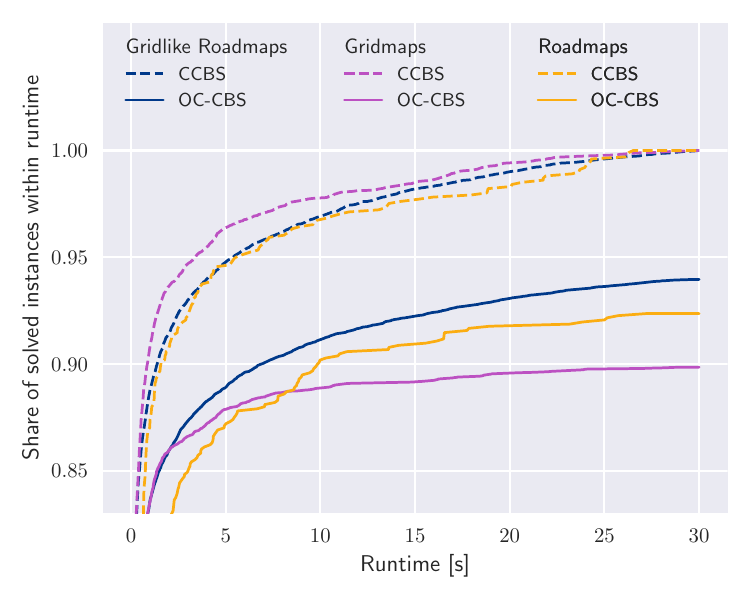}
        \caption{The number of problems solved within $30$ seconds by CCBS and \OCCBS on each of the benchmark sets, normalized by the number of problems solved by CCBS.}
        \label{fig:Comparison:Runtime}
    \end{minipage}
    \hfill
    \begin{minipage}[t]{0.48\linewidth}
        \centering
        \includegraphics[width=\textwidth]{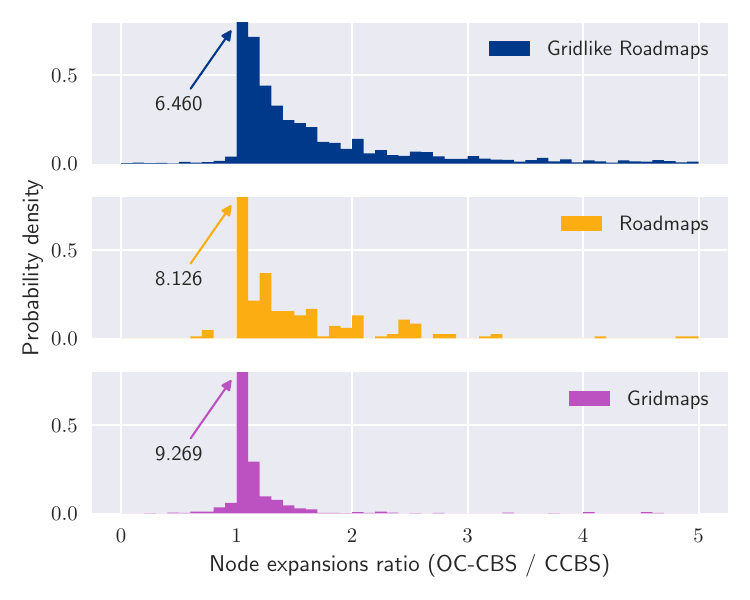}
        \caption{Probability density histogram of the increase in the number of high-level node expansions using \New{\OCCBS} compared to \New{CCBS}. The y-axes are limited to maximum $0.8$; the height of the $(1.0, 1.1)$-bins are denoted specifically.}
        \label{fig:Comparison:Expansions}
    \end{minipage}
\end{figure}

\New{
Figure~\ref{fig:Comparison:SuccessRate} shows the solvers' success rates over agent counts on each map, with the success rate being the share of scenarios that could be solved for the given agent count within the time limit.
\OCCBS lags slightly behind CCBS on all grid maps (top row), 
with the largest difference seen on \emph{Rooms} where the success of \OCCBS starts declining much earlier than CCBS.
We have not investigated why the \emph{Rooms} map in particular triggers such a pronounced performance difference between the solvers. 
However, since \OCCBS and CCBS primarily differ in their treatment of move-wait conflicts, we suspect these instances generate a higher density of such conflicts. Under these conditions, CCBS's unsound branching rule aggressively cuts off areas of the search space that \branchingOurs must exhaustively explore through further branching to maintain exactness. 
This divergence causes \branchingOurs to expand far more nodes (Figure~\ref{fig:Comparison:Expansions}) and require additional runtime.
\begin{figure}[]
    \centering
    \includegraphics[width=1\linewidth]{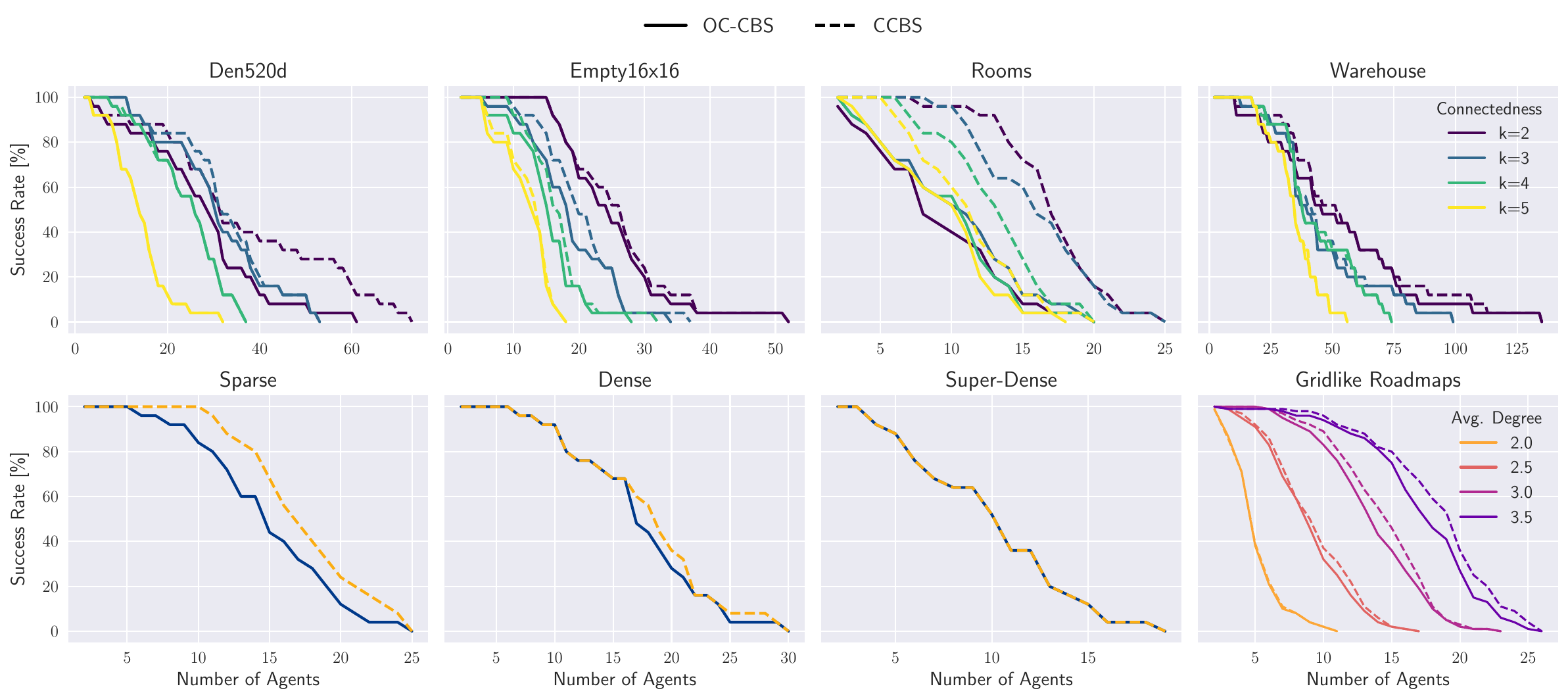}
    \caption{\New{\OCCBS and CCBS's success rate on each map over the agent count. Hatched lines are CCBS. Note that the $x$-axes are differently scaled, due to the large difference in solvable agent count range across the maps.}}
    \label{fig:Comparison:SuccessRate}
\end{figure}

The roadmaps (bottom row, columns 1--3) see a reduced difference between the solvers as the density increases.
Additionally, we see the solvers are able to plan for slightly more agents on \emph{Dense} than on \emph{Sparse}, and much fewer on \emph{Super-Dense}.
As the density increases, so too does the number of paths available for an agent from its start to goal vertex. 
We suspect that this increased freedom reduces the chances of conflicts and makes more possibilities available to resolve them.
The main difference between \OCCBS and CCBS lies in how they handle move-wait conflicts; 
fewer such conflicts reduce the difference in the solvers' performance.
However, the larger graphs come with increased computational overhead. 
The results show the average number of CT node expansions during a scenario's final failed run for \OCCBS and CCBS, respectively: $1.0\times10^5$ and $1.4\times10^5$ on \emph{Sparse},  
$3.2\times10^4$ and $3.7\times10^4$ on \emph{Dense}, and 
$4.9\times10^3$ and $4.9\times10^3$ on \emph{Super-Dense}.
This shows that although a larger map may increase the number of available paths, the computational overhead of handling them leads to fewer iterations of the search being done within a given runtime.

The gridlike roadmaps better isolate the map size from the number of paths available to each agent from its start to goal vertex.
For clarity, Figure~\ref{fig:Comparison:SuccessRate}
, bottom row, column 4, shows the gridlike roadmaps with average degrees $2.0$, $2.5$, $3.0$, and $3.5$.
The average degree serves as a proxy for the number of available paths (see Figure~\ref{fig:gridlike_roadmaps}) without increasing the number of vertices.
As the average degree increases, so too does the number of agents that can be planned by both solvers.

Finally, these results are aggregated in Figure~\ref{fig:Comparison:Runtime}, showing the number of problems that \OCCBS and CCBS can solve within the time limit, for each map type.
To account for different benchmark set sizes, values are normalized by the number of problems CCBS is able to solve.
We see that \OCCBS is slower than CCBS, at worst on gridmaps where around $10$\% fewer problems are solved compared to CCBS.
}
This advantage stems from \New{CCBS}'s more aggressive handling of move-wait conflicts. While this strategy sometimes removes solutions from the search, it often retains at least one optimal solution while reducing the size of the search space.
Consequently, \New{CCBS} generally requires fewer node expansions than \New{\OCCBS}, as illustrated in Figure~\ref{fig:Comparison:Expansions}.
However, \New{CCBS}'s aggressive pruning comes with trade-offs. 
In certain cases, as shown above, all optimal solutions are eliminated, leading to sub-optimal results.
This means that we are effectively comparing an optimal method (\New{\OCCBS}) to a sub-optimal method (\New{CCBS}).

\subsection{Ablation Study on Parameter $\gamma$}
\label{sec:Experiments:Ablation}
The effect of the parameter $\gamma$---used to compute $\delta$ in \New{\OCCBS} (see Definition~\ref{def:proposed_branching})---on solver runtime is investigated by solving a set of benchmark problems with $\gamma\in\left\{0.1, 0.3,\dots, 0.9\right\}$ and recording the number of solved problems within $30$ seconds. 
The benchmark set contains gridlike roadmaps: for every average degree value $2.0, 2.1, \dots, 3.5$, we generate $5$ maps, each with $5$ scenarios. 
To avoid selection bias, this benchmark is separate from the gridlike roadmap set used in Section~\ref{sec:Experiments:Comparison}. 

\begin{figure}[]
    \centering
    \includegraphics[width=0.75\textwidth]{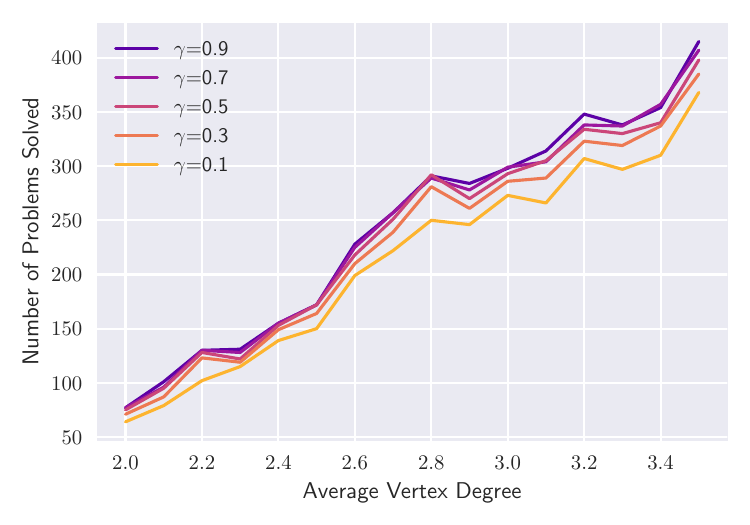}
    \caption{The number of problems solved within a $30$ second time limit for various values of $\gamma$, shown over a range of average vertex degrees on gridlike roadmaps.}
    \label{fig:Ablation}
\end{figure}

The results are reported in Figure~\ref{fig:Ablation}, showing the number of problems that can be solved within $30$ seconds for each value of $\gamma$ value and average vertex degree. 
We note three things:
First, larger $\gamma$ values lead to a decrease in solver runtime on gridlike roadmaps, since more problems are solved as $\gamma$ increases.
Second, it can be seen that more problems are solved within the time limit as the average vertex degree increases. 
As discussed in Section~\ref{sec:experiments:runtime}, this is likely due to the larger number of paths made available to an agent from its start to goal vertex, making these problems less constrained.
\New{
Finally, the difference in performance over the $\gamma$ values grows larger as the average vertex degree increases. 
On the one hand, this is expected if the \emph{proportion} of solved problems across $\gamma$ values remains roughly constant. 
On the other hand, however, a lower average vertex degree should correspond to a more constrained graph and therefore encourage more agent interactions; with more agent interactions, one could expect more move-wait conflicts which are resolved by \branchingOurs using $\gamma$, thereby amplifying the impact $\gamma$ has on the search.
We leave further investigations into these effect for future work.

Note that these results do not provide universal guidance for the best $\gamma$; they should be considered only for the specific map type studied here, gridlike roadmaps.
The effect of $\gamma$ may differ for other map families, obstacle patterns, start-goal configurations, agent geometries, and so on, potentially leading to different conclusions.
}

\section{\New{Conclusions}}
\label{sec:Conclusions}



This work set out to resolve recent findings that CCBS fails to guarantee \NewPrecision{exactness} and solution completeness, leaving a gap in the literature and compromising the guarantees of continuation methods built on it. We formulated a framework for CCBS-like algorithms, established sufficient criteria for \NewPrecision{exactness} and solution completeness, and showed \OCCBS satisfies them. \OCCBS is therefore an \NewPrecision{exact}, solution complete \MAPFR solver under finite-precision arithmetic. Experiments show it competitive with CCBS in runtime, occasionally producing better solutions.

Our implementation further shows that \OCCBS can serve as a drop-in replacement for CCBS in existing systems, instantly restoring any theoretical guarantees that depended on CCBS's now-invalidated correctness. The analytical framework and correctness criteria developed here also provide a reusable basis for evaluating and designing future CCBS-like solvers. We view this contribution both as a correction to the existing CCBS lineage and as a step toward a more rigorous theory of optimal continuous-time multi-agent path finding.

\section{CRediT Authorship Contribution Statement}
\label{sec:AuthorContributions}

\textbf{Alvin Combrink:} Conceptualization, Methodology, Software, Validation, Formal analysis, Investigation, Writing - Original Draft.
\textbf{Sabino Francesco Roselli:} Software, Writing - Review \& Editing, Supervision.
\textbf{Martin Fabian:} Writing - Review \& Editing, Supervision, Funding acquisition.

\section{Declaration of Competing Interest}
\label{sec:CompetingInterestsDeclaration}

Alvin Combrink reports financial support was provided by Sweden’s Innovation Agency. Sabino Francesco Roselli reports financial support was provided by Sweden’s Innovation Agency. Martin Fabian reports financial support was provided by Sweden’s Innovation Agency. If there are other authors, they declare that they have no known competing financial interests or personal relationships that could have appeared to influence the work reported in this paper.

\section{Acknowledgments}
\label{sec:Acknowledgements}
We gratefully acknowledge the Vinnova project CLOUDS (Intelligent algorithms to support Circular soLutions fOr sUstainable proDuction Systems), and the Wallenberg AI, Autonomous Systems and Software Program (WASP) funded by the Knut and Alice Wallenberg Foundation.

\appendix

\section{Implementation Details}

Our implementation of \New{CCBS}\footnote{\GithubOursOriginal} closely follows that of the publicly available CCBS implementation\footnote{\GithubCCBS} from~\cite{CCBS}.
However, minor modifications have been made to ensure a fair comparison between \New{CCBS} and \New{\OCCBS}\footnote{\GithubOurs}. 
For more details than what we describe here, we refer the reader to the provided repository.

First, we do not use any \emph{high-level heuristics} or \emph{disjoint splitting} for \New{CCBS} or \New{\OCCBS}, as these are not the focus of this work. 
Therefore, ensuring compatibility of \New{\OCCBS} with these additional improvements are left for future work.
We do however enable \emph{Cardinal Constraints}.

Second, in the implementation from~\cite{CCBS}, move-wait conflict detection is done using the same method as for move-move conflicts: a binary interval search with a specified precision.
\New{\OCCBS}, however, is highly sensitive to small numerical differences between the detection of move-wait conflicts and the computation of $\IcBar = [\IcBarStart, \IcBarEnd)$.
Specifically, for a move-wait conflict $\langle\langle\move^i, t^i\rangle,\langle\wait^j, t^j\rangle \rangle$ for which $\delta = t^j + \wait_\duration^j - \IcBarStart$ (see Definition~\ref{def:proposed_branching}), the resulting constraint interval used in the constraints on $j$ is $[\IcBarStart + \delta, \IcBarEnd) = [t^j + \wait_\duration^j, \IcBarEnd)$.
This interval overlaps with $j$'s wait action time interval $[t^j, t^j + \wait_\duration^j]$ by only a singular time instant, $t^j + \wait_\duration^j$.
Thus, if the constraint interval is subject to numerical imprecision such that it is computed to $[t^j + \wait_\duration^j + \epsilon, \IcBarEnd)$ for some small $\epsilon>0$, then the resulting constraints using this interval will not forbid $\langle\wait^j, t^j\rangle$ and therefore the detected move-wait conflict may occur again.
Our solution to this is straight-forward: we use $\IcBar$ for both constraint creation and move-wait conflict detection. 
An alternative could be to use a fixed precision value as additional margin on the constraints, however, this could be regarded as a form of discretization which we avoid.  
Besides, since $\IcBar$ is computed analytically (under the assumption that edges are traversed in straight lines, as in all experiments), the conflict detection is not subject to a precision value\New{---as is the case in the binary interval search method---}and is therefore more precise.
Moreover, for the case of general motion functions, \New{\OCCBS}'s numerical sensitivity is easily overcome by using the same computations for conflict detection as for constraint creation (such as by pre-computing intersection intervals with, e.g., a binary interval search).
Therefore, these modifications do not require agents to only traverse edges in straight lines. 
As a consequence of using $\IcBar$ in the conflict detection for \New{\OCCBS}, the same conflict detection is used in \New{CCBS} to avoid any differences in the underlying problem that these two methods solve. 

Finally, the computation of $\IcBar$ is re-implemented for more robustness to numerical imprecision, particularly for near-tangent conflicts where the collision interval is singular or close to singular.

\label{sec:ImplementationDetails}

\section{Validating the Optimality of \New{\OCCBS's Counterexample} Solution}

We validated the optimality of \New{\OCCBS}'s solution to the counterexample in Section~\ref{sec:Experiments:counter example} using a tailor-made Satisfiability Modulo Theory (SMT) model with the SMT solver Z3~\cite{de2008z3}. After roughly $26$~hours of computation, the model returned an equivalent solution with sum-of-costs $9.0$, thereby confirming that \New{\OCCBS}'s solution is an optimal solution. 
The exact implementation and results can be found in our repository\footnote{\GithubSMTimplementation}.

The model requires specifying an exact number of timed actions that each agent executes. 
Thus, the question arises: how many timed actions must be set to ensure that the returned solution is optimal with respect to the \MAPFR problem?
In the following, we will establish that any solution with a given sum-of-costs requires at most a certain number of timed move actions; if any more timed move actions are required by a single agent, then the sum-of-costs of such a solution is necessarily higher than the given sum-of-costs.
We find for this specific problem that any solution requiring more than $5$ timed move actions necessarily has a sum-of-costs $>9.0$. Considering that agents may also execute timed wait actions, we find that the lower bound on the number of timed actions is $11$.
Furthermore, we motivate that there does not exists an upper bound to the number of timed actions.
With these results, it therefore holds that the retained solution (which is equivalent to the solution from \New{\OCCBS}) is optimal.

\newcommand{\durationLB}{d^\mathit{LB}}

Let $\durationLB_i$ denote an optimistic lower bound on the duration of a plan taking agent $i$ from $\Start(i)$ to $\Goal(i)$, which is obtained by not considering potential conflicts with other agents:
\begin{align*}
    \durationLB_1 = 2.5,\qquad \durationLB_2 = 0,\qquad \durationLB_3 = 2.0,\qquad \durationLB_4 = 1.0.
\end{align*}
For solution $\Plans$, let $n$ be the maximum number of timed move actions contained in any single plan $\plan\in\Plans$. 
Since the minimum move action duration over all move actions in the counterexample is $1.0$, a plan with $n$ move actions has duration $\geq n$. 
The sum-of-costs lower bound $\mathit{SOC}^\mathit{LB}_\Plans$ for $\Plans$ occurs when one agent executes $n$ moves while all others follow their optimistically shortest-duration plans:
\begin{align*}
    \mathit{SOC}^\mathit{LB}_\Plans  &= \min_{i=1,..4} \left\{ n + \sum_{j\in\{1,..,4\}\setminus\left\{i\right\}} \durationLB_j \right\} 
    = \min\left\{ 
        \begin{gathered}
            n + \durationLB_2 + \durationLB_3 + \durationLB_4 = n + 3.0 \\
            \durationLB_1 + n + \durationLB_3 + \durationLB_4 = n + 5.5 \\
            \durationLB_1 + \durationLB_2 + n + \durationLB_4 = n + 3.5 \\
            \durationLB_1 + \durationLB_2 + \durationLB_3 + n = n + 4.5
        \end{gathered}
    \right\} = n + 3.0.
\end{align*}
From this, if any solution has sum-of-costs $<9.0$, then $n+3.0 < 9.0$, implying $n \leq 5$.
Since consecutive wait actions can be merged, an agent executing $n$ move actions needs at most $n+1$ wait actions.
Therefore, the maximum number of timed actions per agent is $n+n+1 = 2n+1 = 11$.
Moreover, since wait actions can be split arbitrarily, enforcing that each agent executes exactly $11$ timed actions still allows for an optimal solution to be found. 
\New{The above ensures that allowing each agent a maximum of $11$ timed actions is sufficient to ensure that any solution with a sum-of-costs of $9.0$ or less will be found. }
\label{sec:SMTValidation}


 \bibliographystyle{elsarticle-num} 
 \bibliography{Bib}






\end{document}